\documentclass[twocolumn]{aastex63}
\accepted{for Publication in ApJ}

\usepackage{natbib,aas_macros,amsmath}
\usepackage{multirow,color}
\usepackage{natbib}

\newcommand{\carcsec}{$\!\!\arcsec$}
\newcommand{\m}[1]{\mathrm{#1}}

\newcommand{\redc}[1]{\textcolor{black}{#1}}

\usepackage{booktabs}

\begin{document}
\shortauthors{Harikane et al.}

\shorttitle{
Bright-End of the UVLFs at $z\sim7-14$:
}

\title{
JWST, ALMA, and Keck Spectroscopic Constraints on the UV Luminosity Functions at $\mathbf{z\sim7-14}$:
Clumpiness and Compactness of the Brightest Galaxies in the Early Universe
}


\email{hari@icrr.u-tokyo.ac.jp}
\author[0000-0002-6047-430X]{Yuichi Harikane}
\affiliation{Institute for Cosmic Ray Research, The University of Tokyo, 5-1-5 Kashiwanoha, Kashiwa, Chiba 277-8582, Japan}

\author[0000-0002-7779-8677]{Akio K. Inoue}
\affiliation{Department of Physics, School of Advanced Science and Engineering, Faculty of Science and Engineering, Waseda University, 3-4-1 Okubo, Shinjuku, Tokyo 169-8555, Japan}
\affiliation{Waseda Research Institute for Science and Engineering, Faculty of Science and Engineering, Waseda University, 3-4-1 Okubo, Shinjuku, Tokyo 169-8555, Japan}

\author[0000-0001-7782-7071]{Richard S. Ellis}
\affiliation{Department of Physics and Astronomy, University College London, Gower Street, London WC1E 6BT, UK}

\author[0000-0002-1049-6658]{Masami Ouchi}
\affiliation{National Astronomical Observatory of Japan, 2-21-1 Osawa, Mitaka, Tokyo 181-8588, Japan}
\affiliation{Institute for Cosmic Ray Research, The University of Tokyo, 5-1-5 Kashiwanoha, Kashiwa, Chiba 277-8582, Japan}
\affiliation{Kavli Institute for the Physics and Mathematics of the Universe (WPI), University of Tokyo, Kashiwa, Chiba 277-8583, Japan}

\author[0000-0002-0984-7713]{Yurina Nakazato}
\affiliation{Department of Physics, Graduate School of Science, The University of Tokyo, 7-3-1 Hongo, Bunkyo, Tokyo 113-0033, Japan}

\author[0000-0001-7925-238X]{Naoki Yoshida}
\affiliation{Department of Physics, Graduate School of Science, The University of Tokyo, 7-3-1 Hongo, Bunkyo, Tokyo 113-0033, Japan}

\author[0000-0001-9011-7605]{Yoshiaki Ono}
\affiliation{Institute for Cosmic Ray Research, The University of Tokyo, 5-1-5 Kashiwanoha, Kashiwa, Chiba 277-8582, Japan}

\author[0000-0002-4622-6617]{Fengwu Sun}
\affiliation{Steward Observatory, University of Arizona, 933 N. Cherry Avenue, Tucson, AZ 85721, USA}
\affiliation{Center for Astrophysics $|$ Harvard \& Smithsonian, 60 Garden St., Cambridge, MA 02138, USA}

\author{Riku A. Sato}
\affiliation{Department of Physics, School of Advanced Science and Engineering, Faculty of Science and Engineering, Waseda University, 3-4-1 Okubo, Shinjuku, Tokyo 169-8555, Japan}

\author[0000-0002-2012-4612]{\redc{Giovanni Ferrami}}
\affiliation{School of Physics, University of Melbourne, Parkville, VIC 3010, Australia}
\affiliation{ARC Centre of Excellence for All-Sky Astrophysics in 3 Dimensions (ASTRO 3D)}

\author[0000-0001-7201-5066]{Seiji Fujimoto}
\altaffiliation{Hubble Fellow}
\affiliation{Department of Astronomy, The University of Texas at Austin, Austin, TX, USA}

\author[0000-0003-3954-4219]{Nobunari Kashikawa}
\affiliation{Department of Astronomy, Graduate School of Science, The University of Tokyo, 7-3-1 Hongo, Bunkyo, Tokyo 113-0033, Japan}
\affiliation{Research Center for the Early Universe, The University of Tokyo, 7-3-1 Hongo, Bunkyo-ku, Tokyo 113-0033, Japan}

\author[0000-0003-4368-3326]{Derek J. McLeod}
\affiliation{SUPA\thanks{Scottish Universities Physics Alliance}, Institute for Astronomy, University of Edinburgh, Royal Observatory, Edinburgh EH9 3HJ, UK}

\author[0000-0003-4528-5639]{Pablo G. P\'{e}rez-Gonz\'{a}lez}
\affiliation{Centro de Astrobiolog\'{\i}a (CAB/CSIC-INTA), Ctra. de Ajalvir km 4, Torrej\'on de Ardoz, E-28850, Madrid, Spain}

\author[0000-0002-7712-7857]{Marcin Sawicki}
\affiliation{Department of Astronomy and Physics and the Institute for Computational Astrophysics, Saint Mary's University, 923 Robie Street, Halifax, NS B3H 3C3, Canada}

\author[0000-0001-6958-7856]{Yuma Sugahara}
\affiliation{Waseda Research Institute for Science and Engineering, Faculty of Science and Engineering, Waseda University, 3-4-1 Okubo, Shinjuku, Tokyo 169-8555, Japan}
\affiliation{Department of Physics, School of Advanced Science and Engineering, Faculty of Science and Engineering, Waseda University, 3-4-1 Okubo, Shinjuku, Tokyo 169-8555, Japan}

\author[0000-0002-5768-8235]{Yi Xu}
\affiliation{Institute for Cosmic Ray Research, The University of Tokyo, 5-1-5 Kashiwanoha, Kashiwa, Chiba 277-8582, Japan}
\affiliation{Department of Astronomy, Graduate School of Science, The University of Tokyo, 7-3-1 Hongo, Bunkyo, Tokyo 113-0033, Japan}

\author[0000-0002-7738-5290]{Satoshi Yamanaka}
\affiliation{General Education Department, National Institute of Technology, Toba College, 1-1, Ikegami-cho, Toba, Mie 517-8501, Japan}

\author[0000-0002-1482-5818]{Adam C. Carnall}
\affiliation{SUPA\thanks{Scottish Universities Physics Alliance}, Institute for Astronomy, University of Edinburgh, Royal Observatory, Edinburgh EH9 3HJ, UK}

\author[0000-0002-3736-476X]{Fergus Cullen}
\affiliation{SUPA\thanks{Scottish Universities Physics Alliance}, Institute for Astronomy, University of Edinburgh, Royal Observatory, Edinburgh EH9 3HJ, UK}

\author[0000-0002-1404-5950]{James S. Dunlop}
\affiliation{SUPA\thanks{Scottish Universities Physics Alliance}, Institute for Astronomy, University of Edinburgh, Royal Observatory, Edinburgh EH9 3HJ, UK}

\author[0000-0003-1344-9475]{Eiichi Egami}
\affiliation{Steward Observatory, University of Arizona, 933 N. Cherry Avenue, Tucson, AZ 85721, USA}

\author[0000-0001-9440-8872]{Norman Grogin}
\affiliation{Space Telescope Science Institute, 3700 San Martin Drive, Baltimore, MD 21218, USA}

\author[0000-0001-7730-8634]{Yuki Isobe}
\affiliation{Waseda Research Institute for Science and Engineering, Faculty of Science and Engineering, Waseda University, 3-4-1 Okubo, Shinjuku, Tokyo 169-8555, Japan}

\author[0000-0002-6610-2048]{Anton M. Koekemoer}
\affiliation{Space Telescope Science Institute, 3700 San Martin Drive, Baltimore, MD 21218, USA}

\author[0000-0001-7459-6335]{Nicolas Laporte}
\affiliation{Aix-Marseille Universit\'e, CNRS, CNES, LAM (Laboratoire d’Astrophysique de Marseille), UMR 7326, 13388 Marseille, France}

\author[0000-0003-1700-5740]{Chien-Hsiu Lee}
\affiliation{Hobby-Eberly Telescope, McDonald Observatory, 32 Mt. Fowlkes, Fort Davis, TX 79734, USA}

\author[0000-0002-6668-2011]{Dan Magee}
\affiliation{Department of Astronomy and Astrophysics, UCO/Lick Observatory, University of California, Santa Cruz, CA 95064, USA}

\author[0000-0002-4559-6157]{Hiroshi Matsuo}
\affiliation{National Astronomical Observatory of Japan, 2-21-1 Osawa, Mitaka, Tokyo 181-8588, Japan}
\affiliation{The Graduate University for Advanced Studies, SOKENDAI, 2-21-1 Osawa,
Mitaka, Tokyo 181-8588, Japan}

\author[0000-0001-5063-0340]{Yoshiki Matsuoka}
\affiliation{Research Center for Space and Cosmic Evolution, Ehime University,
Matsuyama, Ehime 790-8577, Japan}

\author[0000-0003-4985-0201]{Ken Mawatari}
\affil{Waseda Research Institute for Science and Engineering, Faculty of Science and Engineering, Waseda University, 3-4-1 Okubo, Shinjuku, Tokyo 169-8555, Japan}

\author[0000-0003-2965-5070]{Kimihiko Nakajima}
\affiliation{National Astronomical Observatory of Japan, 2-21-1 Osawa, Mitaka, Tokyo 181-8588, Japan}

\author[0009-0000-1999-5472]{Minami Nakane}
\affiliation{Institute for Cosmic Ray Research, The University of Tokyo, 5-1-5 Kashiwanoha, Kashiwa, Chiba 277-8582, Japan}
\affiliation{Department of Physics, Graduate School of Science, The University of Tokyo, 7-3-1 Hongo, Bunkyo, Tokyo 113-0033, Japan}

\author[0000-0003-4807-8117]{Yoichi Tamura}
\affiliation{Division of Particle and Astrophysical Science, Graduate School of Science, Nagoya University, Nagoya 464-8602, Japan}

\author[0009-0008-0167-5129]{Hiroya Umeda}
\affiliation{Institute for Cosmic Ray Research, The University of Tokyo, 5-1-5 Kashiwanoha, Kashiwa, Chiba 277-8582, Japan}
\affiliation{Department of Physics, Graduate School of Science, The University of Tokyo, 7-3-1 Hongo, Bunkyo, Tokyo 113-0033, Japan}

\author[0009-0006-6763-4245]{Hiroto Yanagisawa}
\affiliation{Institute for Cosmic Ray Research, The University of Tokyo, 5-1-5 Kashiwanoha, Kashiwa, Chiba 277-8582, Japan}
\affiliation{Department of Physics, Graduate School of Science, The University of Tokyo, 7-3-1 Hongo, Bunkyo, Tokyo 113-0033, Japan}

\begin{abstract}
We present the number densities and physical properties of the bright galaxies spectroscopically confirmed at $z\sim7-14$.
Our sample is composed of \redc{60} galaxies at $z_\mathrm{spec}\sim7-14$, including recently-confirmed galaxies at $z_\mathrm{spec}=12.34-14.32$ with JWST, as well as new confirmations at $z_\mathrm{spec}=6.583-7.643$  with $-24< M_\mathrm{UV}< -21$ mag using ALMA and Keck.
Our JWST/NIRSpec observations have also revealed that very bright galaxy candidates at $z\sim10-13$ identified from ground-based telescope images before JWST are passive galaxies at $z\sim3-4$, emphasizing the necessity of strict screening and spectroscopy in the selection of the brightest galaxies at $z>10$.
The UV luminosity functions derived from these spectroscopic results are consistent with a double power-law function, showing tensions with theoretical models at the bright end.
To understand the origin of the overabundance of bright galaxies, we investigate their morphologies using JWST/NIRCam high-resolution images obtained in various surveys including PRIMER and COSMOS-Web.
We find that $\sim70\%$ of the bright galaxies at $z\sim7$ exhibit clumpy morphologies with multiple sub-components, suggesting merger-induced starburst activity, which is consistent with SED fitting results showing bursty star formation histories.
At $z\gtrsim10$, bright galaxies are classified into two types of galaxies; extended ones with weak high-ionization emission lines, and compact ones with strong high-ionization lines including {\sc Niv]}$\lambda$1486, indicating that at least two different processes (e.g., merger-induced starburst and compact star formation/AGN) are shaping the physical properties of the brightest galaxies at $z\gtrsim10$ and are responsible for their overabundance.
\end{abstract}


\section{Introduction}\label{ss_intro}

Probing the properties of luminous galaxies in the early universe is key to understanding the physical process that governs galaxy formation.
The luminosity function, representing the volume density of galaxies as a function of the luminosity, is one of the most important statistical measurements of galaxies.
It is known that the luminosity functions at low redshifts are described by the Schechter function \citep{1976ApJ...203..297S}, which is derived from the shape of the halo mass function \citep{1974ApJ...187..425P}.
The exponential cutoff at the bright end of the Schechter function is thought to be caused by mass-quenching \citep{2010ApJ...721..193P}. 
However, previous wide-area imaging surveys using ground-based telescopes have reported that the bright end of the UV luminosity functions at $z\sim4-7$ does not follow the Schechter function, but is well described by the double-power-law luminosity function \citep[e.g.,][]{2012MNRAS.426.2772B,2014MNRAS.440.2810B,2015MNRAS.452.1817B,2017MNRAS.466.3612B,2018ApJ...863...63S,2020MNRAS.494.1771A,2023MNRAS.524.4586V}.
Even after subtracting the contributions from quasars, the galaxy luminosity function still shows a bright-end excess beyond the Schechter function at $z\sim4-7$ \citep{2018PASJ...70S..10O,2022ApJS..259...20H}.
Studies using ground-based near-infrared imaging datasets have identified very bright galaxy candidates at $z\sim10-13$ \citep{2020MNRAS.493.2059B,2022ApJ...929....1H}, which also supports the double-power-law function rather than the Schechter function.
However, these studies are based on samples of photometric galaxy candidates selected from imaging datasets, and spectroscopic observations are required to conclude the discussion about the shape of the bright end.

\begin{figure*}
\centering
\begin{center}
\includegraphics[width=0.9\hsize, bb=3 5 495 281,clip]{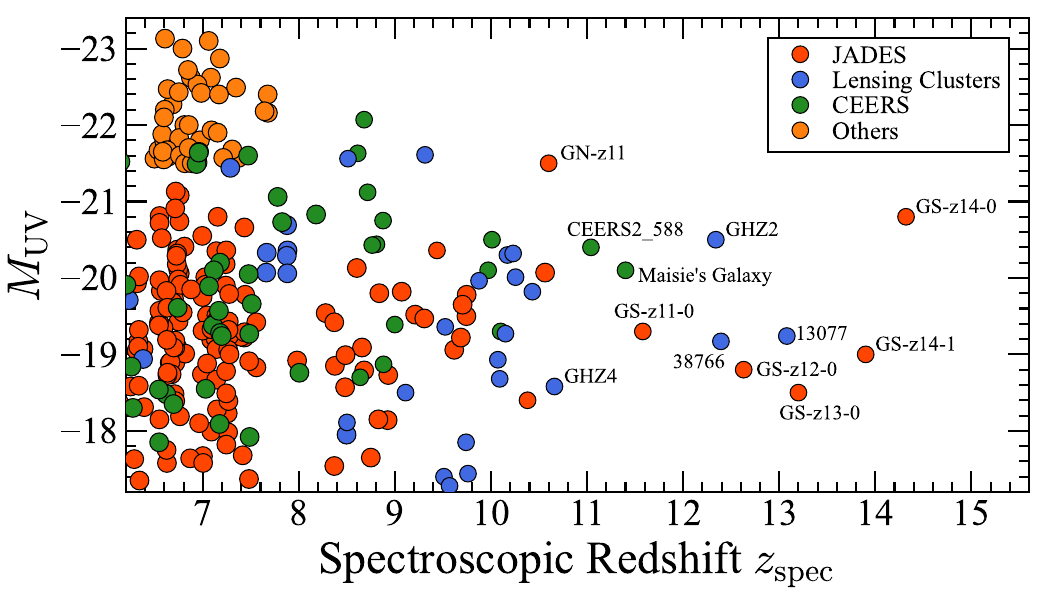}
\end{center}
\vspace{-0.5cm}
\caption{
UV magnitudes as a function of the redshift.
\redc{The circles show the spectroscopic redshifts of galaxies compiled in this study (see Table \ref{tab_sample}) and the literature \citep{2024ApJ...960...56H,2024arXiv240406531D,2023arXiv230811609F}.
The red, blue, green, and orange circles are galaxies in the JADES, lensing cluster (e.g., Abell2744), CEERS, and the other fields, respectively.}
The UV magnitudes of galaxies in \citet[][]{2024arXiv240406531D} are calculated from broad-band fluxes presented in \citet{2024ApJ...964...71H}.
}
\label{fig_Muv_z}
\end{figure*}

Since its first operation in 2022, the James Webb Space Telescope (JWST) has sparked a revolution in high redshift galaxy studies.
Various studies using JWST have reported that the abundance of bright galaxies at $z\gtrsim10$ is higher than theoretical model predictions (e.g., \citealt{2023MNRAS.523.1009B,2023MNRAS.523.1036B,2023ApJ...948L..14C,2023MNRAS.518.6011D,2024arXiv240303171D,2023MNRAS.520.4554D,2022ApJ...940L..55F,2023ApJ...946L..13F,2023arXiv231104279F,2023ApJS..265....5H,2024ApJ...960...56H,2024MNRAS.527.5004M,2022ApJ...940L..14N,2023arXiv231210033R,2023ApJ...951L...1P}, but see also \citealt{2023arXiv231112234W} for a report of relatively low number densities).
Several possibilities are raised and intensively discussed for the origin of this tension between the JWST observations and model predictions (see discussions in \citealt{2023ApJS..265....5H,2024ApJ...960...56H} for a review), such as a high star formation efficiency \citep[e.g.,][]{2023MNRAS.523.3201D,2021MNRAS.506.5512F,2022ApJ...938L..10I}, AGN activity \citep[e.g.,][]{2023ApJ...959...39H,2024arXiv240501629H}, a top-heavy IMF (e.g., \citealt{2005ApJ...626..627O,2022MNRAS.514.4639C,2023ApJ...951L..40S,2024MNRAS.529..628V}, see also \citealt{2023arXiv231212109R}), bursty star formation \citep[e.g.,][]{2023A&A...677L...4P,2023MNRAS.525.3254S,2023MNRAS.526L..47M,2023MNRAS.526.2665S,2023ApJ...955L..35S}, radiation-driven outflows \citep[e.g.,][]{2023MNRAS.522.3986F,2024A&A...684A.207F}, and a flaw in the current cosmology model \citep[e.g.,][]{2023MNRAS.526L..63P,2024ApJ...963....2H}. 
Although some studies have investigated the physical properties of galaxies \citep[e.g.,][]{2024MNRAS.531..997C,2023arXiv230605295E,2024arXiv240413045L,2024arXiv240307103R,2024MNRAS.529.4087T}, including studies for individual bright galaxies such as GN-z11 at $z=10.60$ \citep[e.g.,][]{2023A&A...677A..88B,2023ApJ...952...74T,2024Natur.627...59M,2023arXiv230600953M,2023arXiv230609142S,2024arXiv240416963X}, so far the physical origin of this overabundance of $z\gtrsim10$ galaxies is not clear.

In this study, we investigate the number density and the physical properties of spectroscopically confirmed bright galaxies at $z\sim7-14$.
We will discuss the shape of the bright end of the UV luminosity function and the morphologies and star formation histories of the bright galaxies using the JWST, ALMA, and Keck datasets.
These results are useful to understand the physical origin of the overabundance of $z\gtrsim10$ galaxies and the process that governs galaxy formation in the early universe.
Moreover, this study will be an important reference for future wide-area imaging surveys using Euclid \citep[e.g.,][]{2024arXiv240513505W}, Nance Grace Roman Space Telescope, and GREX-PLUS \citep{2022SPIE12180E..1II}, which will allow us to search for very bright galaxies at $z\gtrsim10$.

This paper is organized as follows.
We describe our galaxy sample and spectroscopic and photometric datasets in Section \ref{ss_data}.
Section \ref{ss_LF} presents the calculation of the effective survey volume and the results of the UV luminosity functions based on the spectroscopically confirmed galaxies.
In Sections \ref{ss_mor} and \ref{ss_sed}, we show the morphologies of the bright galaxies and the results of the SED fitting.
In Section \ref{ss_dis}, we discuss the physical origin of the overabundance of bright galaxies at $z\sim7$ and $z\sim12-14$, and the impact of the low-redshift interlopers in the galaxy selection using the wide-area survey datasets.
Section \ref{ss_summary} summarizes our findings.
Throughout this paper, we use the Planck cosmological parameter sets of the TT, TE, EE+lowP+lensing+BAO result \citep{2020A&A...641A...6P}: $\Omega_\m{m}=0.3111$, $\Omega_\Lambda=0.6899$, $\Omega_\m{b}=0.0489$, $h=0.6766$, and $\sigma_8=0.8102$.
With this cosmological parameter set, the angular size of $1.\carcsec0$ corresponds to 5.338 kpc at $z=7.0$.
All magnitudes are in the AB system \citep{1983ApJ...266..713O}.

\begin{figure*}
\centering
\begin{center}
\includegraphics[width=0.7\hsize, bb=8 13 567 355]{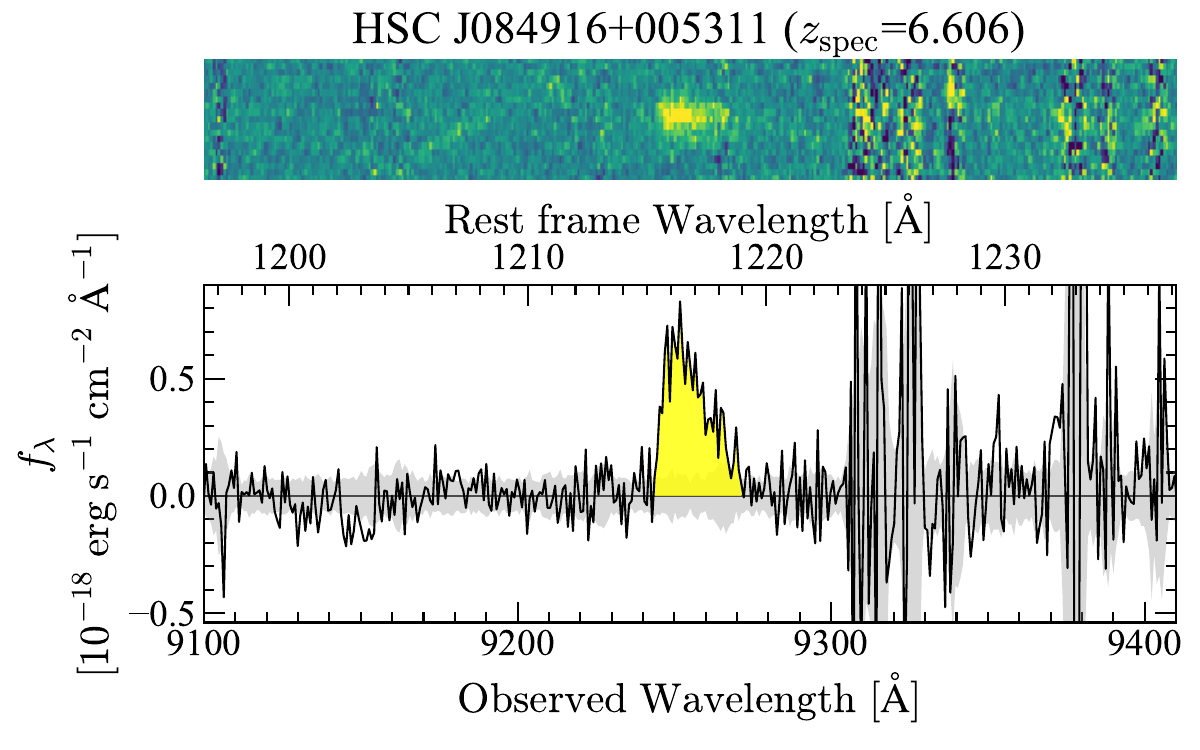}
\end{center}
\begin{center}
\includegraphics[width=0.7\hsize, bb=8 13 567 355]{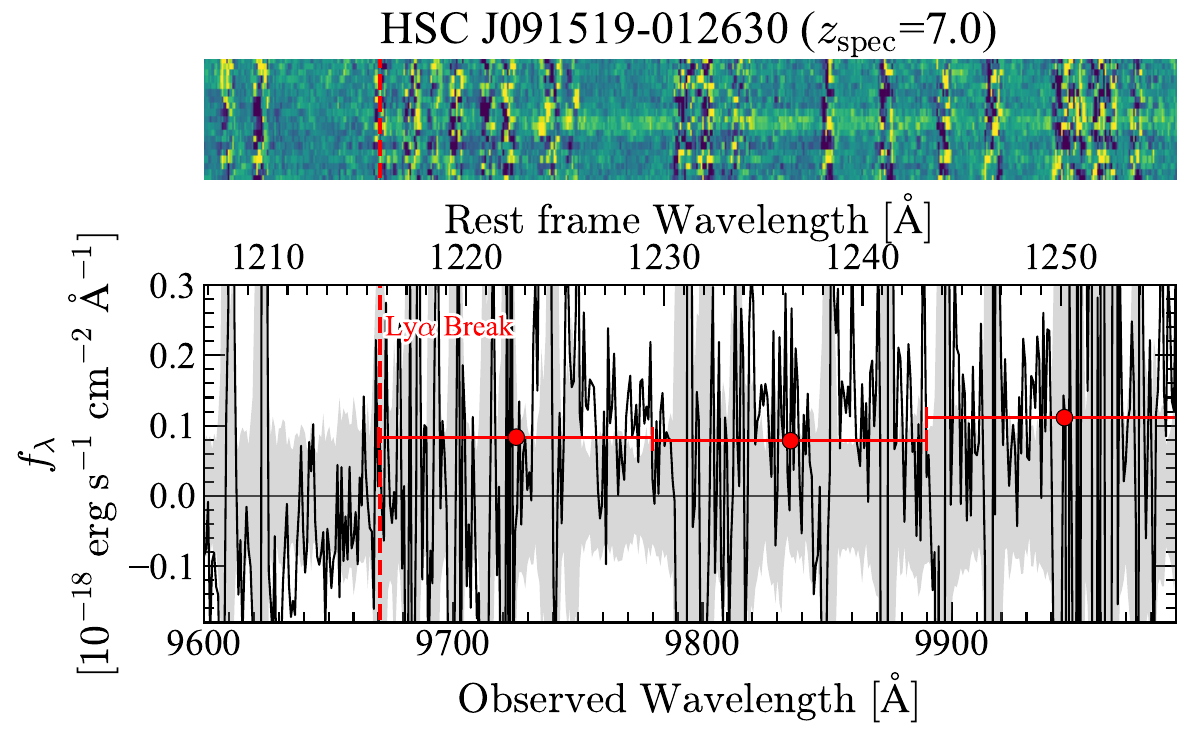}
\end{center}
\vspace{-0.5cm}
\caption{
Keck/LRIS spectra of HSC J084916+005311 at $z_\m{spec}=6.606$ (top) and HSC J091519-012630 at $z_\m{spec}=7.0$ (bottom).
For each object, the top panel shows the two-dimensional spectrum (yellow is positive), and the bottom panel shows the one-dimensional spectrum.
For HSC J091519-012630, we plot the averaged spectra over 110 $\m{\AA}$ bins with the red-filled circles to show the continuum.
The Ly$\alpha$ line is clearly detected in HSC J084916+005311, and the continuum and a break around 9670 $\m{\AA}$ are identified in HSC J091519-012630.
}
\label{fig_spec_Keck}
\end{figure*}

\section{Galaxy Sample and Observational Dataset}\label{ss_data}

\subsection{Galaxy Sample}\label{ss_sample}

In this study, we use a sample of \redc{60} galaxies spectroscopically confirmed at $z_\m{spec}=6.538-14.32$.
The sample is composed of 50 bright galaxies at $z_\m{spec}\sim7-8$ with UV magnitudes brighter than $M_\m{UV}<-21.0$ mag including four new confirmations with Keck and ALMA, and \redc{10} galaxies recently confirmed at $z_\m{spec}\sim\redc{10}-14$ with JWST.
Table \ref{tab_sample} summarizes the properties of galaxies in our sample, and Figure \ref{fig_Muv_z} shows the spectroscopic redshifts of our sample as well as other studies including \citet{2024ApJ...960...56H}.
In conjunction with the results in \citet{2024ApJ...960...56H}, we can investigate luminosity functions in a wide redshift range of $7\lesssim z \lesssim 14$.
In addition to the confirmed galaxies at $z_\m{spec}\gtrsim 7$, we use the results of JWST spectroscopic follow-ups targeting galaxy candidates at $z\sim10-13$ that are found to be low-redshift interlopers.
We describe the sample in detail below.

\begin{figure*}
\centering
\begin{minipage}{0.8\hsize}
\centering
\begin{minipage}{0.3\hsize}
\begin{center}
\includegraphics[width=0.99\hsize, bb=6 -30 230 251]{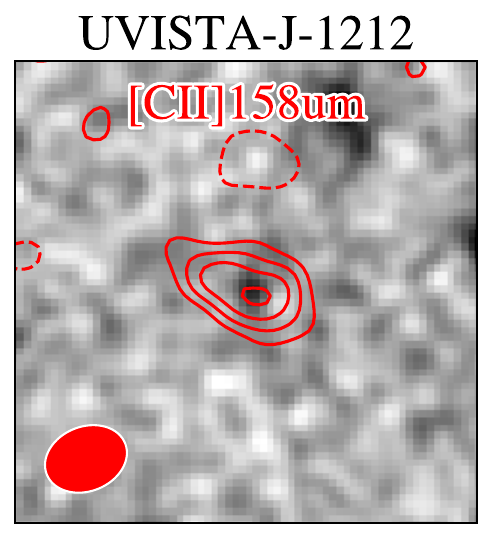}
\end{center}
\end{minipage}
\begin{minipage}{0.68\hsize}
\begin{center}
\includegraphics[width=0.99\hsize, bb=5 2 500 286]{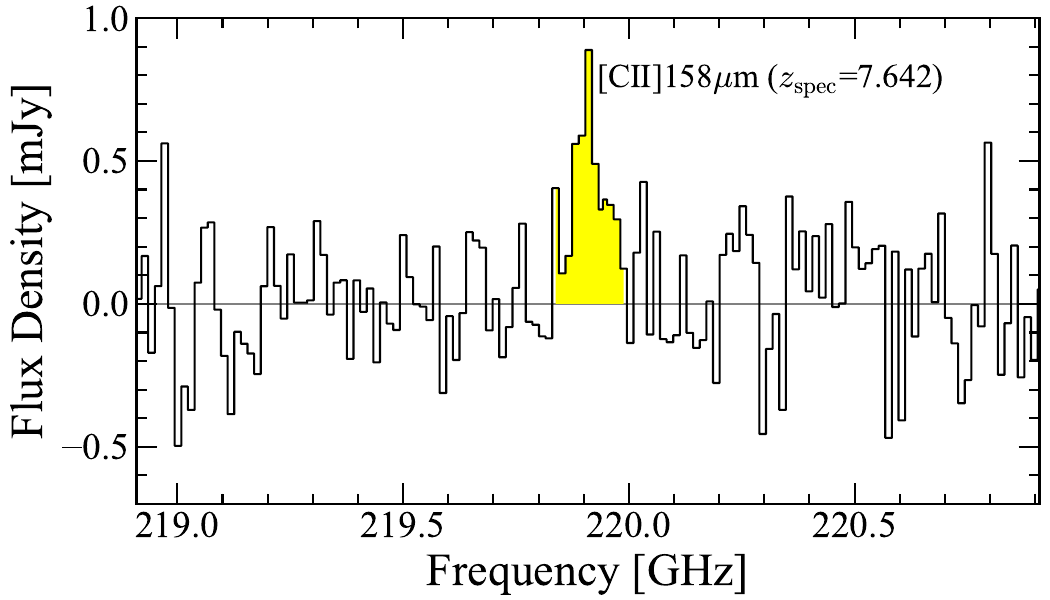}
\end{center}
\end{minipage}
\end{minipage}
\centering
\begin{minipage}{0.8\hsize}
\begin{minipage}{0.3\hsize}
\begin{center}
\includegraphics[width=0.99\hsize, bb=6 -30 230 251]{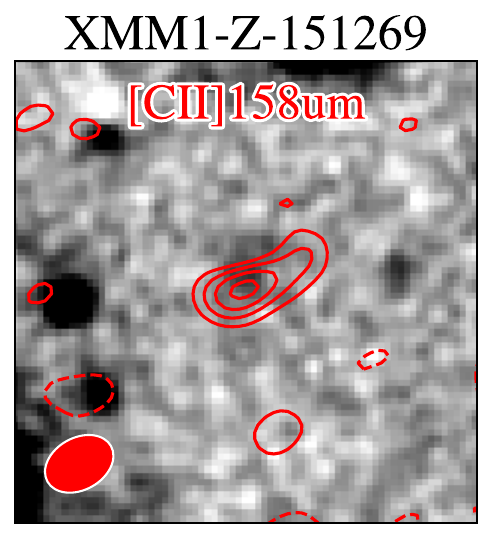}
\end{center}
\end{minipage}
\begin{minipage}{0.68\hsize}
\begin{center}
\includegraphics[width=0.99\hsize, bb=5 2 500 286]{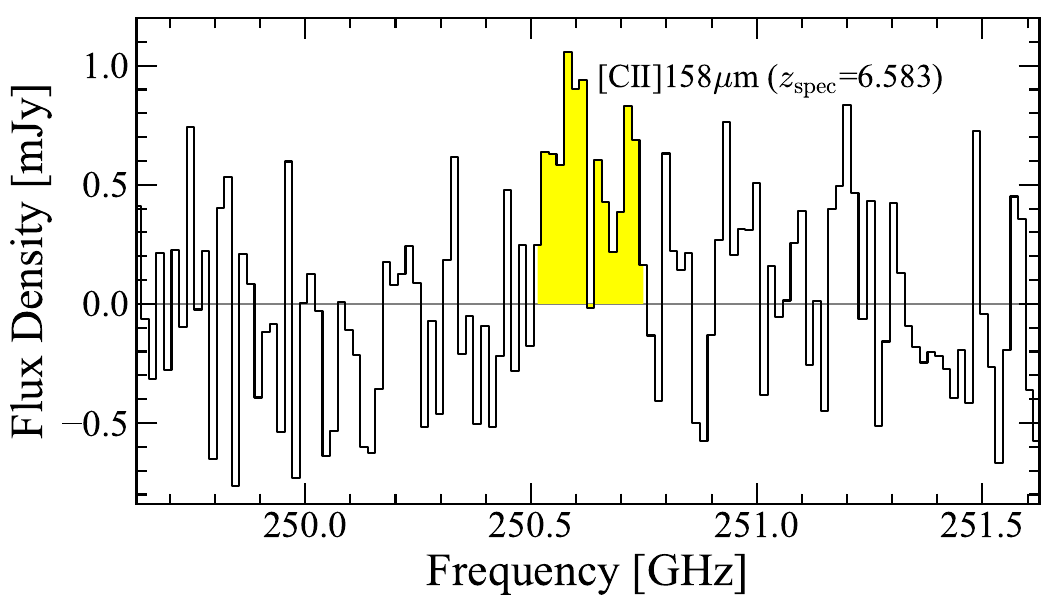}
\end{center}
\end{minipage}
\end{minipage}
\caption{
ALMA data of UVISTA-J-1212 at $z_\m{spec}=7.642$ (top) and XMM1-Z-151269 at $z_\m{spec}=6.583$ (bottom).
The left panels show the {\sc[Cii]}158$\mu$m maps made with the CASA task {\tt immoments}, by integrating over 140 and 220 km s$^{-1}$, comparable to the {\sc[Cii]} line widths of UVISTA-J-1212 and XMM1-Z-151269, respectively.
The red contours are drawn at $1\sigma$ intervals from $\pm2\sigma$.
The backgrounds are rest-UV images (ground-based $J$-band images).
The images are $4\arcsec\times4\arcsec$, and the red ellipses at the lower left corner indicate the
synthesized beam sizes of ALMA.
The right panels show ALMA spectra around the {\sc[Cii]} line after continuum subtraction.
These spectra are extracted from a $1.\carcsec4$-diameter circular aperture.
The {\sc[Cii]} line is detected at the $\sim6\sigma$ significance level in both objects.
It is conceivable that XMM1-Z-151269 is a merger given the possible double-peak {\sc[Cii]} emission, but deeper and higher-resolution data is needed for a definitive conclusion.
}
\label{fig_spec_ALMA}
\end{figure*}

\begin{figure*}
\centering
\begin{center}
\includegraphics[width=0.7\hsize, bb=24 11 928 315]{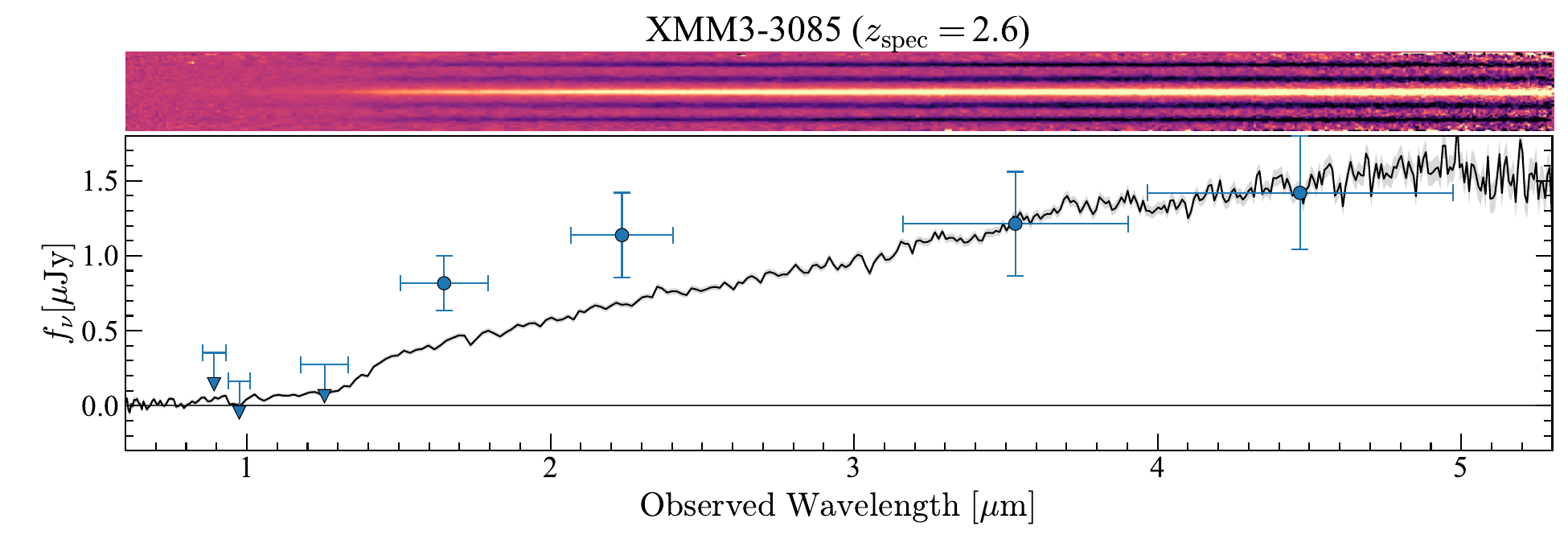}
\end{center}
\begin{center}
\includegraphics[width=0.7\hsize, bb=24 11 928 315]{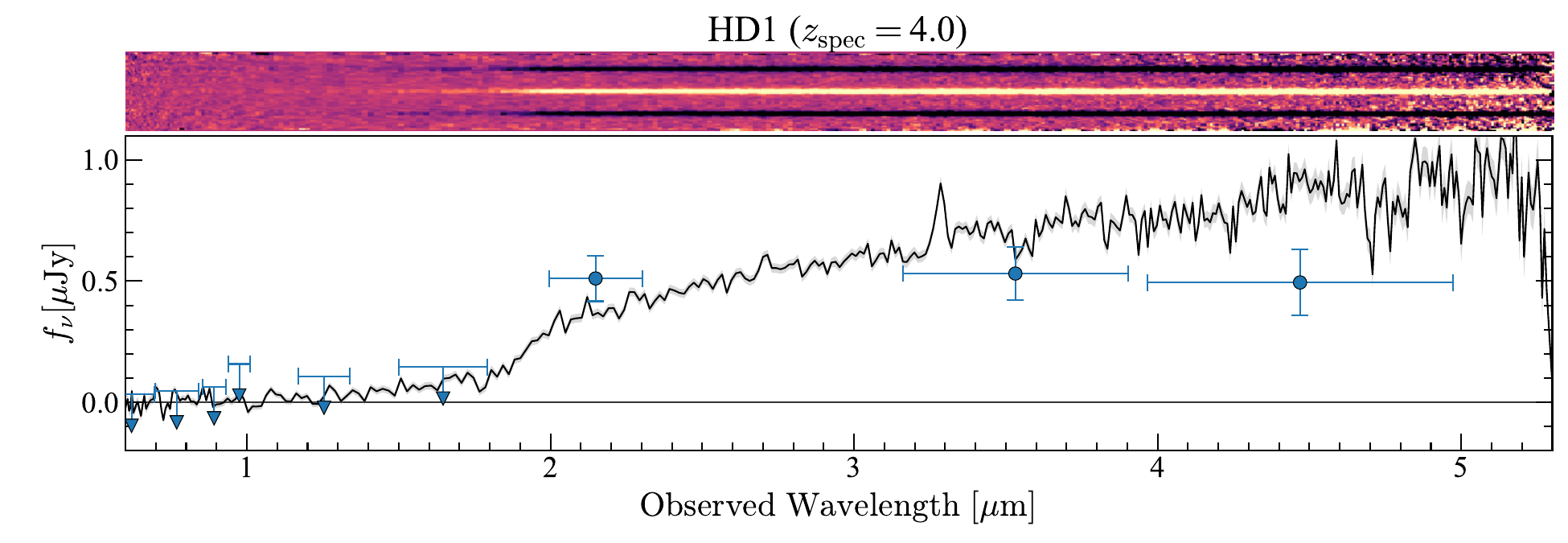}
\end{center}
\begin{center}
\includegraphics[width=0.7\hsize, bb=24 11 928 315]{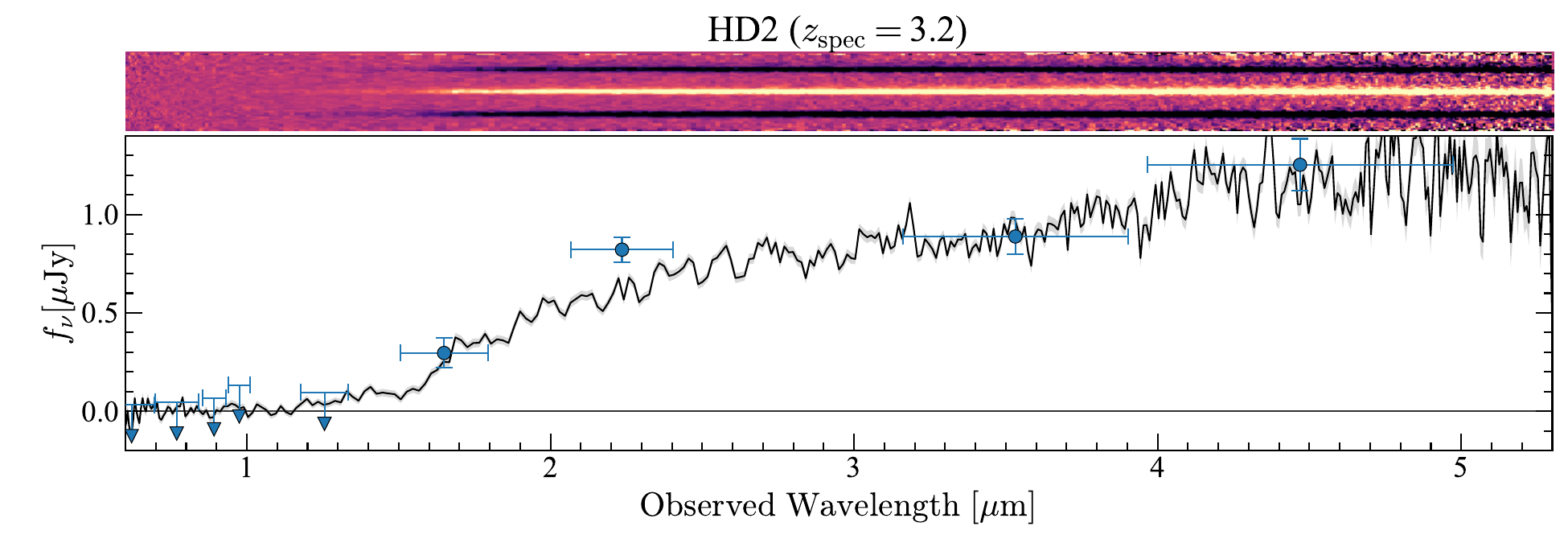}
\end{center}
\vspace{-0.5cm}
\caption{
JWST/NIRSpec spectra of bright galaxy candidates at $z\sim10-13$, XMM3-3085 in \citet{2020MNRAS.493.2059B} and HD1 and HD2 in \citet{2022ApJ...929....1H}.
The spectra show the Balmer breaks around $1-2\ \m{\mu m}$, indicating that these sources are low-redshift interlopers at $z_\m{spec}\sim3-4$.
The blue circles represent photometric data points in \citet{2022ApJ...929....1H} for HD1 and HD2, and those measured using the final data release of the VIDEO survey \citep{2023MNRAS.524.4586V} for XMM3-3085.
These sources were selected as $z\sim10-13$ galaxies due to the photometric scatters seen in the discrepancies between the photometric data points and the spectrum.
}
\label{fig_spec_JWST}
\end{figure*}

\subsubsection{Keck/LRIS Spectroscopy}\label{ss_Keck}

We conducted Keck/Low Resolution Imaging Spectrometer (LRIS) spectroscopy targeting bright galaxy candidates at $z\sim7$ identified in the HSC Wide field in \citet{2022ApJS..259...20H} from the Subaru/Hyper Suprime-Cam (HSC) survey datasets \citep{2018PASJ...70S...8A,2018PASJ...70S...4A,2019PASJ...71..114A,2022PASJ...74..247A}.
Observations were conducted in the Multi-Object Spectroscopy (MOS) mode on 2023 April 14th and 2024 February 8th and 9th (S23A-001N and S24A-001N, PI: Y. Harikane).
We used the 600/10000 grating with the central wavelength of 9000 $\m{\AA}$ and the D680 dichroic, resulting in the wavelength resolution of $R\sim1500$ at 9000 $\m{\AA}$.
The slit width was 0.\carcsec 8 and the seeing size was $\sim1$\arcsec in an FWHM.
The exposure time was $\sim1$ hour per each target.
We reduced the data using {\tt PypeIt} \citep{pypeit:zenodo}.

In the Keck/LRIS spectroscopy, we  targeted five $z\sim7$ galaxy candidates, and determined spectroscopic redshifts of two bright galaxies, HSC J084916$+$005311 and HSC J091519$-$012630, to be $z_\m{spec}=6.606$ and $z_\m{spec}=7.0$, respectively.
Figure \ref{fig_spec_Keck} presents the spectra of the two galaxies.
The spectrum of HSC J084916$+$005311 shows a very bright and asymmetric emission line around 9250 $\m{\AA}$, consistent with the Ly$\alpha$ emission line at $z=6.606$.
The line width of the Ly$\alpha$ emission after instrumental broadening correction is $\sim500$ km s$^{-1}$.
The spectrum of HSC J091519$-$012630 shows a continuum break around $\sim9670$ $\m{\AA}$, which is interpreted as the Lyman-$\alpha$ break at $z_\m{spec}\sim7.0$.
The redshifts obtained here are consistent with those determined in ALMA {\sc[Cii]} observations ($z_\m{[CII]}=6.600$ and $6.955$ for HSC J084916$+$005311 and HSC J091519$-$012630, respectively, Sun et al. in prep.), which also supports our redshift determinations.
The confirmed two galaxies are very bright with UV absolute magnitudes of $-23.9\leq M_\m{UV}\leq -23.1$ mag.

\subsubsection{ALMA Spectroscopy}\label{ss_ALMA}

Two galaxies identified as $z\gtrsim7$ galaxy candidates, UVISTA-1212 and XMM1-Z-151269, were observed in an ALMA large program Reionization Era Bright Emission Line Survey (REBELS; 2019.1.01634.L, PI: R. Bouwens; \citealt{2022ApJ...931..160B}) after the submission of the Bouwens et al's survey paper. 
\redc{The REBELS program observed 40 UV-bright ($M_\m{UV}\lesssim-22$ mag) galaxies at $z>6.5$ with {\sc [Cii]}158$\mu$m or {\sc [Oiii}]88$\mu$m with a spatial resolution of $\sim1.\carcsec2-1.\carcsec6$.}
We reduced and calibrated the archival data obtained in the REBELS program using the Common Astronomy Software (CASA; \citealt{2007ASPC..376..127M}) pipeline version 6.4.1.12 in the standard manner with scripts provided by the ALMA observatory.
Using the task {\tt tclean}, we produced images and cubes with the natural weighting without taper to maximize point-source sensitivities.
The beam sizes were $\sim1.\carcsec 3-1.\carcsec 6$.
The data analysis by the PI team will be presented in Schouws et al. in prep. (see also JWST GO-6480).

The right panels of Figure \ref{fig_spec_ALMA} display the obtained ALMA spectra of the two galaxies extracted with a 1.\carcsec4-diameter circular aperture.
The emission line is clearly detected around the frequencies of 219.9 and 250.6 GHz in UVISTA-1212 and XMM1-Z-151269 at the 6.0 and 5.9$\sigma$ significance levels, respectively.
We calculate these signal-to-noise ratios using 0.\arcsec6-diameter circular aperture in the same manner as \citet{2020ApJ...896...93H}.
As shown in the left panels of Figure \ref{fig_spec_ALMA}, these emission lines are cospatial with the rest-frame UV emission in the $J$-band images.
These emission lines in UVISTA-1212 and XMM1-Z-151269 are interpreted as the {\sc[Cii]}158$\mu$m lines at $z_\m{spec}=7.642$ and $6.583$, respectively, consistent with photometric redshift estimates in the literature \citep{2020MNRAS.493.2059B,2022ApJ...931..160B}. 
The {\sc[Cii]} line profile of XMM1-Z-151269 shows two peaks, suggesting the possibility of a merger, but deeper and higher-resolution data is needed for a definitive conclusions.

\subsubsection{JWST Spectroscopy}\label{ss_spec_jwst}

We conducted JWST/NIRSpec spectroscopy for very bright galaxy candidates at $z\gtrsim10$ identified in the ground-based images, XMM3-3085 at $z_\m{phot}\sim11$ \citep{2020MNRAS.493.2059B} and HD1 and HD2 at $z_\m{phot}\sim12-13$ \citep{2022ApJ...929....1H}, whose best-fit photometric redshifts are $z>10$ with $\Delta \chi^2>4$.
Observations for XMM3-3085 were conducted on 2024 January 8th with Prism using the S400A1 fixed slit  (GO-2792; PI: Y. Harikane).
The total integration time was 3545 seconds.
Observation for HD1 and HD2 were conducted on 2023 January 6th and 2022 August 16th, respectively, with Prism using the S400A1 fixed slit  (GO-1740; PI: Y. Harikane).
The total integration times were 2873 and 1801 seconds for HD1 and HD2, respectively. 
We used the level-3 product obtained from the Mikulski Archive for Space Telescopes (MAST) for XMM3-3085, and reduced data in \citet{2024MNRAS.534.3552S} for HD1 and HD2.
Note that MAST level-3 products for HD1 and HD2 are almost identical to the reduced data.

Figure \ref{fig_spec_JWST} shows the obtained NIRSpec spectra of the three galaxies.
We find that these three candidates are not $z>10$ galaxies as suggested by previously-obtained best-fit photometric redshifts in \citet{2020MNRAS.493.2059B} and \citet{2022ApJ...929....1H}, but are passive galaxies at $z\sim3-4$.
The spectrum of XMM3-3085 does not display a clear continuum break like the Lyman break but shows a continuum detection below $\sim1.4\ \mu$m, not consistent with the $z_\m{phot}\sim11$ solution.
From spectral fitting using {\tt Prospector} \citep{2021ApJS..254...22J}, we estimate the spectroscopic redshift of XMM3-3085 to be $z_\m{spec}=2.6$.
Similarly, HD1 and HD2 show continuum detections below $\sim1.7\ \mu$m, not consistent with the $z_\m{phot}\sim12-13$ solutions, and the spectroscopic redshifts are measured to be $z_\m{spec}=4.0$ and $3.2$.
These lower redshift solutions align with alternative solutions suggested in previous studies \citep{2020MNRAS.493.2059B,2022ApJ...929....1H,2023A&A...671A..29K}.
In Figure \ref{fig_spec_JWST}, we also compare the obtained NIRSpec spectra with the photometric data points used to select these galaxy candidates.
Since the measured photometric fluxes deviate from the spectrum in some bands, it is likely that these low-redshift passive galaxies are scattered into the galaxy selection at $z\gtrsim10$ due to photometric errors, especially in the Spitzer bands where the background subtraction is not straightforward (see the data points at $>3\ \mu$m in HD1).
Another bright galaxy candidate at $z\gtrsim12$, HD3, was also observed in the program GO-1740 and turned out to be a low-redshift interloper.
More detailed analyses of HD1, HD2, and HD3 using the NIRSpec spectra are presented in \citet{2024MNRAS.534.3552S}.

\begin{figure}
\centering
\begin{center}
\includegraphics[width=0.95\hsize, bb=5 2 500 286,clip]{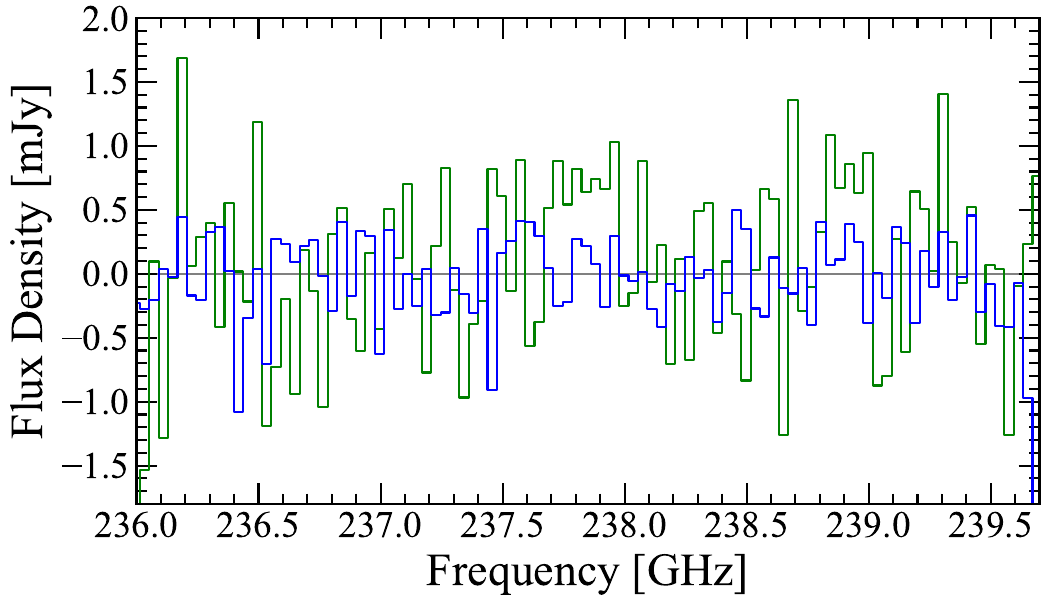}
\end{center}
\caption{
ALMA spectrum of HD1.
The green line shows the Cycle 7 data, where a $4\sigma$ line-like signal was reported around the frequency of 237.8 GHz.
No signal is identified in the Cycle 8 data (the blue line).
}
\label{fig_spec_ALMA_HD1}
\end{figure}

Previous ALMA Band 6 spectroscopy for HD1 in Cycle 7 DDT showed a $3.8\sigma$ line-like tentative signal around 237.8 GHz, which can be interpreted as the {\sc[Oiii]}88$\mu$m line at $z=13.3$ \citep{2022ApJ...929....1H}.
ALMA Band 4 data also showed a $4\sigma$ tentative feature that can be consistent with the {\sc[Cii]}158$\mu$m line, but statistical tests demonstrated that these $\sim4\sigma$-level signals are fully consistent with being random noise features \citep{2023A&A...671A..29K}. 
To investigate the previously reported line-like signal, we conducted additional ALMA Band 6 observations covering 237.8 GHz in Cycle 8 (2021.1.00207.S; PI: Y. Harikane).
Figure \ref{fig_spec_ALMA_HD1} shows the spectra obtained in Cycle 7 DDT and Cycle 8, resulting in no significant detection around 238 GHz in the Cycle 8 data, which is consistent with the low-redshift solution from the JWST/NIRSpec spectroscopy.
Similarly, ALMA Band 7 observations for XMM3-3085 (2021.1.00341.S, 2022.1.00522.S; PI: Y. Harikane) do not show any significant emission line, which also agrees with the JWST/NIRSpec spectroscopic result.

These spectroscopic observations reveal that the three very bright galaxy candidates at $z\gtrsim10$ selected from the ground-based images before JWST are low-redshift interlopers at $z\sim3-4$.
The possibilities of these low-redshift solutions were already discussed in the discovery papers \citep{2020MNRAS.493.2059B,2022ApJ...929....1H}, but our observations highlight the importance of the spectroscopy when discussing very bright galaxies at $z>10$.
These results suggest that passive galaxies with Balmer breaks at intermediate redshifts can be selected as Lyman break galaxies at high redshifts due to photometric scatters, and are important contaminants that should be taken into account in the galaxy selection, in addition to galaxies with strong emission lines that boost the broad-band and medium-band fluxes and mimic a Lyman break-like SED \citep{2023Natur.622..707A,2022arXiv220802794N,2023ApJ...943L...9Z}.
The implications for the UV luminosity functions and future bright galaxy selections using wide-area survey datasets are discussed in Sections \ref{ss_uvlf_result} and \ref{ss_dis_lowz}, respectively.
As discussed in Section \ref{ss_dis_lowz}, these passive low-redshift interlopers are erroneously selected as high redshift galaxies because of 1) large photometric scatters originating from relatively shallow ground-based and Spitzer datasets, and 2) their very bright magnitudes.
Note that these interlopers are not significant in JWST-selected photometric candidates because they are usually faint compared to the galaxies discussed here.
Indeed, high spectroscopic success rates are reported in JWST-selected candidates \cite[e.g.,][]{2023Natur.622..707A,2023ApJ...951L..22A,2023ApJ...949L..25F,2023arXiv230811609F}.

\subsubsection{Literature at $z\sim7-8$}

In addition to the four galaxies spectroscopically confirmed in Sections \ref{ss_Keck} and \ref{ss_ALMA}, we have compiled bright galaxies with spectroscopic confirmations in the literature.
We include 24 bright ($M_\m{UV}<-21.5$ mag) galaxies at $z_\m{spec}\sim7-8$ in the COSMOS and UDS fields from the REBELS program \citep{2022ApJ...931..160B}, and from \citet{2023ApJ...954..103S}.
In the COSMOS field, we also take four galaxies at $z_\m{spec}\sim7$ from \citet{2021MNRAS.502.6044E,2022MNRAS.512.4248E}.
From the Subaru/HSC survey, we include one very bright ($M_\m{UV}=-23.6$ mag) galaxy, HSC J023526$-$031737 at $z_\m{spec}=6.913$ (\citealt{2018PASJ...70S..10O,2022ApJS..259...20H}, M. Sawicki et al. in prep.). 
In addition, we use six galaxies at $z_\m{spec}\sim7-8$ with $M_\m{UV}<-21.0$ mag in the JWST CEERS and GLASS fields from \cite{2023ApJS..269...33N}, and 11 galaxies from other spectroscopic studies (see Table \ref{tab_sample} for their references).

\subsubsection{Literature at $z\sim\redc{10}-14$}

To extend our analysis to higher redshifts, we include 10 galaxies at $z_\m{spec}\sim10-14$ recently confirmed with JWST.
We use GHZ2, a bright galaxy initially identified in the JWST/NIRCam images \citep[e.g.,][]{2022ApJ...940L..14N,2022ApJ...938L..15C} and recently confirmed at $z_\m{spec}=12.34$ with NIRSpec and MIRI spectroscopy \citep{2024arXiv240310238C,2024arXiv240310491Z}.
We also include JADES-GS-z14-0 ($z_\m{spec}=14.32$, hereafter GS-z14-0) and JADES-GS-z14-1 ($z_\m{spec}=13.90$, hereafter GS-z14-1), which are firstly identified in the JADES Origins Field \citep{2023arXiv231012340E} by \citet{2023arXiv231210033R} and recently confirmed with NIRSpec by \citet{2024arXiv240518485C}.
\redc{Finally, we add seven galaxies at $z\sim10-11$ recently confirmed in \citet{2024arXiv241010967N}.}
Although these galaxies are relatively faint ($-21\lesssim M_\m{UV}\lesssim -19$ mag) compared to  galaxies at $z\sim7-8$ in this sample, they are useful to obtain meaningful constraints on the number densities of galaxies at $z\sim10-14$.

In total, our sample consists of \redc{60} galaxies at $z_\m{spec}=6.538-14.32$ (Table \ref{tab_sample}). 
These galaxies are selected in multiple survey fields with various methods, and we carefully estimate the survey volume in Section \ref{ss_volume}.

\startlongtable
\begin{deluxetable*}{ccccccc}
\tablecaption{List of Spectroscopically-Confirmed Galaxies Compiled in This Study}
\label{tab_sample}
\tabletypesize{\footnotesize}
\setlength{\tabcolsep}{5.pt}
\tablehead{\colhead{Name} & \colhead{R.A.} & \colhead{Decl.} & \colhead{$z_\m{spec}$} & \colhead{$M_\m{UV}$} & \colhead{Spec-$z$ Ref.} & \colhead{Phot. Ref.} \\
\colhead{(1)}& \colhead{(2)}& \colhead{(3)}& \colhead{(4)}& \colhead{(5)}& \colhead{(6)}& \colhead{(7)}} 
\startdata
\multicolumn{7}{c}{High Redshift Galaxies at $z_\mathrm{spec}>6.5$}\\
JADES-GS-z14-0 & 03:32:19.90 & $-$27:51:20.3 & \redc{14.18} & $-20.8$ & Car24\redc{ab,Sc24} & Rob23,He24 \\
JADES-GS-z14-1 & 03:32:17.83 & $-$27:53:09.3 & 13.90 & $-19.0$ & Car24a & Rob23 \\
GHZ2 & 00:13:59.74 & $-$30:19:29.1 & 12.34 & $-20.5$ & Cas24,Za24 & Nai22,Cas22,Do23,Har23,Bou23\\
\redc{GHZ4} & 00:14:03.30 & -30:21:05.6 & 10.66$^\dagger$ & -18.58 & Nap24 & Cas22,Har23\\
\redc{GHZ7} & 00:13:48.33 & -30:19:14.6 & 10.43 & -19.82 & Nap24 & Cas22\\
\redc{GHZ8} & 00:13:48.34 & -30:19:18.5 & 10.23 & -20.32 & Nap24 & Cas22\\
\redc{GHZ9} & 00:13:54.90 & -30:20:43.9 & 10.15 & -19.27 & Nap24 & Cas22\\
\redc{GLASS-z11-17225} & 00:14:01.75 & -30:20:35.5 & 10.09 & -18.68 & Nap24 & McLeod24\\
\redc{GHZ1} & 00:14:02.86 & -30:22:18.7 & 9.875 & -19.96 & Nap24 & Nai22,Cas22,Do23,Har23,Bou23\\
\redc{GLASS-83338} & 00:13:49.13 & -30:19:00.8 & 9.523 & -19.36 & Nap24 & At23\\
CEERS\_01023 & 14:20:45.22 & $+$53:02:01.1 & 7.779 & $-21.1$ & Nak23 & Nak23\\
UVISTA-Y-002 & 10:02:12.56 & $+$02:30:45.7 & 7.677 & $-22.2$ & Bou22 & Ste17,Ste19,Bow20\\
UVISTA-Y-001 & 09:57:47.90 & $+$02:20:43.7 & 7.675 & $-22.4$ & Bou22 & Ste17,Ste19,Bow20\\
UVISTA-1212 & 10:02:31.81 & $+$02:31:17.1 & 7.643 & $-22.2$ & This Study & Bow20\\
CEERS\_00698 & 14:20:12.08 & $+$53:00:26.8 & 7.471 & $-21.6$ & Nak23 & Nak23\\
UVISTA-Y-879 & 09:57:54.69 & $+$02:27:54.9 & 7.370 & $-21.6$ & Bou22 & Bow20\\
XMM3-Z-110958 & 02:25:07.94 & $-$05:06:40.7 & 7.346 & $-22.5$ & Bou22 & Bou22\\
UVISTA-Y-003 & 10:00:32.32 & $+$01:44:31.3 & 7.306 & $-21.7$ & Bou22,En22b,Row24 & St17,19,Bow20\\
GLASS\_10021 & 00:14:26.04 & $-$30:25:06.7 & 7.286 & $-21.4$ & Nak23 & Nak23\\
SXDF-NB1006-2 & 02:18:56.54 & $-$05:19:58.9 & 7.212 & $-21.6$ & Sh12,In16,Re23 & Sh12\\
XMM1-Z-276466 & 02:16:25.09 & $-$04:57:38.5 & 7.177 & $-22.9$ & Bou22 & Bou22\\
UVISTA-65666 & 10:01:40.69 & $+$01:54:52.5 & 7.168 & $-22.4$ & Fu16,Has19 & Bow14,Bow17a\\
COS-zs7-1 & 10:00:23.76 & $+$02:20:37.0 & 7.154 & $-21.9$ & Sta17,La17 & RB16\\
UVISTA-Y-004 & 10:00:58.49 & $+$01:49:56.0 & 7.090 & $-21.9$ & Bou22 & Ste17,Ste19,Bow20\\
XMM3-Z-432815 & 02:26:46.19 & $-$04:59:53.5 & 7.084 & $-22.6$ & Bou22,En22b & Bou22\\
UVISTA-304416 & 10:00:43.36 & $+$02:37:51.3 & 7.060 & $-23.1$ & Sc23 & Bow17a\\
HSC J091519$-$012630 & 09:15:19.55 & $-$01:26:30.5 & 7.0 & $-23.9$ & This Study & Har22a\\
UVISTA-238225 & 10:01:52.31 & $+$02:25:42.3 & 6.982 & $-22.4$ & Bou22 & Bow14,Bow17a\\
XMM1-Z-1664 & 02:17:15.24 & $-$05:07:45.7 & 6.970 & $-21.8$ & Bou22 & Bou22\\
z70-1 & 10:02:15.52 & $+$02:40:33.4 & 6.965 & $-21.3$ & Zh20 & It18\\
CEERS\_00716 & 14:20:19.28 & $+$52:59:35.7 & 6.961 & $-21.6$ & Nak23 & Nak23\\
CEERS\_01142 & 14:20:14.57 & $+$52:57:31.4 & 6.957 & $-21.6$ & Nak23 & Nak23\\
LAE-7 & 10:03:05.20 & $+$02:09:14.7 & 6.945 & $-21.5$ & Hu21 & Hu21\\
UVISTA-Z-1595 & 10:01:04.60 & $+$02:38:56.7 & 6.943 & $-22.5$ & Bou22 & Bou22\\
CEERS\_00717 & 14:20:19.54 & $+$52:58:19.8 & 6.934 & $-21.5$ & Nak23 & Nak23\\
HSC J023536$-$031737 & 02:35:36.58 & $-$03:17:37.7 & 6.913 & $-23.6$ & SaP & On18,Har22a\\
COS-788571 & 09:59:21.68 & $+$02:14:53.0 & 6.884 & $-21.5$ & En21b & En21\\
XMM3-Z-1122596 & 02:27:13.11 & $-$04:17:59.2 & 6.875 & $-22.6$ & Bou22,En22b & Bou22\\
COS-3018555981 & 10:00:30.18 & $+$02:15:59.7 & 6.854 & $-22.0$ & Sm18,La17 & Sm15\\
COS-87259 & 09:58:58.27 & $+$01:39:20.2 & 6.853 & $-21.7$ & En22a & En21\\
COS-862541 & 10:03:05.25 & $+$02:18:42.7 & 6.845 & $-22.7$ & Bou22,En22b & En21\\
COS-955126 & 09:59:23.63 & $+$02:23:32.7 & 6.813 & $-21.5$ & En21b & En21\\
COS-2987030247 & 10:00:29.86 & $+$02:13:02.4 & 6.808 & $-22.0$ & Sm18,La17 & Sm15\\
COS-1009842 & 10:00:23.38 & $+$02:31:14.7 & 6.761 & $-21.6$ & En21b & En21\\
UVISTA-Z-019 & 10:00:29.89 & $+$01:46:46.4 & 6.753 & $-21.8$ & Sc23 & Sc23\\
UVISTA-Z-007 & 09:58:46.21 & $+$02:28:45.8 & 6.750 & $-22.4$ & Sc23 & Sc23\\
XMM1-88152 & 02:19:35.13 & $-$05:23:19.2 & 6.750 & $-21.8$ & Bou22 & En21\\
COS-369353 & 10:01:59.07 & $+$01:53:27.8 & 6.729 & $-21.7$ & Bou22 & En21\\
UVISTA-304384 & 10:01:36.85 & $+$02:37:49.1 & 6.685 & $-22.3$ & Bou22 & Bow14,En21\\
COS-469110 & 10:00:04.37 & $+$01:58:35.7 & 6.645 & $-21.6$ & Bou22 & En21\\
UVISTA-169850 & 10:02:06.47 & $+$02:13:24.2 & 6.633 & $-22.5$ & Bou22 & Bow14,En21\\
COSMOS24108 & 10:00:47.34 & $+$02:28:42.9 & 6.629 & $-21.7$ & Pe16 & Pe18\\
HSC J084916$+$005311 & 08:49:16.59 & $+$00:53:11.0 & 6.606 & $-23.1$ & This Study & Har22a\\
CR7 & 10:00:57.99 & $+$01:48:15.5 & 6.604 & $-22.2$ & So15,Ma17 & So15\\
UVISTA-104600 & 10:00:42.13 & $+$02:01:56.8 & 6.598 & $-21.8$ & Bou22 & Bow14,En21\\
Himiko & 02:17:57.58 & $-$05:08:44.9 & 6.595 & $-22.1$ & Ou09 & Ou13\\
COLA1 & 10:02:35.40 & $+$02:12:13.5 & 6.593 & $-21.6$ & Hu16,Ma18 & Hu16\\
XMM1-Z-151269 & 02:18:47.47 & $-$05:10:20.3 & 6.583 & $-21.6$ & This Study & Bou22\\
COS-1304254 & 10:02:54.05 & $+$02:42:12.0 & 6.577 & $-21.9$ & Bou22 & En21\\
UVISTA-Z-1373 & 09:57:36.99 & $+$02:05:11.3 & 6.538 & $-21.7$ & Bou22 & Bou22\\
\hline
\multicolumn{7}{c}{Low Redshift Interlopers}\\
HD1 & 10:01:51.31 & $+$02:32:50.0 & 4.0 & \nodata & This Study & Har22b\\
HD2 & 02:18:52.44 & $-$05:08:36.1 & 3.2 & \nodata & This Study & Har22b\\
XMM3-3085 & 02:26:59.11 & $-$05:12:17.8 & 2.6 & \nodata & This Study & Bow20\\
\enddata
\tablecomments{(1) Name. 
(2) Right ascension.
(3) Declination.
(4) Spectroscopic redshift.
(5) Absolute UV magnitude. 
(6,7) References for spectroscopic redshifts and photometry (At23: \citealt{2023MNRAS.524.5486A},
Bou22: \citealt{2022ApJ...931..160B},
Bou23: \citealt{2023MNRAS.523.1009B}, 
Bow14: \citealt{2014MNRAS.440.2810B},
Bow17a: \citealt{2017MNRAS.466.3612B},
Bow20: \citealt{2020MNRAS.493.2059B},
Cas22: \citealt{2022ApJ...938L..15C}, 
Cas24: \citealt{2024arXiv240310238C},
Car24a: \citealt{2024arXiv240518485C},
\redc{Car24b}: \citealt{2024arXiv240920533C},
Do23: \citealt{2023MNRAS.518.6011D}, 
En21a: \citealt{2021MNRAS.500.5229E},
En21b: \citealt{2021MNRAS.502.6044E},
En22a: \citealt{2022MNRAS.512.4248E},
En22b: \citealt{2022MNRAS.517.5642E},
Fu16: \citealt{2016ApJ...822...46F},
Har22a: \citealt{2022ApJS..259...20H}, 
Har22b: \citealt{2022ApJ...929....1H}, 
Har23: \citealt{2023ApJS..265....5H}, 
He24: \citealt{2024arXiv240518462H}
Hu16: \citealt{2016ApJ...825L...7H},
Hu21: \citealt{2021NatAs...5..485H},
In16: \citealt{2016Sci...352.1559I},
It18: \citealt{2018ApJ...867...46I},
La17: \citealt{2017ApJ...851...40L},
Ma17: \citealt{2017ApJ...851..145M},
Ma18: \citealt{2018AA...619A.136M},
Nai22: \citealt{2022ApJ...940L..14N}, 
Nak23: \citealt{2023ApJS..269...33N}, 
\redc{Nap24: \citealt{2024arXiv241010967N},}
Ou09: \citealt{2009ApJ...696.1164O},
Ou13: \citealt{2013ApJ...778..102O},
Pe16: \citealt{2016ApJ...829L..11P},
Pe18: \citealt{2018AA...619A.147P},
RB16: \citealt{2016ApJ...823..143R},
Re23: \citealt{2023ApJ...945...69R},
Rob23: \citealt{2023arXiv231210033R},
Row24: \citealt{2024arXiv240506025R},
SaP: Sawicki et al. in prep.,
Sh12: \citealt{2012ApJ...752..114S},
Sc23: \citealt{2023ApJ...954..103S},
\redc{Sc24}: \citealt{2024arXiv240920549S},
Sm15: \citealt{2015ApJ...801..122S},
Sm18: \citealt{2018Natur.553..178S},
So15: \citealt{2015ApJ...808..139S},
Ste17: \citealt{2017ApJ...851...43S},
Ste19: \citealt{2019ApJ...883...99S},
Za24: \citealt{2024arXiv240310491Z}, 
Zh20: \citealt{2020ApJ...891..177Z}).\\
\redc{$^\dagger$ This spectroscopic redshift is tentative \citep[see][]{2024arXiv241010967N}. Thus we do not use this source in the number density estimate.}
}
\end{deluxetable*}

\subsection{Imaging Dataset}

\subsubsection{JWST/NIRCam}

We will use JWST/NIRCam and HST images to investigate the photometric and morphological properties of bright galaxies at $z\sim7-14$ compiled in this study and in \citet{2024ApJ...960...56H}.
JWST/NIRCam images were taken in the COSMOS, UDS, CEERS, Abell2744, GOODS-North, and GOODS-South fields.
Public Release IMaging for Extragalactic Research (PRIMER; Dunlop et al., in preparation) survey conducted imaging observations over a total of $\sim400$ arcmin$^2$ in the COSMOS and UDS fields taken with eight NIRCam filters, F090W, F115W, F150W, F200W, F277W, F356W, F410M, and F444W. 
The PRIMER data were reduced using the PRIMER Enhanced NIRCam Image Processing Library (PENCIL; Magee et al., in preparation) software.
The PENCIL is built on top of STScI's JWST Calibration Pipeline (v1.12.5) but also includes additional processing steps not included in the standard calibration pipeline, such as the subtraction of $1/f$ noise striping patterns and the subtraction of wisps artifacts in the short wavelength filters.
Additionally, the COSMOS-Web survey \citep{2023ApJ...954...31C} mapped a 0.6 deg$^2$ area in the COSMOS field with four filters, F115W, F150W, F277W, and F444W.
The COSMOS-Web data were reduced using the JWST Calibration Pipeline (versions 1.12.5) and the Calibration Reference Data System context file of {\tt jwst\_1193.pmap} with custom modifications described in \citet{2023ApJS..265....5H}.
The CEERS field was observed in the CEERS survey \citep{2023ApJ...946L..13F} with seven NIRCam filters, F115W, F150W, F200W, F277W, F356W, F410M, and F444W.
We use reduced images released by the CEERS team \citep[see][for the data reduction]{2023ApJ...946L..12B}.\footnote{https://ceers.github.io/releases.html}
The NIRCam images of the Abell2744 cluster field were taken in two surveys, the GLASS survey \citep{2022arXiv220607978T} and the UNCOVER survey \citep{2022arXiv221204026B}.
Reduced images provided by the UNCOVER team are used in this study.\footnote{https://jwst-uncover.github.io/DR2.html}
Finally, the JADES program \citep{2023arXiv230602465E} conducted NIRCam observations in the GOODS-North and GOODS-South fields.
We use imagesthat were reduced with {\tt grizli} \citep{grizli} and are provided in the DAWN JWST Archive \citep[versions 7.2 and 7.3 for GOODS-North and GOODS-South, respectively, see also][]{2023ApJ...947...20V}.

\subsubsection{HST}

We also use HST/ACS and WFC3 images in the COSMOS and UDS fields.
The Cosmic Assembly Near-infrared Deep Extragalactic Legacy Survey (CANDELS: \citealt{2011ApJS..197...35G}, \citealt{2011ApJS..197...36K}) obtained images over a total of $\sim300$ arcmin$^2$ in the COSMOS and UDS fields with $V_{606}$, $I_{814}$, $J_{125}$, and $H_{160}$ filters.
We use the reduced imaging data provided by the 3D-HST team \citep{2012ApJS..200...13B,2014ApJS..214...24S}.
In addition, the COSMOS-Drift And SHift (COSMOS-DASH) survey \citep{2019ApJ...880...57M,2022ApJ...925...34C} conducted $H_{160}$ imaging observations covering an area of 0.49 deg$^2$.
Data products provided in the MAST are used in this study.

\section{UV Luminosity Function}\label{ss_LF}

\subsection{Effective Volume Estimate}\label{ss_volume}

Using the sample of spectroscopically confirmed galaxies constructed in Section \ref{ss_sample}, we calculate the UV luminosity functions at $z\sim7-14$ in the bright magnitude range, which are not investigated in previous studies such as \citet{2024ApJ...960...56H} and \citet{2023arXiv230811609F}.
Because our samples is composed of 50 galaxies at $z\sim7-8$ and 3 galaxies at $z\sim12-14$, we divide our sample into the four redshift subsamples at $z_\m{spec}=6.5-7.5$, $7.5-8.5$, $11.0-13.5$, and $13.5-15.0$ to calculate the number densities at $z\sim7$, $8$, $12$, and $14$.
Since the galaxies in our spectroscopic sample are confirmed with various instruments whose target selection and detection completeness are not well-known, we carefully estimate the effective volume for the luminosity functions.
\redc{As detailed below, we use the two methods to estimate the luminosity functions.
If the all photometric candidates in a magnitude bin are spectroscopically observed, we calculate the best estimate of the number density with errors using the effective volume published in the literature.
If not all of the candidates are observed, we put a lower limit on the number density using the number of the confirmed sources and the survey area.}

In the magnitude ranges of $-23.5<M_\m{UV}<-22.0$ mag at $z_\m{spec}=6.5-7.5$ and $-22.5<M_\m{UV}<-22.0$ at $z_\m{spec}=7.5-8.5$, we use galaxies in the COSMOS field where the spectroscopic completeness is high thanks to some intensive spectroscopic surveys \citep[e.g.,][]{2021MNRAS.502.6044E,2022ApJ...931..160B}.
We count the number of galaxies spectroscopically confirmed in the area of $150.8<\m{R.A.}<149.3$ deg and $1.70<\m{decl.}<2.75$ deg, corresponding to the survey area of 1.5 deg$^2$ (comparable to one in \citealt{2014MNRAS.440.2810B,2020MNRAS.493.2059B}), and calculate the survey volume in the redshift range of $z=6.5-7.5$ or $z=7.5-8.5$.
At the magnitude bins of $M_\m{UV}=-23.2$ mag ($-23.45<M_\m{UV}<-22.95$ mag) and $M_\m{UV}=-22.7$ mag ($-23.95<M_\m{UV}<-23.45$ mag), we have found that all of the galaxy candidates reported in \citet{2017MNRAS.466.3612B} are spectroscopically observed.
Thus in these two bins, we calculate the number densities rather than lower limits, assuming a 100\% completeness.
In the other bins, we obtain the lower limits of the number densities, since there are remaining candidates that are not yet spectroscopically observed.

In the brightest magnitude bin at $z\sim7$, $-24.5<M_\m{UV}<-23.5$ mag, we use HSC J023526$-$031737 at $z_\m{spec}=6.913$, which was first photometrically selected in \citet{2018PASJ...70S..10O}, and obtain a lower limit of the number density using the survey area of 102.7 deg$^2$ in \citet{2018PASJ...70S..10O}.
Similarly, in the faint magnitude bins of $-22.0<M_\m{UV}<-21.0$ mag at $z_\m{spec}=6.5-7.5$ and $-21.5<M_\m{UV}<-21.0$ mag at $z_\m{spec}=7.5-8.5$, we calculate lower limits of the number densities with galaxies in the JWST CEERS and GLASS fields from \cite{2023ApJS..269...33N}.
We use the survey area of 72 arcmin$^2$, which corresponds to the effective coverage of NIRSpec pointings in CEERS and GLASS.
Regarding the GLASS, we do not consider the gravitational lensing, resulting in a larger survey area and obtaining a conservative lower limit.

We also estimate the number densities of galaxies at $z\sim12$ and 14 using the recently spectroscopically confirmed galaxies at $z=12-14$.
At $z\sim12$, since the confirmed galaxy, GHZ2 at $z=12.34$ is the only galaxy in the brightest magnitude bin at $z\sim12$ in \citet{2023ApJS..265....5H}, we use the survey area calculated therein to estimate the number density.
At $z\sim14$, the two confirmed galaxies, GS-z14-0 ($M_\m{UV}=-20.8$ mag) and GS-14-1 ($M_\m{UV}=-19.0$ mag) are originally selected in \citet{2023arXiv231210033R}.
Since the brightest bin in \citet{2023arXiv231210033R} includes only GS-z14-0, we adopt the number density therein for the estimate at $M_\m{UV}=-20.8$ mag.
In the fainter magnitude bin ($M_\m{UV}=19.0$ mag), we obtain a lower limit of the number density using the survey volume in \citet{2023arXiv231210033R}.

We also obtain upper limits on the number densities of the brightest galaxies at $z\sim10$ and $12$ based on the spectroscopic results of the $z\sim10-12$ bright galaxy candidates, XMM3-3085, HD1, and HD2, which are found to be low-redshift passive galaxies (Section \ref{ss_spec_jwst}).
At $z\sim10$, we estimate the survey volume from the inverse of the number density in \citet{2020MNRAS.493.2059B}.
At $z\sim12$, we use the survey volume calculated in \citet{2022ApJ...929....1H}.

The 1$\sigma$ uncertainty of the number density is calculated by taking the Poisson confidence limit \citep{1986ApJ...303..336G} and cosmic variance into account.
We estimate the cosmic variance in the number densities following the procedures in \citet{2004ApJ...600L.171S}.
As the large-scale bias parameter needed for the cosmic variance calculation, we adopt $b=7$ obtained by the clustering analysis of galaxies at $z\sim7$ in \citet{2016ApJ...821..123H}, \redc{which is broadly comparable with recent JWST estimates for higher redshift galaxies \citep{2024MNRAS.533.2391D}.
Note that the value of the bias parameter does not change our conclusion because the Poisson error is much larger than the cosmic variance due to the small number of galaxies.
For example, if we adopt $b=10$, the error for the number density at $z\sim12$ and $M_\m{UV}=-20.5$ mag changes only by 3\%.}
In this way, the 1$\sigma$ uncertainty presented in this study includes both the Poisson uncertainty and the cosmic variance.

\begin{deluxetable}{cc}
\tablecaption{Spectroscopic Constraints on the Luminosity Function at Each Redshift}
\label{tab_LF}
\tablehead{\colhead{$M_\m{UV}$} & \colhead{$\Phi$} \\
\colhead{(ABmag)}& \colhead{($\m{Mpc^{-3}}\ \m{mag^{-1}}$)}}
\startdata
\multicolumn{2}{c}{$z\sim7\ (z=6.5-7.5)$}\\
$-23.7$ & $>2.3^{+5.3}_{-1.9}\times10^{-9}$\\
$-23.2$ & $1.6^{+3.7}_{-1.3}\times10^{-7}$\\
$-22.7$ & $4.8^{+4.6}_{-2.6}\times10^{-7}$\\
$-22.2$ & $>1.1^{+0.4}_{-0.4}\times10^{-6}$\\
$-21.7$ & $>2.4^{+3.3}_{-1.7}\times10^{-5}$\\
$-21.2$ & $>1.2^{+2.8}_{-1.1}\times10^{-5}$\\
\hline
\multicolumn{2}{c}{$z\sim8\ (z=7.5-8.5)$}\\
$-22.2$ & $>5.3^{+5.2}_{-2.9}\times10^{-7}$\\
$-21.2$ & $>1.3^{+3.1}_{-1.2}\times10^{-5}$\\
\hline
\multicolumn{2}{c}{$z\sim9\ (z=8.5-9.5)$}\\
$-22.0$ & $6.6^{+7.1}_{-4.7}\times10^{-6}$$^\dagger$\\
$-21.0$ & $>5.1^{+7.0}_{-3.8}\times10^{-6}$$^\dagger$\\
$-20.0$ & $>2.9^{+3.2}_{-2.2}\times10^{-5}$$^\dagger$\\
$-19.0$ & $>3.5^{+3.7}_{-2.4}\times10^{-5}$$^\dagger$\\
\hline
\multicolumn{2}{c}{$z\sim10\ (z=9.5-11.0)$}\\
$-23.5$ & $<6.8\times10^{-8}$\\
$-21.6$ & $1.0^{+2.3}_{-0.9}\times10^{-6}$$^\dagger$\\
$-20.6$ & $\redc{>8.7^{+20.5}_{-8.4}\times10^{-6}}$\\
$-19.6$ & $\redc{>2.6^{+2.8}_{-1.8}\times10^{-5}}$\\
$-18.6$ & $1.9^{+4.7}_{-1.9}\times10^{-4}$$^\dagger$\\
$-17.6$ & $6.3^{+15.8}_{-6.3}\times10^{-4}$$^\dagger$\\
\hline
\multicolumn{2}{c}{$z\sim12\ (z=11.0-13.5)$}\\
$-23.5$ & $<1.1\times10^{-7}$\\
$-20.5$ & $8.6^{+19.9}_{-7.4}\times10^{-6}$\\
$-20.1$ & $>8.8^{+9.1}_{-5.5}\times10^{-6}$$^\dagger$\\
$-18.7$ & $>6.6^{+6.0}_{-4.6}\times10^{-5}$$^\dagger$\\
\hline
\multicolumn{2}{c}{$z\sim14\ (z=13.5-15.0)$}\\
$-20.8$ & $3.7^{+8.7}_{-3.6}\times10^{-5}$\\
$-19.0$ & $>3.7^{+8.7}_{-3.6}\times10^{-5}$\\
\hline
\multicolumn{2}{c}{$z\sim16$}\\
$-21.9$ & $<9.8\times10^{-6}$$^\dagger$\\
\hline
\multicolumn{2}{c}{$z\sim7\ (z=6.5-7.5, \mathrm{individual})$$^*$}\\
$-22.4$ & $5.2^{+6.9}_{-3.4}\times10^{-7}$\\
$-21.9$ & $>1.8^{+0.6}_{-0.6}\times10^{-6}$\\
$-21.4$ & $>3.6^{+3.7}_{-2.3}\times10^{-5}$\\
\enddata
\tablecomments{Errors and upper and lower limits are $1\sigma$.\\
$^\dagger$ Taken from \citet{2024ApJ...960...56H}.\\
$^*$ Measurements when multiple sub-components are split into separate `galaxies' (see Section \ref{ss_LFsub}).}
\end{deluxetable}

\begin{figure*}
\centering
\begin{minipage}{1\hsize}
\begin{minipage}{0.49\hsize}
\begin{center}
\includegraphics[width=0.99\hsize, bb=7 9 430 358,clip]{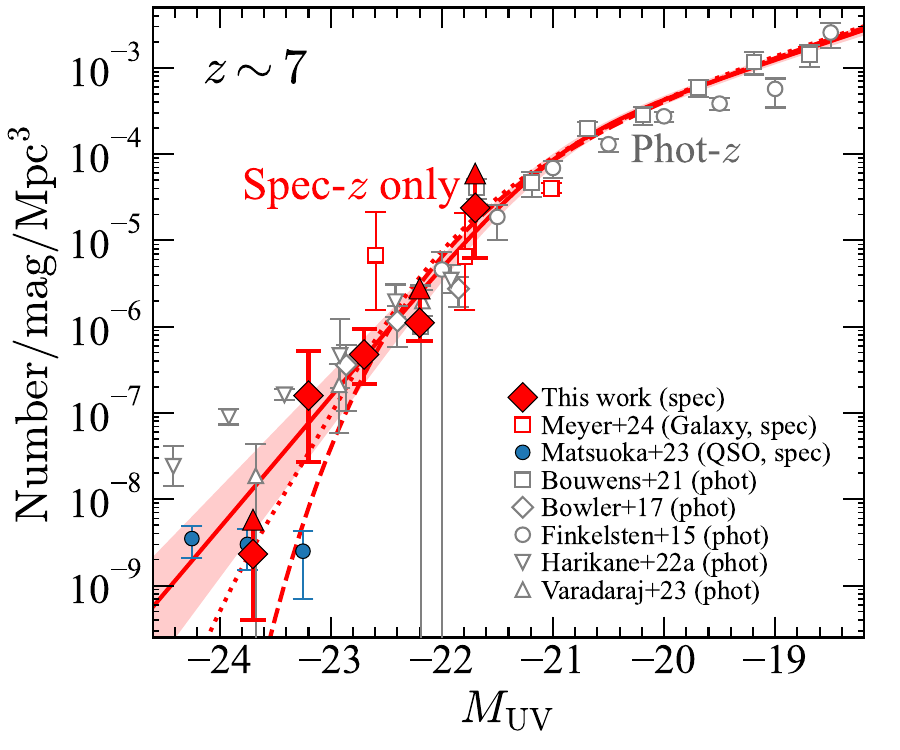}
\end{center}
\end{minipage}
\begin{minipage}{0.49\hsize}
\begin{center}
\includegraphics[width=0.99\hsize, bb=7 9 430 358,clip]{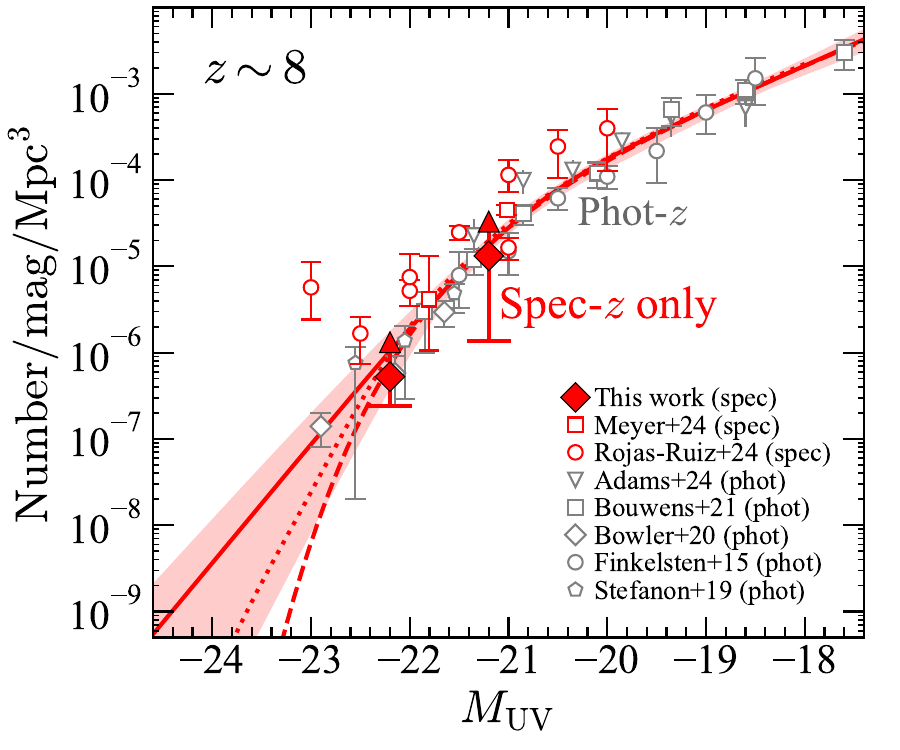}
\end{center}
\end{minipage}
\end{minipage}
\vspace{0.2cm}
\\
\centering
\begin{minipage}{1\hsize}
\begin{minipage}{0.49\hsize}
\begin{center}
\includegraphics[width=0.99\hsize, bb=7 9 430 358,clip]{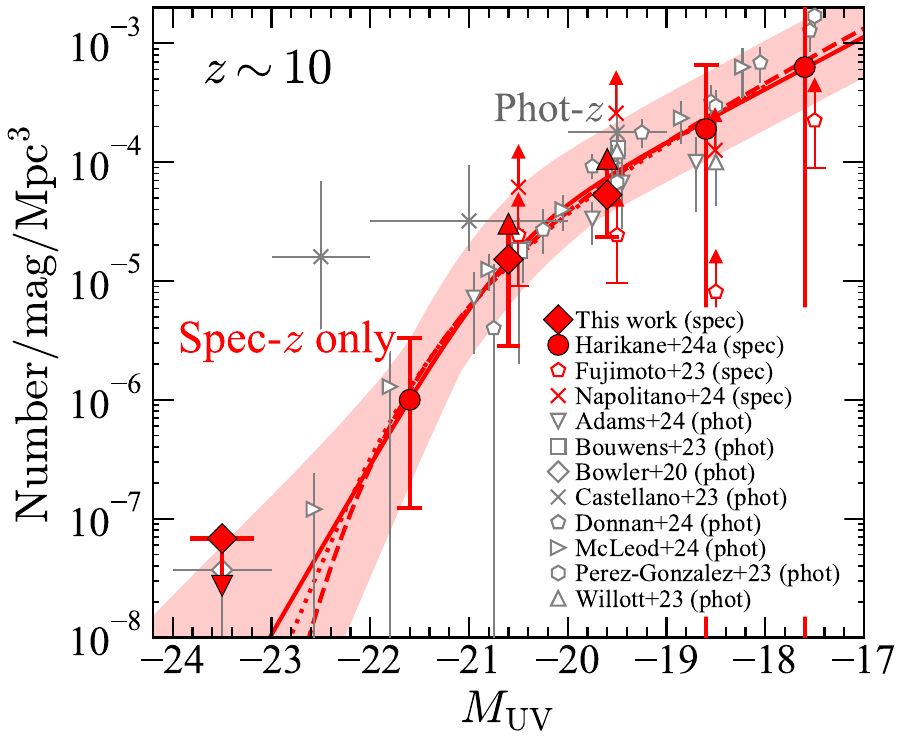}
\end{center}
\end{minipage}
\begin{minipage}{0.49\hsize}
\begin{center}
\includegraphics[width=0.99\hsize, bb=7 9 430 358,clip]{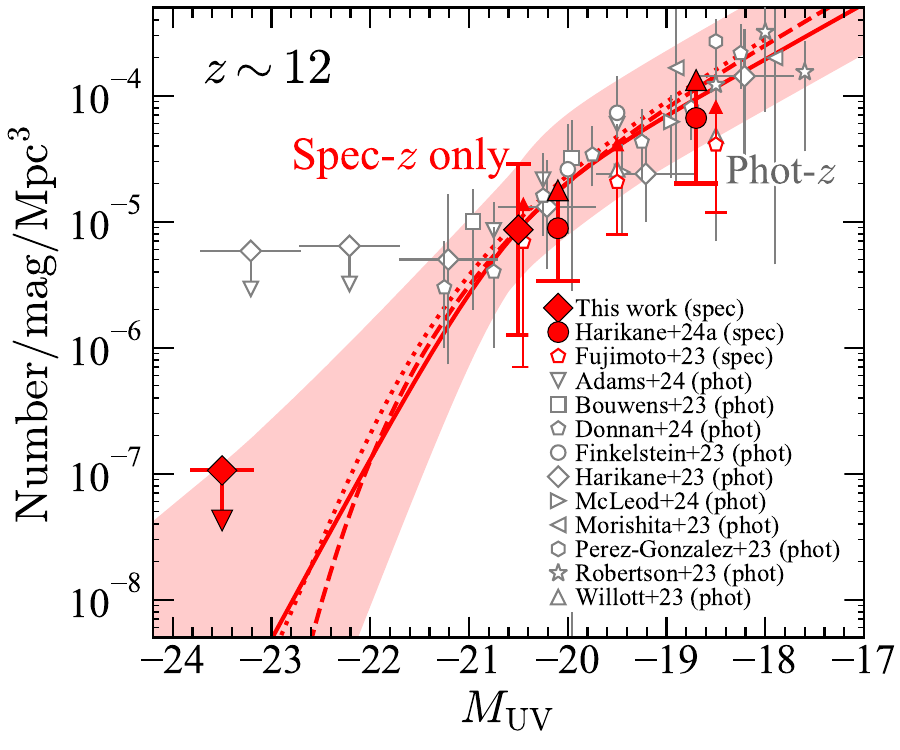}
\end{center}
\end{minipage}
\end{minipage}
\caption{
UV luminosity functions at $z\sim 7$ (upper-left), $z\sim8$ (upper-right), $z\sim10$ (lower-left), and $z\sim12$ (lower-right).
The red diamonds represent the number densities of galaxies with spectroscopic redshifts derived in this study (Table \ref{tab_LF}).
The errors include the cosmic variance (see text).
The red filled circles at $z\sim10$ and $12$ are the spectroscopic constraints from \citet{2024ApJ...960...56H}, and the blue circles at $z\sim7$ are number densities of spectroscopically-confirmed quasars (QSOs) in \citet{2023ApJ...949L..42M}.
\redc{The other red symbols show spectroscopic constraints in the literature \citep{2024arXiv240505111M,2024arXiv240800843R,2023arXiv230811609F,2024arXiv241010967N}.}
The data point at $z\sim12$ and $M_\m{UV}=-20.5$ mag in \citet{2024arXiv241010967N} is shifted by $+0.05$ mag for clarity.
The gray symbols are estimates based on photometric samples by previous studies \citep{2024ApJ...965..169A,2021AJ....162...47B,2023MNRAS.523.1009B,2017MNRAS.466.3612B,2020MNRAS.493.2059B,2023ApJ...948L..14C,2024arXiv240303171D,2015ApJ...810...71F,2023arXiv231104279F,2022ApJS..259...20H,2023ApJS..265....5H,2023ApJ...946L..35M,2024MNRAS.527.5004M,2023ApJ...951L...1P,2023arXiv231210033R,2019ApJ...883...99S,2023MNRAS.524.4586V,2023arXiv231112234W}.
The red solid and dashed lines are our best-fit double power-law and Schechter functions, respectively, and the shaded region shows the $1\sigma$ uncertainties for the double power-law fit (Table \ref{tab_LFpar}).
At $z\sim10$ and $z\sim12$, although the bright-end slope is fixed to $\beta=-4.60$ in the fitting, the allowed $\beta$ range ($z\sim10$: $\beta<-3.0$, $z\sim12$: $\beta<-2.4$) is plotted to show the uncertainty of the constraint on the bright end.
\redc{The red dotted lines are the lensed Schechter function calculated in \citet{2023MNRAS.523L..21F} assuming the size-luminosity relation in \citet{2015ApJS..219...15S}.}
At $z\sim7$, the spectroscopic constraints at the bright end prefer the double power-law \redc{or lensed Schechter function to the original} Schechter function.
}
\label{fig_uvlf}
\end{figure*}

\begin{figure}
\begin{center}
\includegraphics[width=0.99\hsize, bb=7 9 430 358,clip]{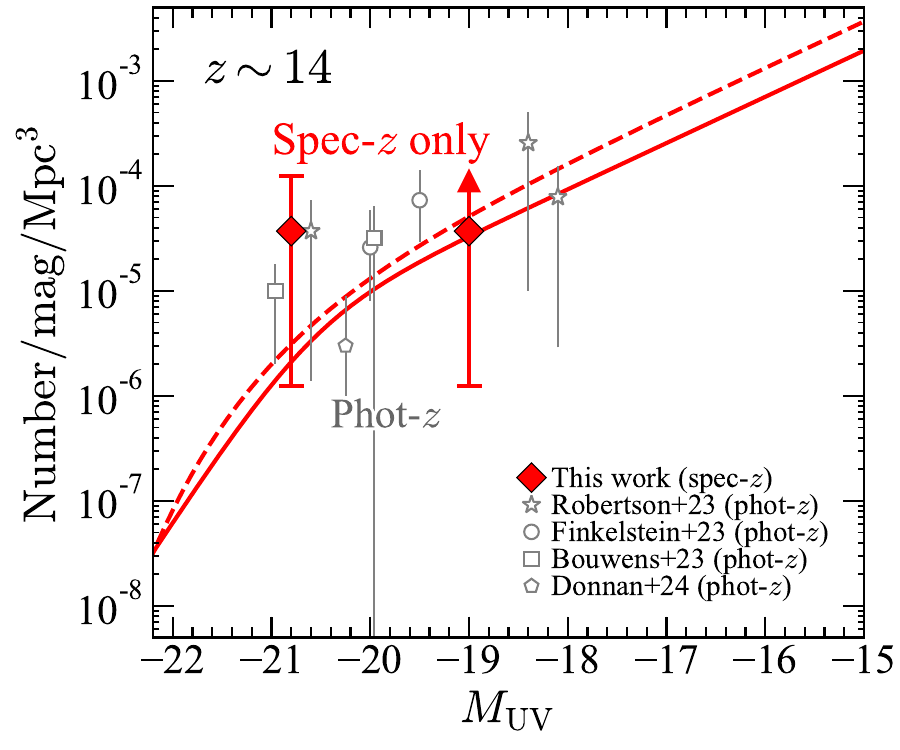}
\end{center}
\caption{
Same as Figure \ref{fig_uvlf} but at $z\sim14$.
The red solid and dashed lines are double power-law and Schechter functions, respectively, whose parameters are extrapolated from the $z\sim10-12$ results (Table \ref{tab_LFpar}).
}
\label{fig_uvlf_z14}
\end{figure}

\begin{figure*}
\centering
\begin{minipage}{1\hsize}
\begin{minipage}{0.49\hsize}
\begin{center}
\includegraphics[width=0.99\hsize, bb=7 9 430 358,clip]{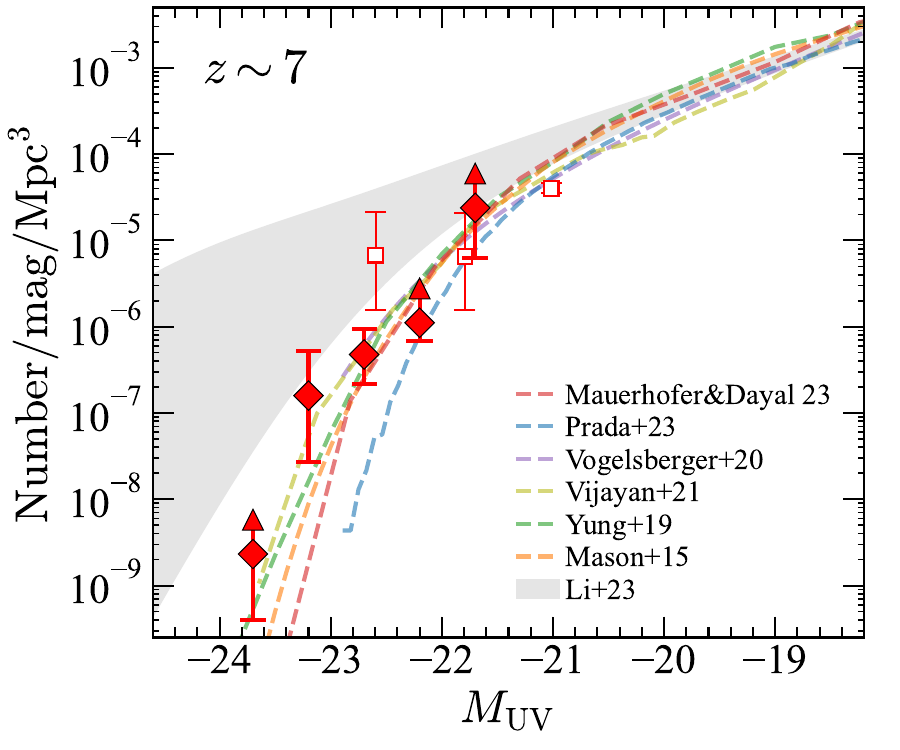}
\end{center}
\end{minipage}
\begin{minipage}{0.49\hsize}
\begin{center}
\includegraphics[width=0.99\hsize, bb=7 9 430 358,clip]{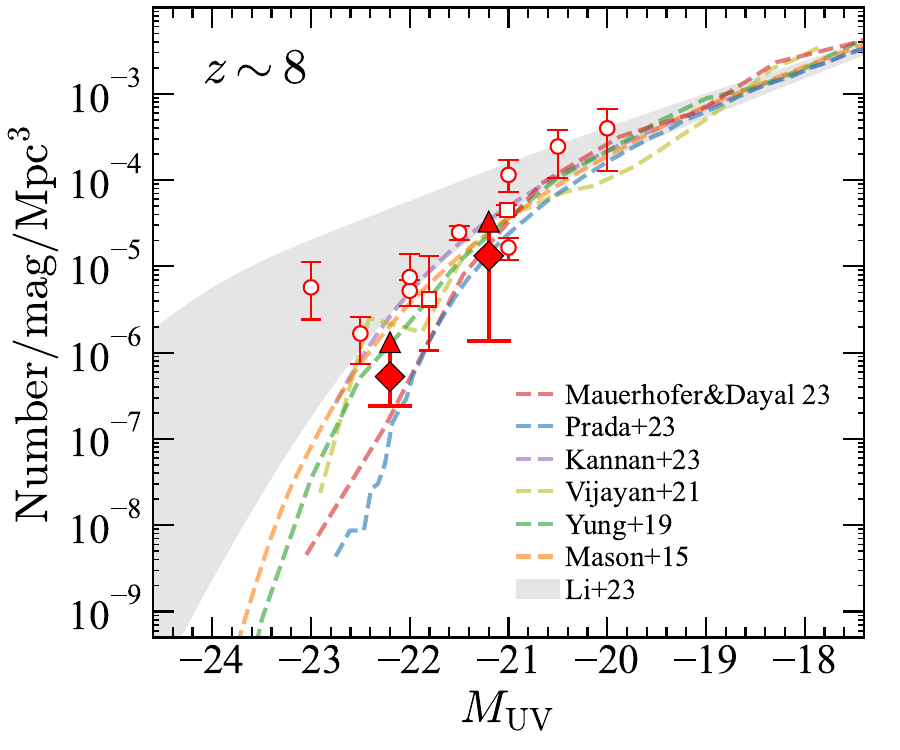}
\end{center}
\end{minipage}
\end{minipage}
\vspace{0.2cm}
\\
\centering
\begin{minipage}{1\hsize}
\begin{minipage}{0.49\hsize}
\begin{center}
\includegraphics[width=0.99\hsize, bb=7 9 430 358,clip]{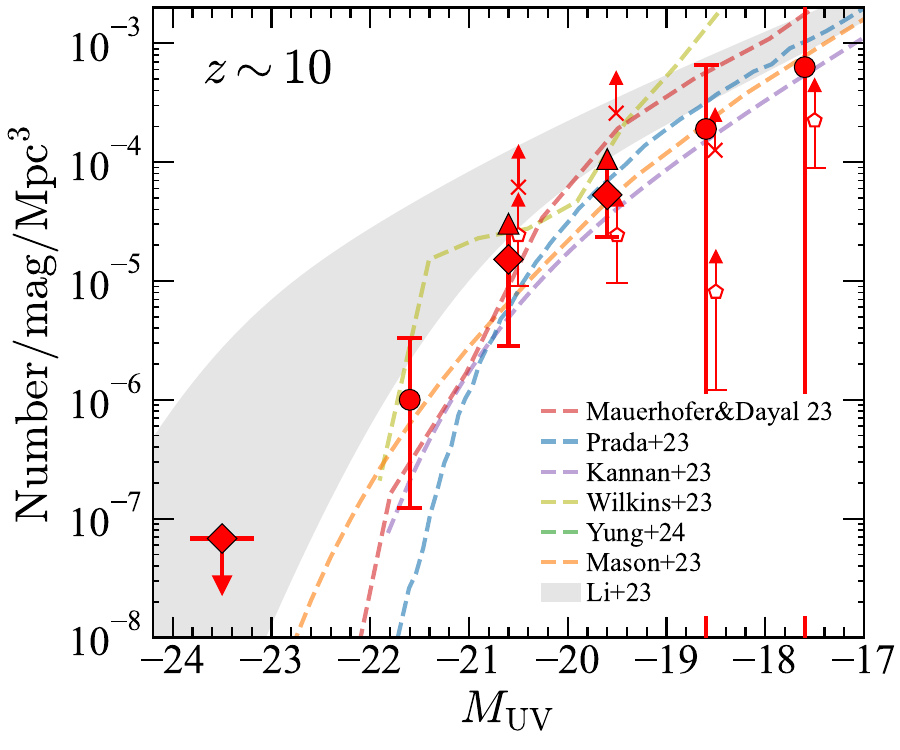}
\end{center}
\end{minipage}
\begin{minipage}{0.49\hsize}
\begin{center}
\includegraphics[width=0.99\hsize, bb=7 9 430 358,clip]{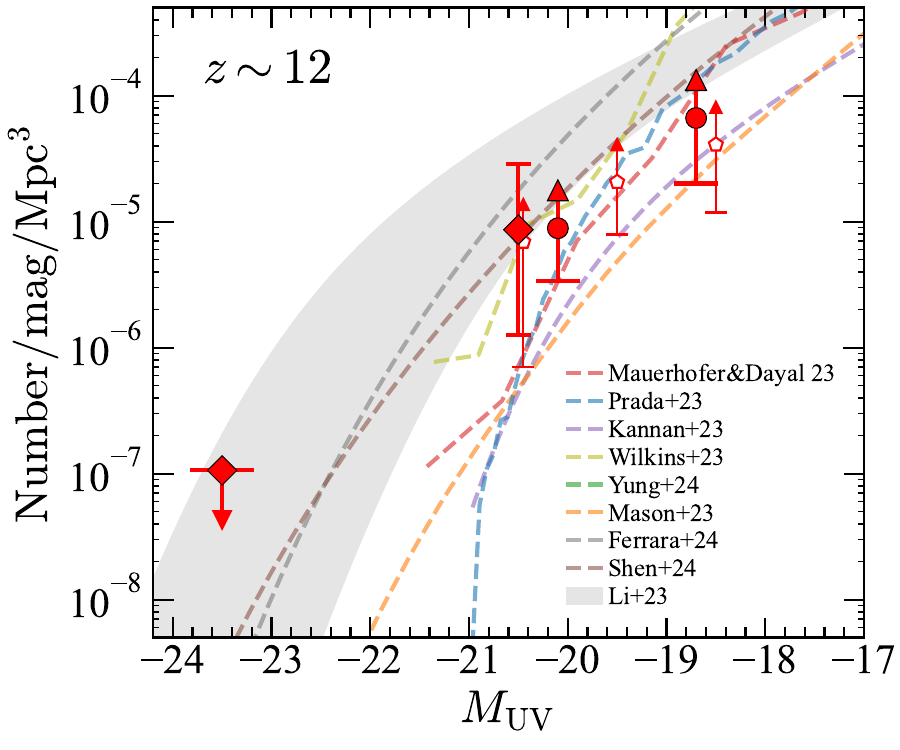}
\end{center}
\end{minipage}
\end{minipage}
\caption{
Comparison of the luminosity functions with theoretical predictions in the literature at $z\sim 7$ (upper-left), $z\sim8$ (upper-right), $z\sim10$ (lower-left), and $z\sim12$ (lower-right).
The red symbols show observational results based on the spectroscopically-confirmed galaxies obtained in this study (filled diamond), \citet[][filled circle]{2024ApJ...960...56H}, \citet[][open square]{2024arXiv240505111M}, \redc{\citet[][open circle]{2024arXiv240800843R}, \citet[][cross]{2024arXiv241010967N}}, and \citet[][open pentagon]{2023arXiv230811609F}.
The dashed lines and shaded region show predictions of theoretical and empirical models in \citet{2023MNRAS.526.2196M}, \citet{2023arXiv230411911P}, \citet{2020MNRAS.492.5167V}, \citet{2021MNRAS.501.3289V}, \cite{2019MNRAS.483.2983Y,2024MNRAS.527.5929Y}, \citet[their model with dust extinction]{2015ApJ...813...21M,2023MNRAS.521..497M}, \citet{2023MNRAS.524.2594K}, \citet{2023MNRAS.519.3118W}, \citet{2024A&A...684A.207F}, \citet[][]{2023arXiv231114662L}, \redc{and \citet{2024MNRAS.533.3923S}}.
For models in \citet[][]{2023arXiv231114662L}, a range of a maximum efficiency parameter of $\epsilon_\m{max}=0.2-1.0$ is plotted as the grey shaded region.
Spectroscopic constraints for bright galaxies with $-24<M_\m{UV}<-23$ mag at $z\sim7$ ($-21<M_\m{UV}<-20$ mag at $z\sim12$) are higher than the number densities of some model predictions.
}
\label{fig_uvlf_model}
\end{figure*}

\begin{figure}
\begin{center}
\includegraphics[width=0.99\hsize, bb=7 9 430 358,clip]{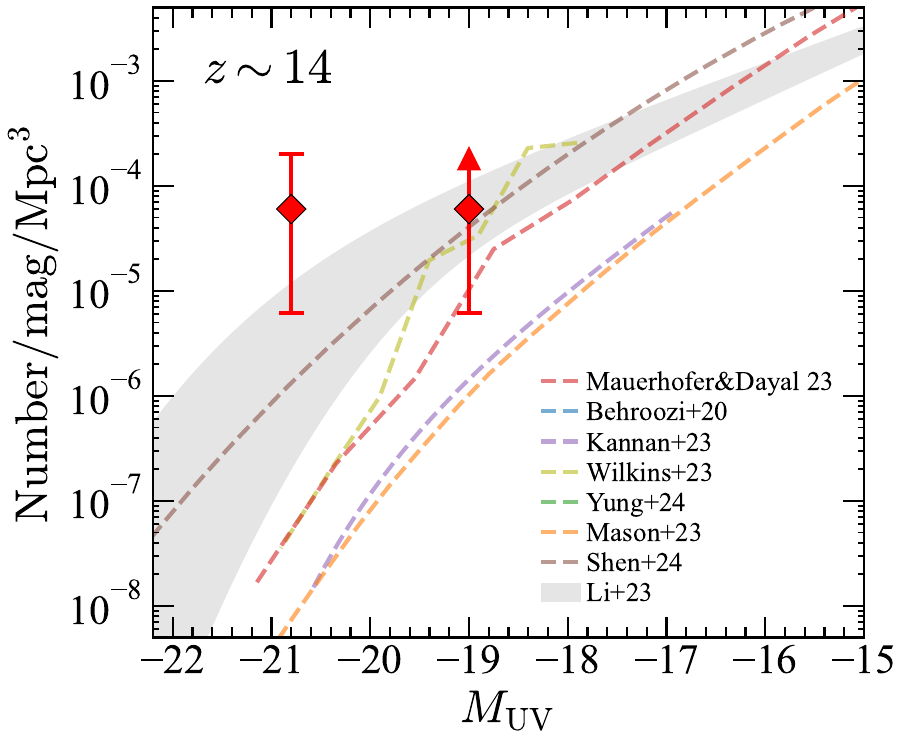}
\end{center}
\caption{
Same as Figure \ref{fig_uvlf_model} but at $z\sim14$.
Our constraints on the number density are higher than model predictions, especially at $M_\m{UV}=-21$ mag.
}
\label{fig_uvlf_model_z14}
\end{figure}

\subsection{Results}\label{ss_uvlf_result}

Figures \ref{fig_uvlf} and \ref{fig_uvlf_z14} show our constraints on the number densities of galaxies at $z\sim7$, $8$, $10$, $12$, and $14$ and Table \ref{tab_LF} summarizes them.
Our spectroscopic constraints are consistent with previous estimates of the number density in the literature based on photometric samples.
Our results are also consistent with the number densities of spectroscopically confirmed {\sc[Oiii]}$\lambda$5007 emitters at $z\sim7-8$ identified in the JWST FRESCO survey \citep{2024arXiv240505111M}, and spectroscopic lower limits obtained in \citet{2023arXiv230811609F}.

We fit our results at $z\sim7-12$ with the double-power-law function,
\begin{eqnarray}
&&\Phi(M_{\rm UV}) = \frac{\ln 10}{2.5} \phi^* \nonumber \\
&&\times \left[10^{0.4(\alpha+1)(M_{\rm UV} - M_{\rm UV}^*)} + 10^{0.4(\beta+1)(M_{\rm UV} - M_{\rm UV}^*)} \right]^{-1},
\label{eq_dpl}
\end{eqnarray}
and the Schechter function,
\begin{eqnarray}
\Phi(M_{\rm UV}) 
	&=& \frac{\ln 10}{2.5} \phi^* 10^{-0.4 (M_{\rm UV} - M_{\rm UV}^*) (\alpha +1)} \nonumber \\
	&& \times \exp \left( - 10^{-0.4 (M_{\rm UV} - M_{\rm UV}^*)} \right),
\label{eq_schechter}
\end{eqnarray}
\redc{where $\phi^*$ is the overall normalization, $M_{\rm UV}^*$ is the characteristic magnitude,  and $\alpha$ and $\beta$ are the faint and bright-end slopes, respectively.}
In fitting, we use the results of this study and \citet{2021AJ....162...47B} at $z\sim7-8$, and this study and \citet{2024ApJ...960...56H} at $z\sim10-12$, whose samples are not overlapping each other.
We also fix the faint end slope $\alpha=-2.10$ and the bright end slope $\beta=-4.60$ in the fitting at $z\sim10$ and $12$, and the characteristic UV magnitude to $M_\m{UV}^*=-20.60$ mag in the fit at $z\sim12$, based on the lower redshift results.
\redc{To take the lower and upper limits into account in the fit, we follow a $\chi^2$ minimization procedure presented in \citet{2012PASP..124.1208S} and define $\chi^2$ as follows
\begin{align}
\chi^2&=\sum_{i}\left(\frac{\Phi_{i}-\Phi_{\m{model},i}}{\sigma_{\Phi_{i}}}\right)^2\notag\\
&-2\sum_{j}\m{ln}\int^{\Phi_{\m{upper},j}}_{-\infty}d\Phi\ \m{exp}\left[-\frac{1}{2}\left(\frac{\Phi-\Phi_{\m{model},j}}{\sigma_{\Phi_{j}}}\right)^2\right]\notag\\
&-2\sum_{k}\m{ln}\int^{\infty}_{\Phi_{\m{lower},k}}d\Phi\ \m{exp}\left[-\frac{1}{2}\left(\frac{\Phi-\Phi_{\m{model},k}}{\sigma_{\Phi_{k}}}\right)^2\right],
\end{align}
where $\Phi$, $\sigma_\Phi$, and $\Phi_\m{model}$ are the observed number density, its uncertainty, and the number density from the fitted model, respectively.
The indices $i$, $j$, and $k$ correspond to the magnitudes bins with the best estimate, upper limit, and lower limit, and $\Phi_{\m{upper}}$ and $\Phi_{\m{lower}}$ are the $1\sigma$ upper and lower limits, respectively.}

The best-fit functions are plotted in Figure \ref{fig_uvlf} and the estimated parameters are summarized in Table \ref{tab_LFpar}.
At $z\sim7$, the bright end of the luminosity function is well described with the double-power-law function rather than the Schechter function with a $2\sigma$ level.
\redc{The bright end at $z\sim7$ is also consistent with the expectation of the lensed Schechter function in \citet{2023MNRAS.523L..21F}}.
In the other redshift bins, both the double-power-law and Schechter functions can reasonably reproduce our spectroscopic constraints of the number densities.
Wide photometric and spectroscopic surveys are needed to constrain the shape of the bright-end luminosity functions at $z\gtrsim8$.
In Figure \ref{fig_uvlf_z14}, we also plot the double-power-law and Schechter functions at $z\sim14$ whose parameters are estimated from the extrapolations using the best-fit results at $z\sim10-12$ (see Table \ref{tab_LFpar}).

\subsection{Comparison with Model Predictions}\label{ss_uvlf_model}

In Figures \ref{fig_uvlf_model} and \ref{fig_uvlf_model_z14}, we compare the spectroscopic constraints in this study and \citet{2024ApJ...960...56H} with theoretical model predictions.
At $z\sim7$, most of the models agree with the number densities of galaxies fainter than $M_\m{UV}\simeq-23.0$ mag, but more than half of the models predict lower number densities than the observations at the magnitude brighter than $M_\m{UV}\simeq-23.0$ mag.
At $z\sim8$ and $10$, the observed number densities can be reproduced by most of the models compared here.
At $z\sim12$, the number density of GHZ2 at $M_\m{UV}=-20.5$ mag is higher than most of the models except for \citet{2023MNRAS.519.3118W}, \citet{2024A&A...684A.207F}, and \citet{2023arXiv231114662L}, similar to the lower limit at $M_\m{UV}=-20.1$ mag obtained in \citet{2024ApJ...960...56H}.

The constraints at $z\sim14$ also show tension with the model predictions (Figure \ref{fig_uvlf_model_z14}).
Especially, the number density at $M_\m{UV}=-20.9$ mag is more than 100 times higher than most of the predictions.
This number density is based on the most distant galaxy recently confirmed, GS-z14-0 at $z=14.32$.
The redshift determination is considered to be reliable, given the unambiguous confirmation of the redshift via a clearly detected Lyman break in \citet{2024arXiv240518485C}.
Although there is a lower redshift galaxy ($z_\m{spec}=3.475$) that is close to GS-z14-0 with a separation of 0.\carcsec4, the lensing magnification factor is estimated to be small, less than $\mu=1.2$ \citep[see][]{2024arXiv240518485C}.
Thus this number density estimate can be considered to be reliable, unless the cosmic variance is much stronger than our estimate and the JADES Origins Field is significantly biased.
Larger area datasets are needed to understand the real effect of the cosmic variance.
Physical origins for these discrepancies between the observations and model predictions at $z\sim12-14$ are discussed in Section \ref{ss_dis_bright}.

\subsection{SFR Density}\label{ss_uvlf_sfrd}

\begin{figure}
\begin{center}
\includegraphics[width=0.99\hsize, bb=6 7 426 319,clip]{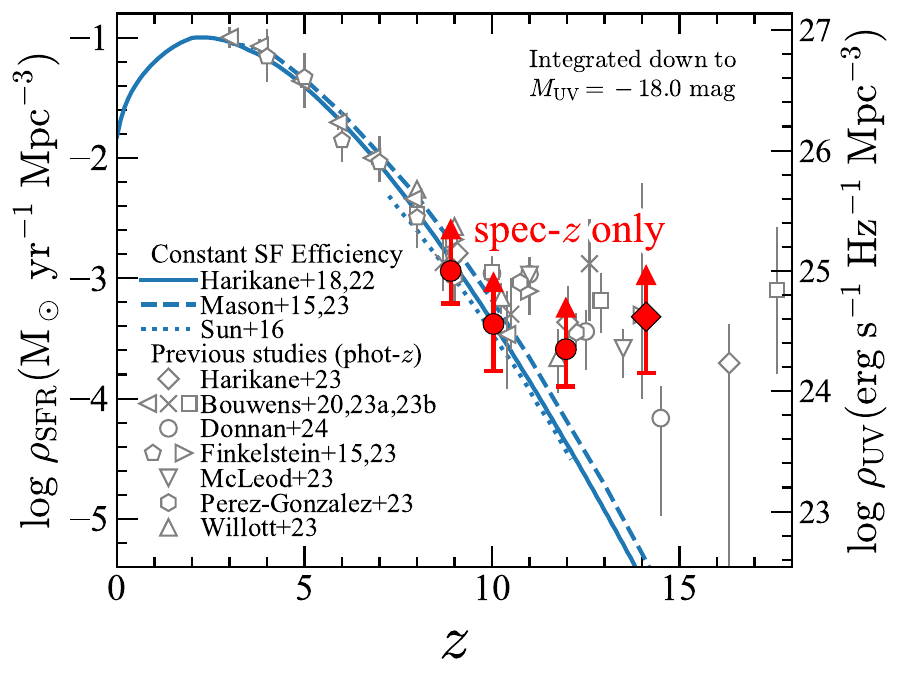}
\end{center}
\caption{
Cosmic SFR density evolution.
The red diamond represents a lower limit on the cosmic SFR density at $z\sim14$ obtained in this study integrated down to $M_\m{UV}=-18.0\ \m{mag}$ (corresponding to $\m{SFR}_\m{UV}=0.8\ M_\odot\ \m{yr^{-1}}$, based on the \citet{1955ApJ...121..161S} IMF with a conversion factor of $\m{SFR}/L_\mathrm{UV}=1.15\times10^{-28}\ M_\odot\ \m{yr}^{-1}/(\m{erg\ s^{-1}\ Hz^{-1}})$).
The error includes both the $1\sigma$ Poisson error and the cosmic variance.
The red circles are spectroscopic lower limits obtained in \citet{2024ApJ...960...56H}.
The blue curves are predictions of the constant star formation (SF) efficiency models of \citet[][solid]{2018PASJ...70S..11H,2022ApJS..259...20H}, \citet[][dashed]{2015ApJ...813...21M,2023MNRAS.521..497M}, and \citet[][dotted]{2016MNRAS.460..417S}.
The obtained lower limits of the SFR densities at $z\sim12-14$ are higher than the model predictions.
The gray open symbols are estimates of previous studies using photometric samples \citep{2020ApJ...902..112B,2023MNRAS.523.1009B,2023MNRAS.523.1036B,2024arXiv240303171D,2015ApJ...810...71F,2023arXiv231104279F,2024MNRAS.527.5004M,2023ApJS..265....5H,2023ApJ...951L...1P,2023arXiv231112234W}.
}
\label{fig_cSFR}
\end{figure}

Based on the two recently-confirmed galaxies in \citet{2024arXiv240518485C}, we calculate the lower limit of the cosmic star formation rate (SFR) density at $z\sim14$.
We use the survey volume discussed in Section \ref{ss_volume}, and convert the observed UV luminosities of the two galaxies to SFRs using the following equation assuming the \citet{1955ApJ...121..161S} IMF,
\begin{equation}\label{eq_SFR_LUV}
\m{SFR}\ (M_\odot\ \m{yr^{-1}})=1.15\times10^{-28} L_\m{UV}\ (\m{erg\ s^{-1} Hz^{-1}}).
\end{equation}
\redc{We assume the \citet{1955ApJ...121..161S} IMF for comparison with the literature}.
Figure \ref{fig_cSFR} shows our spectroscopic lower limit based on the two galaxies brighter than $M_\m{UV}=-18.0$ mag, corresponding to the SFR of $\m{SFR}_\m{UV}=0.8\ M_\odot\ \m{yr^{-1}}$, and Table \ref{tab_cSFR} summarizes the measurements in this study and in \citet{2024ApJ...960...56H}.
We also plot estimates based on the photometric samples in the literature \citep{2020ApJ...902..112B,2023MNRAS.523.1009B,2023MNRAS.523.1036B,2024arXiv240303171D,2015ApJ...810...71F,2023arXiv231104279F,2024MNRAS.527.5004M,2023ApJS..265....5H,2023ApJ...951L...1P,2023arXiv231112234W}.
Since some of these studies calculate the SFR densities with different integration limits from $M_\m{UV}=-18.0$ mag, we have corrected their results based on the difference between the SFR density integrated down to their limit and that down to $M_\m{UV}=-18.0$ mag using their fiducial luminosity function, in the same manner as \citet{2023MNRAS.523.1009B}.
Our lower limit at $z\sim14$ is consistent with these photometric estimates and is more than 10 times higher than the model predictions assuming a constant star formation efficiency \citep{2018PASJ...70S..11H,2022ApJS..259...20H,2015ApJ...813...21M,2023MNRAS.521..497M,2016MNRAS.460..417S}.

\begin{deluxetable*}{cccccc}
\tablecaption{Fit Parameters for Luminosity Functions}
\label{tab_LFpar}
\tablehead{\colhead{Redshift} & \colhead{Fitted Function} & \colhead{$M_\m{UV}^*$} & \colhead{$\m{log}\phi^*$} & \colhead{$\alpha$} & \colhead{$\beta$} \\
\colhead{}& \colhead{}& \colhead{(ABmag)}& \colhead{($\m{Mpc^{-3}}$)} &  \colhead{}& \colhead{}}
\startdata
$z\sim7$ & DPL & $-21.01^{+0.30}_{-0.26}$& $-3.74^{+0.23}_{-0.22}$& $-2.08^{+0.12}_{-0.11}$& $-4.81^{+0.50}_{-0.56}$\\
& Schechter & $-20.89^{+0.27}_{-0.23}$& $-3.51^{+0.22}_{-0.20}$& $-1.97^{+0.14}_{-0.12}$\\
$z\sim8$ & DPL & $-20.88^{+0.25}_{-0.67}$& $-4.10^{+0.22}_{-0.65}$& $-2.27^{+0.16}_{-0.25}$& $-4.45^{+0.31}_{-2.04}$\\
& Schechter & $-20.82^{+0.36}_{-0.55}$& $-3.92^{+0.34}_{-0.51}$& $-2.16^{+0.24}_{-0.21}$\\
$z\sim10$ & DPL & $\redc{-20.61^{+0.71}_{-0.90}}$& $\redc{-4.50^{+0.66}_{-0.68}}$& $-2.10$(fixed)& $-4.60$(fixed)\\
& Schechter & $\redc{-20.65^{+0.65}_{-1.10}}$& $\redc{-4.43^{+0.56}_{-0.79}}$& $-2.10$(fixed)\\
$z\sim12$ & DPL & $-20.60$(fixed)& $-4.82^{+0.52}_{-0.40}$& $-2.10$(fixed)& $-4.60$(fixed)\\
& Schechter & $-20.60$(fixed)& $-4.67^{+0.51}_{-0.43}$& $-2.10$(fixed)\\
$z\sim14$ & DPL & $-20.60$(fixed)& $-5.14^\dagger$& $-2.10$(fixed)& $-4.60$(fixed)\\
& Schechter & $-20.60$(fixed)& $-4.86^\dagger$& $-2.10$(fixed)\\
\hline
$z\sim7$$^*$ & DPL & $-21.08^{+0.21}_{-0.22}$& $-3.80^{+0.17}_{-0.20}$& $-2.10^{+0.10}_{-0.11}$& $-5.56^{+0.70}_{-0.92}$\\
\enddata
\tablecomments{Errors are $1\sigma$.\\
$^\dagger$ Extrapolated from the results at $z\sim10$ and 12.\\
$^*$ Paremeters when multiple sub-components are split into separate `galaxies' (see Section \ref{ss_LFsub}).}
\end{deluxetable*}

\begin{deluxetable}{ccc}
\tablecaption{Spectroscopic Constraints on Cosmic UV Luminosity Density and SFR Density}
\label{tab_cSFR}
\tablehead{\colhead{Redshift} & \colhead{$\m{log}\rho_\m{UV}$} & \colhead{$\m{log}\rho_\m{SFR,UV}$} \\
\colhead{}& \colhead{($\m{erg\ s^{-1}\ Hz^{-1}\ Mpc^{-3}}$)}& \colhead{($\m{M_\odot\ yr^{-1}\ Mpc^{-3}}$)}}
\startdata
$z\sim9$$^\dagger$ & $ >25.00_{-0.27}^{+0.23}$ &$ >-2.94_{-0.27}^{+0.23}$\\
$z\sim10$$^\dagger$ & $ >24.56_{-0.39}^{+0.38}$ &$ >-3.38_{-0.39}^{+0.38}$\\
$z\sim12$$^\dagger$ & $ >24.35_{-0.31}^{+0.23}$ &$ >-3.59_{-0.31}^{+0.23}$\\
$z\sim14$ & $ >24.74_{-0.47}^{+0.39}$ &$ >-3.32_{-0.47}^{+0.39}$\\
\enddata
\tablecomments{Errors are $1\sigma$. $\rho_\m{SFR,UV}$ is the SFR density based on the \citet{1955ApJ...121..161S} IMF without dust extinction correction.\\
$^\dagger$ Taken from \citet{2024ApJ...960...56H}.}
\end{deluxetable}

\begin{figure*}
\centering
\begin{minipage}{1\hsize}
\begin{minipage}{0.49\hsize}
\begin{center}
\includegraphics[width=1\hsize, bb=0 0 576 163]{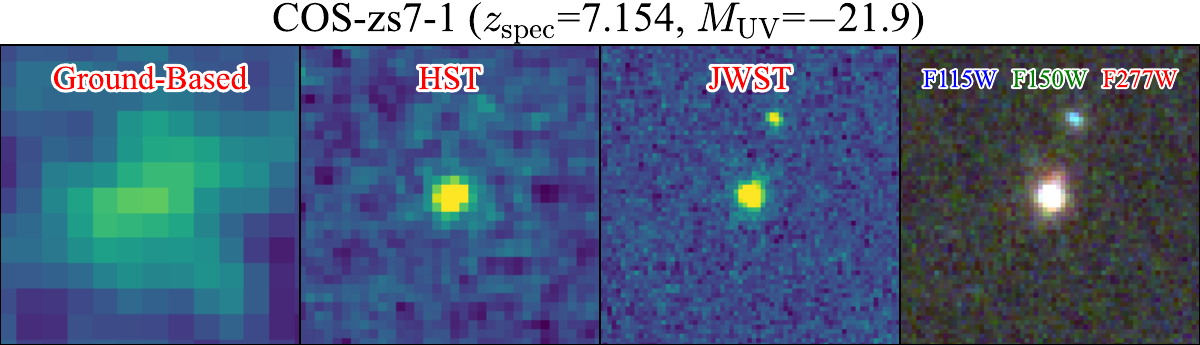}
\end{center}
\end{minipage}
\begin{minipage}{0.49\hsize}
\begin{center}
\includegraphics[width=1\hsize, bb=0 0 576 163]{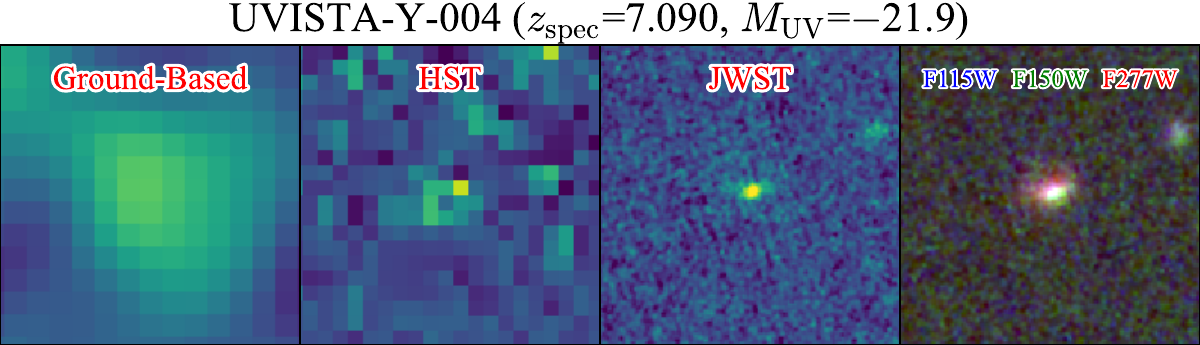}
\end{center}
\end{minipage}
\end{minipage}
\vspace{0.1cm}
\\
\centering
\begin{minipage}{1\hsize}
\begin{minipage}{0.49\hsize}
\begin{center}
\includegraphics[width=1\hsize, bb=0 0 576 163]{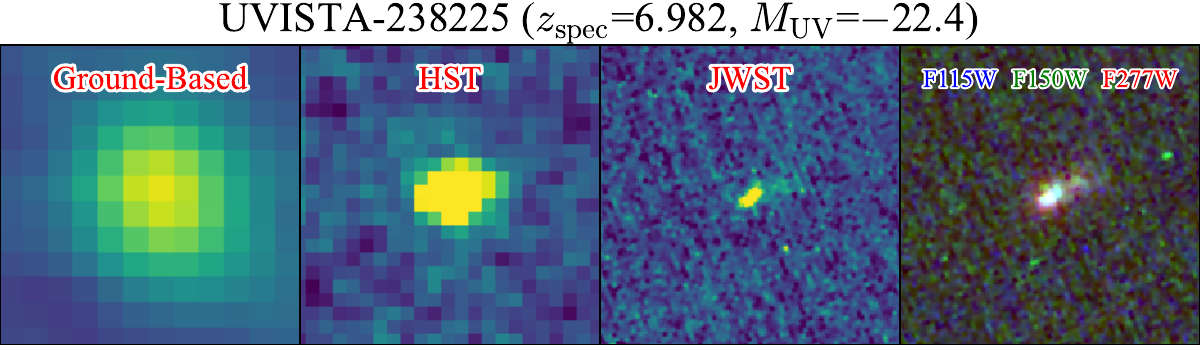}
\end{center}
\end{minipage}
\begin{minipage}{0.49\hsize}
\begin{center}
\includegraphics[width=1\hsize, bb=0 0 576 163]{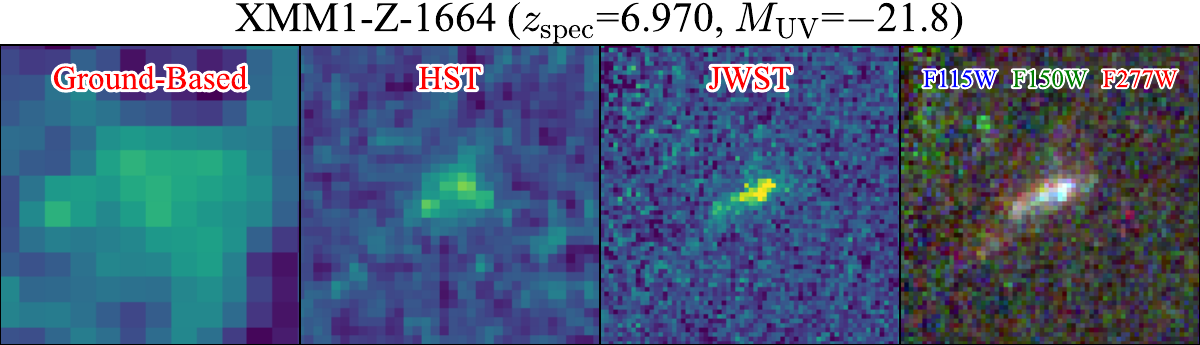}
\end{center}
\end{minipage}
\end{minipage}
\vspace{0.1cm}
\\
\centering
\begin{minipage}{1\hsize}
\begin{minipage}{0.49\hsize}
\begin{center}
\includegraphics[width=1\hsize, bb=0 0 576 163]{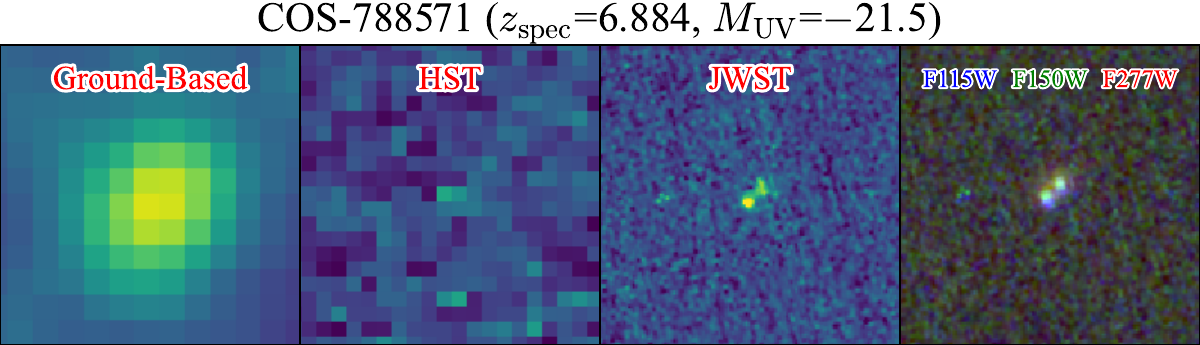}
\end{center}
\end{minipage}
\begin{minipage}{0.49\hsize}
\begin{center}
\includegraphics[width=1\hsize, bb=0 0 576 163]{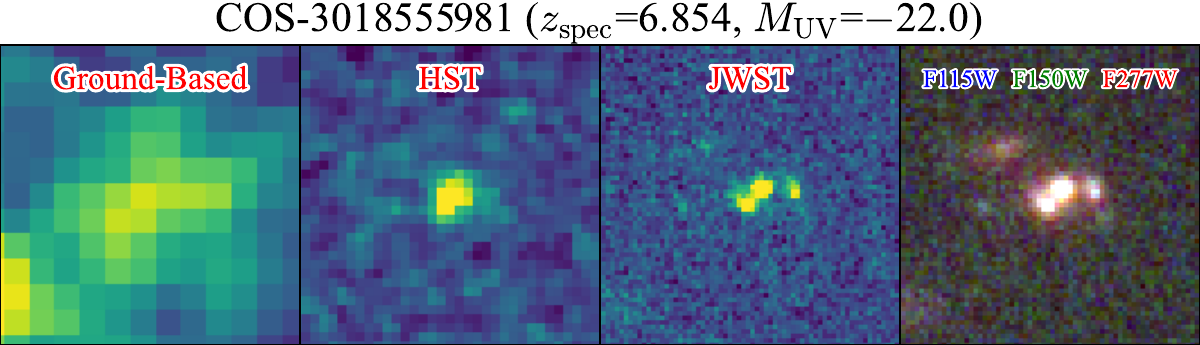}
\end{center}
\end{minipage}
\end{minipage}
\vspace{0.1cm}
\\
\centering
\begin{minipage}{1\hsize}
\begin{minipage}{0.49\hsize}
\begin{center}
\includegraphics[width=1\hsize, bb=0 0 576 163]{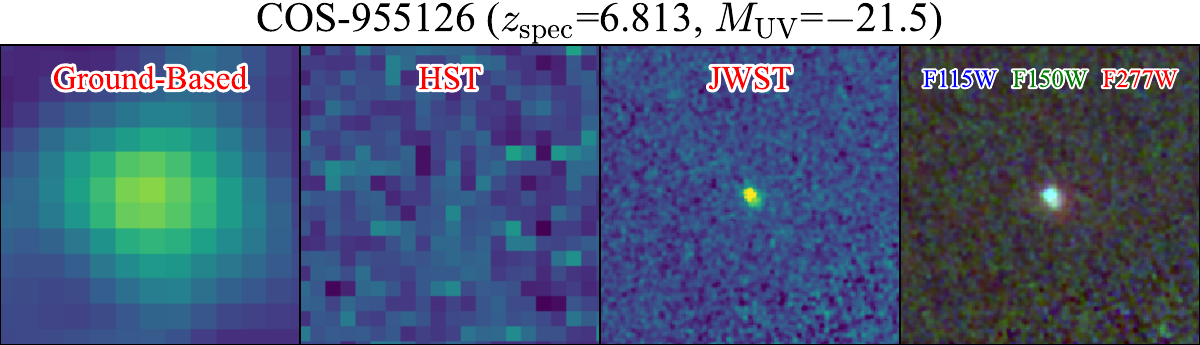}
\end{center}
\end{minipage}
\begin{minipage}{0.49\hsize}
\begin{center}
\includegraphics[width=1\hsize, bb=0 0 576 163]{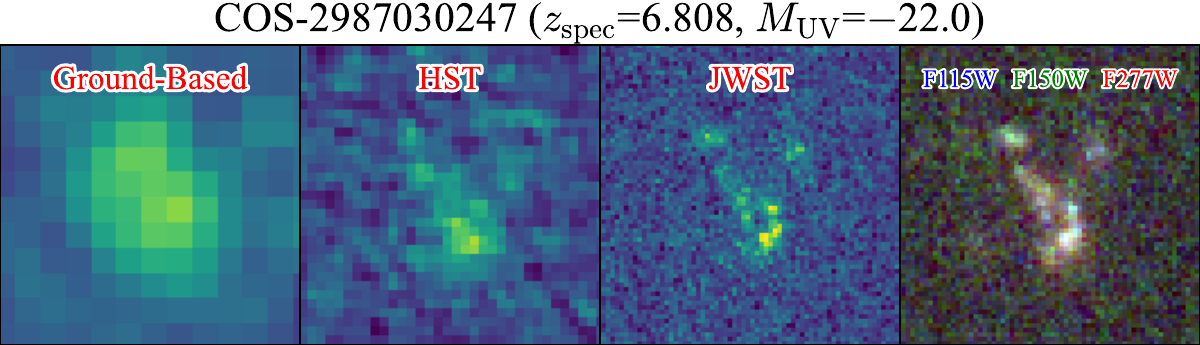}
\end{center}
\end{minipage}
\end{minipage}
\vspace{0.1cm}
\\
\centering
\begin{minipage}{1\hsize}
\begin{minipage}{0.49\hsize}
\begin{center}
\includegraphics[width=1\hsize, bb=0 0 576 163]{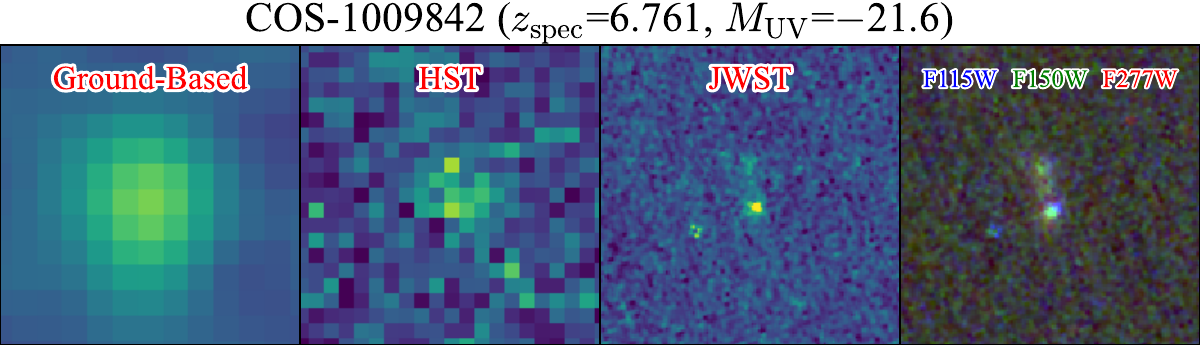}
\end{center}
\end{minipage}
\begin{minipage}{0.49\hsize}
\begin{center}
\includegraphics[width=1\hsize, bb=0 0 576 163]{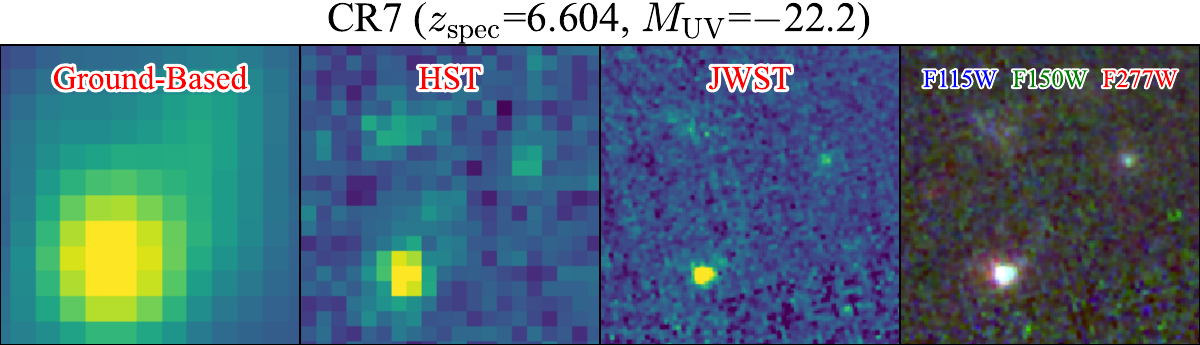}
\end{center}
\end{minipage}
\end{minipage}
\vspace{0.1cm}
\\
\centering
\begin{minipage}{1\hsize}
\begin{minipage}{0.49\hsize}
\begin{center}
\includegraphics[width=1\hsize, bb=0 0 576 163]{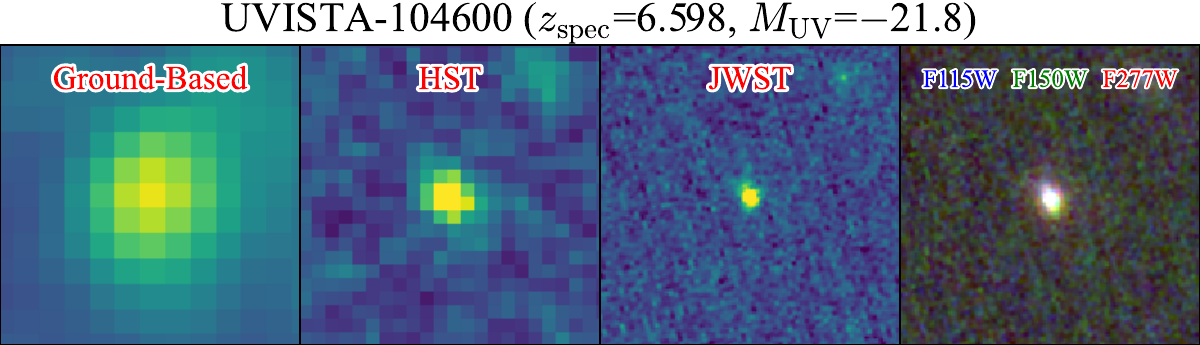}
\end{center}
\end{minipage}
\begin{minipage}{0.49\hsize}
\begin{center}
\includegraphics[width=1\hsize, bb=0 0 576 163]{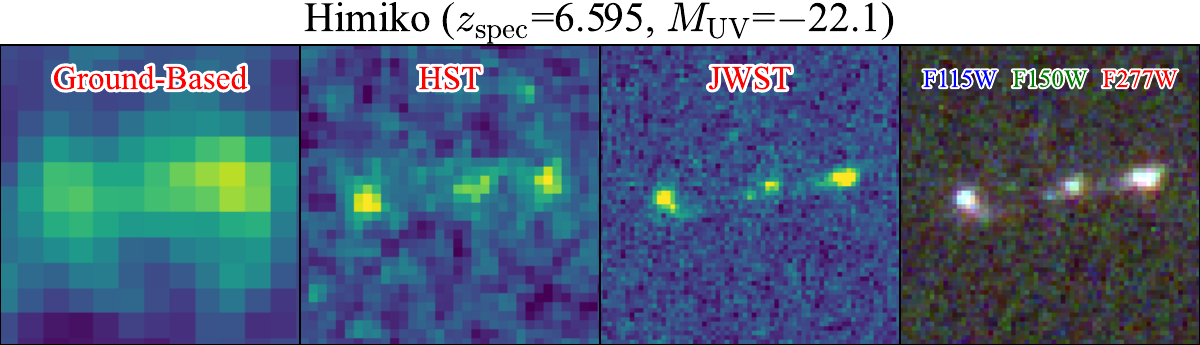}
\end{center}
\end{minipage}
\end{minipage}
\caption{
$2\arcsec\times2\arcsec$ cutout images of selected 12 galaxies used in this study with the UV magnitudes of $M_\m{UV}\leq-21.5$.
From left to right we show the ground-based $H$-band, the HST/WFC3 $J_{125}$, $JH_{140}$, or $H_{160}$, the JWST/NIRCam F115W, and the JWST/NIRCam false color images.
The false color images are made from the F115W, F150W, and F277W data.
The high-resolution JWST image with the PSF FWHM of $\sim0.\carcsec07$ allows us to identify individual sub-components that are blended in the HST and ground-based images whose PSFs are $\sim0.\carcsec2$ and $\sim1\arcsec$, respectively.
Some objects (e.g., COS-788571 and COS-955126) are not clearly detected in the HST images because these images are taken in the COSMOS-DASH program whose survey depth is shallow, $\sim25$ mag.
}
\label{fig_snapshot}
\end{figure*}

\begin{figure*}
\centering
\begin{minipage}{1\hsize}
\begin{minipage}{0.15\hsize}
\begin{center}
\includegraphics[width=1\hsize, bb=0 0 144 162]{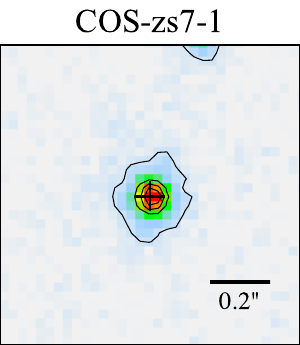}
\end{center}
\end{minipage}
\begin{minipage}{0.15\hsize}
\begin{center}
\includegraphics[width=1\hsize, bb=0 0 144 162]{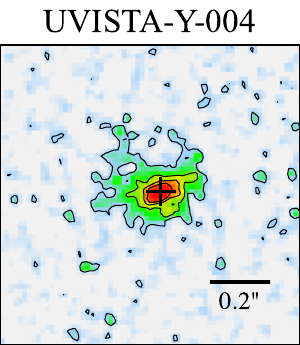}
\end{center}
\end{minipage}
\begin{minipage}{0.15\hsize}
\begin{center}
\includegraphics[width=1\hsize, bb=0 0 144 162]{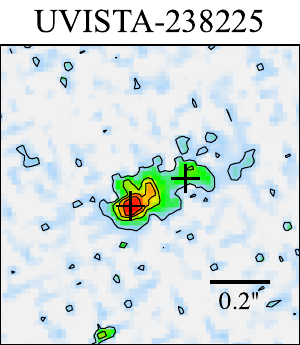}
\end{center}
\end{minipage}
\begin{minipage}{0.15\hsize}
\begin{center}
\includegraphics[width=1\hsize, bb=0 0 144 162]{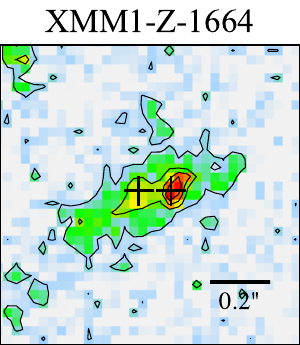}
\end{center}
\end{minipage}
\begin{minipage}{0.15\hsize}
\begin{center}
\includegraphics[width=1\hsize, bb=0 0 144 162]{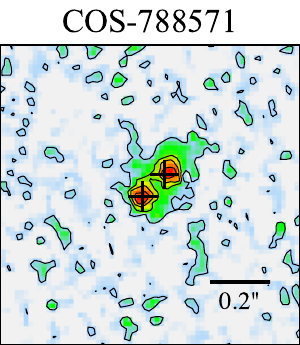}
\end{center}
\end{minipage}
\begin{minipage}{0.15\hsize}
\begin{center}
\includegraphics[width=1\hsize, bb=0 0 144 162]{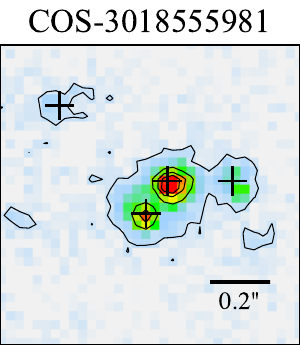}
\end{center}
\end{minipage}
\end{minipage}
\vspace{0.1cm}
\\
\centering
\begin{minipage}{1\hsize}
\begin{minipage}{0.15\hsize}
\begin{center}
\includegraphics[width=1\hsize, bb=0 0 144 162]{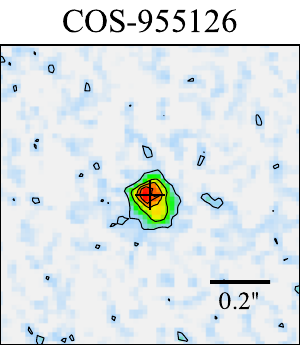}
\end{center}
\end{minipage}
\begin{minipage}{0.15\hsize}
\begin{center}
\includegraphics[width=1\hsize, bb=0 0 144 162]{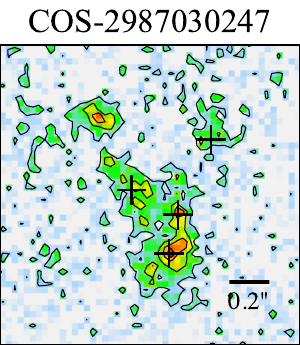}
\end{center}
\end{minipage}
\begin{minipage}{0.15\hsize}
\begin{center}
\includegraphics[width=1\hsize, bb=0 0 144 162]{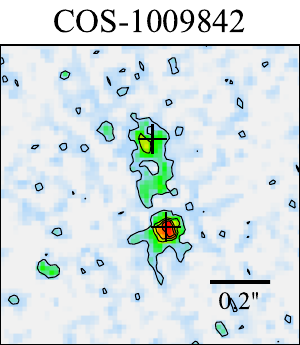}
\end{center}
\end{minipage}
\begin{minipage}{0.15\hsize}
\begin{center}
\includegraphics[width=1\hsize, bb=0 0 144 162]{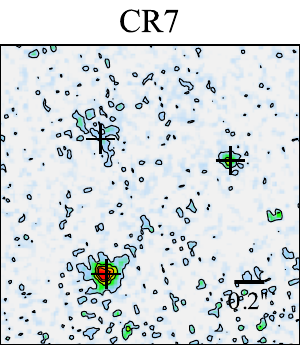}
\end{center}
\end{minipage}
\begin{minipage}{0.15\hsize}
\begin{center}
\includegraphics[width=1\hsize, bb=0 0 144 162]{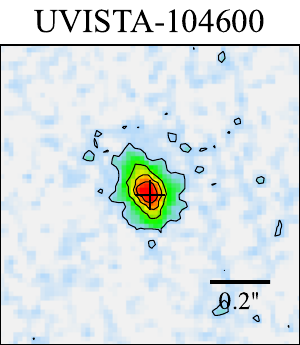}
\end{center}
\end{minipage}
\begin{minipage}{0.15\hsize}
\begin{center}
\includegraphics[width=1\hsize, bb=0 0 144 162]{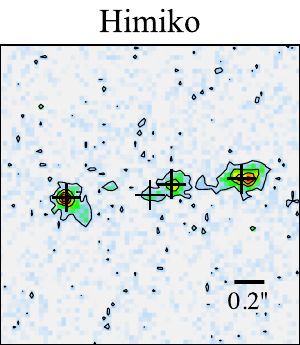}
\end{center}
\end{minipage}
\end{minipage}
\caption{
JWST/NIRCam F115W cutout images of the selected bright galaxies at $z\sim7$.
The contours are drawn from 10\% to 70\% of the peak surface brightness with a 20\% interval to highlight any extended emission.
The black crosses show the positions of the sub-components (see Section \ref{ss_sub}).
We find that more than half of the sources have multiple sub-components. 
The scale of $0.\carcsec2$, corresponding to 1.068 kpc at $z=7.0$, is displayed as a scale bar in the lower right of each image.
}
\label{fig_jwstsnap}
\end{figure*}

\section{Morphology}\label{ss_mor}

\subsection{Multiple Sub-Components in Bright Galaxies at $z\sim7$}\label{ss_sub}

We investigate morphologies of bright galaxies spectroscopically confirmed at $z\gtrsim7$ using high-resolution HST and/or JWST images.
Among the 50 galaxies at $z\sim7-8$ used in this study, 23 galaxies are observed and clearly detected with either HST/WFC3 and/or JWST/NIRCam.
From the 23 galaxies, we select 12 galaxies that are observed with both HST and JWST and are brighter than $M_\m{UV}=-21.5$ mag.
Figure \ref{fig_snapshot} shows the selected 12 galaxies at $z_\m{spec}=6.595-7.154$.
We find that some galaxies show clumpy structures spatially extended up to $\sim1\arcsec$ ($\sim5\ \m{kpc}$).
In addition, the JWST/NIRCam F115W image with a high spatial resolution of $\sim0.\carcsec07$ allows us to identify multiple sub-components/clumps that are not identified with the HST images whose spatial resolution is $\sim0.\carcsec2$ (e.g., COS-3018555981).
Other rest-UV images (e.g., F150W) also show similar clumpy structures in these galaxies.

To quantitatively discuss the sub-components in the galaxies, we run SExtractor \citep{1996A&AS..117..393B} on the JWST and HST images with a parameter set of {\tt DETECT\_MINAREA$=$5}, {\tt DETECT\_THRESH=3}, {\tt ANALYSIS\_THRESH=3}, {\tt DEBLEND\_NTHRESH$=$32}, and {\tt DEBLEND\_MINCOUNT$=$0.01}, following \citet{2017MNRAS.466.3612B}.
We then visually inspect the sub-components detected at $>5\sigma$ significance levels to remove any spurious detections and foreground objects.
In Figure \ref{fig_jwstsnap}, we show the surface brightness maps of the 12 galaxies in the JWST/NIRCam F115W images and the positions of the sub-components detected with SExtractor.
We find that 8 of the 12 galaxies in Figure \ref{fig_jwstsnap} have more than one component, indicating that $66\pm14\%$ of the bright galaxies exhibit multiple sub-components.
Even if we include galaxies observed only with HST, the fraction of galaxies with multiple sub-components is still high, $60\pm12\%$.
Such a high fraction of galaxies with multiple sub-components at $z\sim7$ is comparable to or slightly higher than those of similarly bright galaxies ($M_\m{UV}\sim-22$ mag) at $z\sim6-7$ ($\sim40-60\%$ e.g., \citealt{2013ApJ...773..153J,2013AJ....145....4W,2017MNRAS.466.3612B,2022PASJ...74...73S}, see also \citealt{2024MNRAS.52711372A} for fainter galaxies), but our estimate is the first result based on the spectroscopically confirmed galaxies at $z\sim7$.
Our fraction at $z\sim7$ is higher than those at $z\sim4-5$ ($\sim20-30\%$) with similar luminosities \citep[e.g.,][]{2022PASJ...74...73S}, although this difference could be due to the difference in the spatial resolutions of the datasets used (e.g., JWST vs. other telescopes).

Two possible mechanisms are proposed to form these multiple sub-components in galaxies; galaxy mergers \citep[e.g.,][]{2008A&A...492...31D} and the violent disk instability \citep[e.g.,][]{2009ApJ...703..785D,2013MNRAS.435..999D}.
Recently, \citet{2024arXiv240208911N} discuss that galaxies with multiple sub-components extended up to 5 kpc can be made by a merger, where multiple clumps form in gas debris in tidal tails induced by the merger event.
Since the merger fraction is expected to be high for the bright $z\sim7$ galaxies in this study which are hosted by massive dark matter halos, this merger-induced clump formation is a plausible scenario for the origin of the observed multiple sub-components (see also discussions in \citealt{2024MNRAS.52711372A}).
The possible increasing trend of the clumpy galaxy fraction towards higher redshifts is also consistent with the redshift evolution of the halo merger rate predicted by simulations \citep[e.g.,][]{2010MNRAS.406.2267F,2015MNRAS.449...49R}. 
The other possibility is the violent disk instability, where clumps are predicted to form in unstable regions where the Toomre $Q$ parameter \citep{1964ApJ...139.1217T} is below a critical value in thick and gas-rich galaxy disks \citep[e.g.,][]{1998Natur.392..253N,2009ApJ...703..785D,2010MNRAS.404.2151C,2010Natur.463..781T,2010MNRAS.407.1223R,2014MNRAS.442.1230R,2019MNRAS.488.4400I}.
However, the scale of the observed clumpy structures (up to 5 kpc) is much larger than a typical scale of disks ($\sim1$ kpc), where clumps are made via the violent disk instability \citep[e.g.,][]{2024arXiv240218543F}. 
IFU observations reveal complex velocity structures supporting the merger-induced clump formation in \redc{at least five} galaxies studied here \citep[e.g.,][]{2017ApJ...851..145M,2018ApJ...854L...7C,2019PASJ...71...71H,2020MNRAS.498.3043M,2023ApJ...945...69R,2024arXiv240317133S,2024arXiv241107695S,2024arXiv241108627M}, rather than the violent disk instability, although recent ALMA {\sc[Cii]}158$\mu$m observations reveal a rotation of cold gas in one galaxy, UVISTA-Y-003 \citep{2024arXiv240506025R}.
Further IFU observations for more galaxies are needed for a definitive conclusion, but the currently-available IFU data, the spatial extent of the clumpy structure, and theoretical simulations suggest that the majority of the observed multiple subcomponents are made by mergers.

\subsection{Luminosity Function at $z\sim7$ with Sub-Components}\label{ss_LFsub}

In the luminosity function measurements in Section \ref{ss_LF}, we have classified each galaxy with multiple sub-components as a single object due to the close separation of the clumps ($\lesssim5$ kpc), in the same manner as \citet{2017MNRAS.466.3612B}.
If we interpret the sub-components as clumps made by the violent disk instability, it is reasonable to treat these multiple sub-components as a single galaxy.
If the sub-components are formed in tidal tails of a merger event, it is not clear whether we should regard these sub-components as a single object or not.

In either case, the measurements in Section \ref{ss_LF} are useful to compare with previous observational and theoretical studies that have similarly classified each galaxy with multiple sub-components as a single object.
Previous observational studies for bright galaxies rely on ground-based imaging data whose spatial resolution is poor ($\sim1\arcsec$) and do not resolve these galaxies into multiple sub-components \citep[e.g.,][]{2018PASJ...70S..10O,2022ApJS..259...20H,2023MNRAS.524.4586V}.
In most of the theoretical models \citep{2015ApJ...805...79M,2023MNRAS.521..497M,2024MNRAS.527.5929Y,2019MNRAS.483.2983Y,2023MNRAS.526.2196M,2023arXiv230411911P}, star formation processes in galaxies are calculated based on halo merger trees, and a galaxy with multiple sub-components is treated as a single galaxy in a single halo.
Only cosmological simulations with high resolutions can detect sub-components in these galaxies.
Thus the comparison conducted in Figure \ref{fig_uvlf_model} is fair for most of the model predictions where these sub-components are treated as a single galaxy.

\begin{figure}
\centering
\begin{center}
\includegraphics[width=0.99\hsize, bb=7 9 430 358,clip]{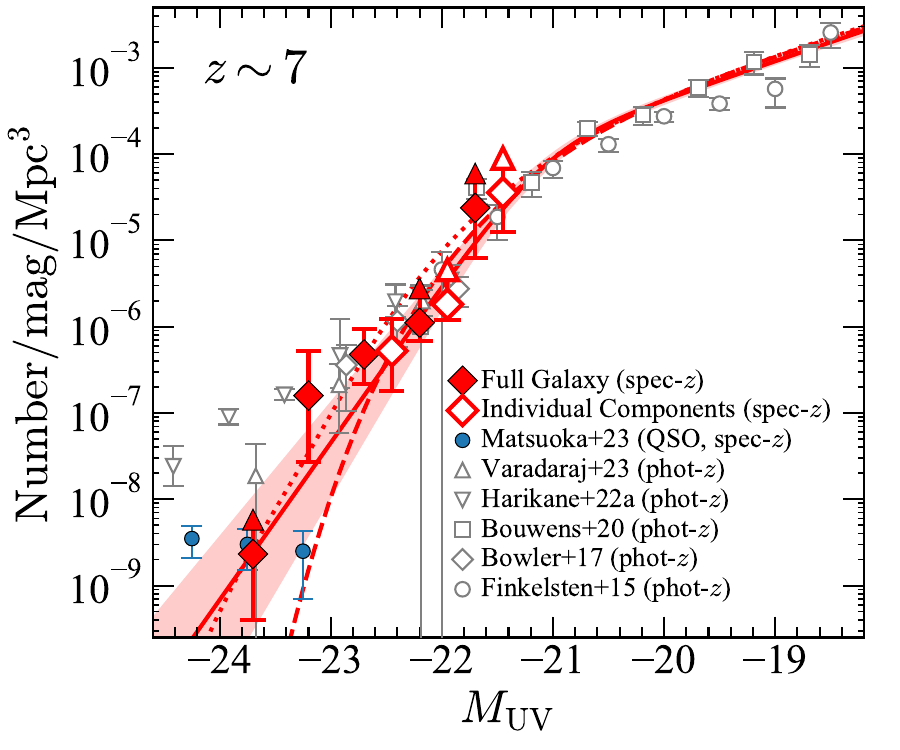}
\end{center}
\caption{
Same as the upper-left panel of Figure \ref{fig_uvlf}, but with the measurements calculated
from the individual sub-components of each ground-based selected galaxy (the open red diamonds).
The red-filled diamonds show the results when multi-component galaxies are instead plotted as single objects (as shown in Figure \ref{fig_uvlf}).}
\label{fig_uvlf_z7jwst}
\end{figure}

Nevertheless, it is useful to present the luminosity function measurements with multiple sub-components treated as individual galaxies.
Using the sub-components identified by the SExtractor in the JWST and HST images (see Section \ref{ss_sub}) and assuming the MAG\_AUTO magnitude as a total magnitude for each component, we re-calculate the UV luminosity function of $z\sim7$ galaxies.
Figure \ref{fig_uvlf_z7jwst} presents the derived UV luminosity function.
The luminosity function calculated from the individual sub-components shows a slightly steeper decline at the bright end than the previous measurement where the sub-components are treated as a single galaxy.
Note that it is not clear whether the luminosity function calculated from the individual sub-components follows the Schechter function or not, because the brightest galaxies in our study with $-24<M_\m{UV}<-23$ mag, where the tension with the Schechter function can be seen, are not yet observed with either HST or JWST.

\begin{figure*}
\centering
\begin{minipage}{1\hsize}
\begin{center}
\begin{minipage}{0.2\hsize}
\begin{center}
\includegraphics[width=1\hsize, bb=0 0 144 175]{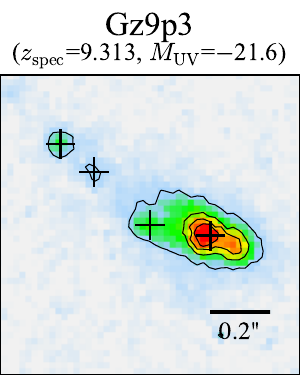}
\end{center}
\end{minipage}
\begin{minipage}{0.2\hsize}
\begin{center}
\includegraphics[width=1\hsize, bb=0 0 144 175]{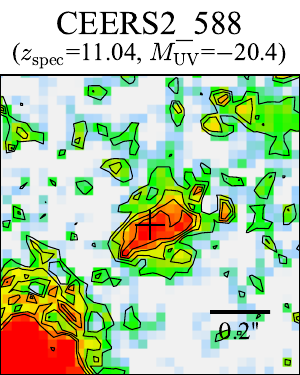}
\end{center}
\end{minipage}
\begin{minipage}{0.2\hsize}
\begin{center}
\includegraphics[width=1\hsize, bb=0 0 144 175]{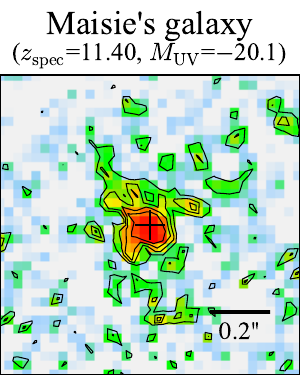}
\end{center}
\end{minipage}
\begin{minipage}{0.2\hsize}
\begin{center}
\includegraphics[width=1\hsize, bb=0 0 144 175]{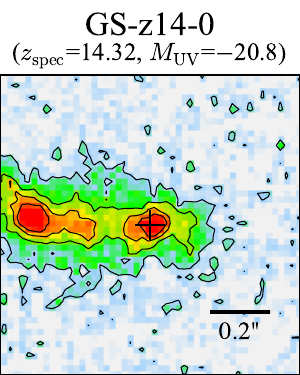}
\end{center}
\end{minipage}
\end{center}
\end{minipage}
\caption{
Same as Figure \ref{fig_jwstsnap} but for extended galaxies at $z>9$.
JWST/NIRCam F200W cutout images are used, and the displayed scale of $0.\carcsec2$ corresponds to 0.747 kpc at $z=12$.
}
\label{fig_jwstsnap_highz1}
\end{figure*}

\begin{figure*}
\centering
\begin{minipage}{1\hsize}
\begin{center}
\begin{minipage}{0.2\hsize}
\begin{center}
\includegraphics[width=1\hsize, bb=0 0 144 175]{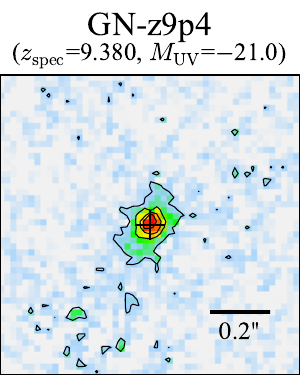}
\end{center}
\end{minipage}
\begin{minipage}{0.2\hsize}
\begin{center}
\includegraphics[width=1\hsize, bb=0 0 144 175]{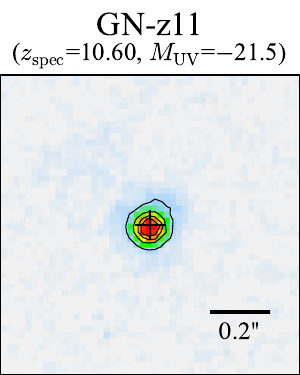}
\end{center}
\end{minipage}
\begin{minipage}{0.2\hsize}
\begin{center}
\includegraphics[width=1\hsize, bb=0 0 144 175]{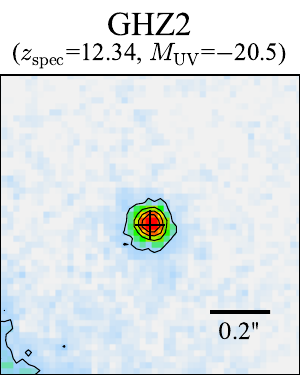}
\end{center}
\end{minipage}
\end{center}
\end{minipage}
\caption{
Same as Figure \ref{fig_jwstsnap} but for compact galaxies at $z>9$.
}
\label{fig_jwstsnap_highz2}
\end{figure*}

\begin{figure*}
\centering
\begin{center}
\includegraphics[width=0.6\hsize, bb=5 11 351 350,clip]{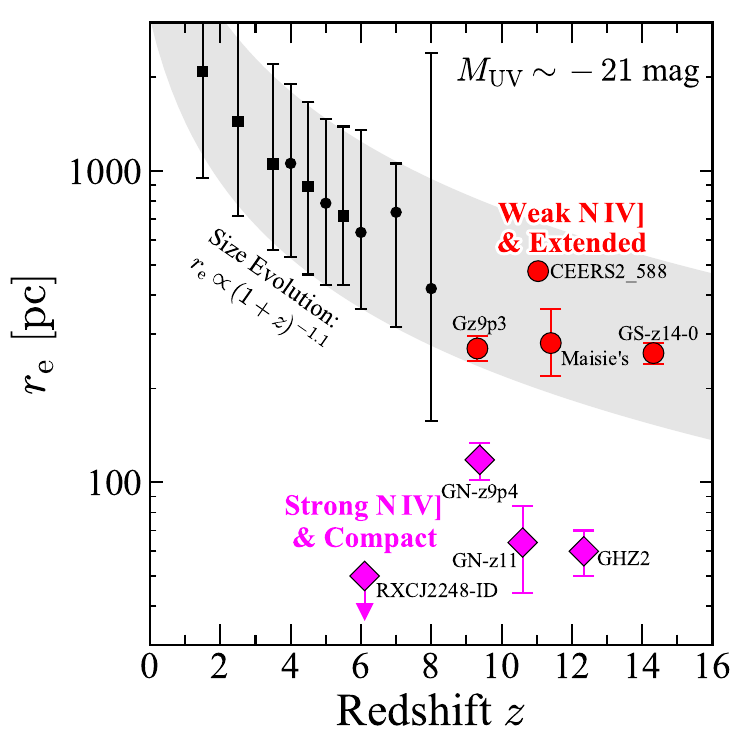}
\end{center}
\caption{
Rest-frame UV effective radius of galaxies as a function of the redshift.
The black symbols are size measurements for bright galaxies with $M_\m{UV}\sim-21$ mag at $z=0-8$ in \citet[][square: star forming galaxies, circle: Lyman break galaxies]{2015ApJS..219...15S}, and the gray shaded region shows the size evolution for galaxies with $M_\m{UV}\sim-21$ mag and its $1\sigma$ dispersion in \citet{2015ApJS..219...15S}.
The red circles are bright galaxies at $z>9$ whose sizes follow the redshift evolution measured at $z\sim0-8$ (Gz9p3: \citealt{2024NatAs.tmp...53B}, CEERS2\_588: \citealt{2023ApJ...946L..13F,2024ApJ...960...56H}, Maisie's galaxy: \citealt{2023ApJ...946L..13F,2023Natur.622..707A}, and GS-z14-0: \citealt{2023arXiv231210033R,2024arXiv240518485C}), and the magenta diamonds are compact galaxies whose effective radii are smaller than $\sim100$ pc (RXCJ2248-ID: \citealt{2024MNRAS.529.3301T}, GN-z9p4: \citealt{2024arXiv240608408S}, GN-z11: \citealt{2023A&A...677A..88B,2023ApJ...952...74T}, GHZ2: \citealt{2023ApJ...951...72O,2024arXiv240310238C}).
Bright galaxies at $z\gtrsim10$ can be classified into two types of galaxies; extended galaxies with weak high-ionization emission lines (the red circles), and compact galaxies with strong high-ionization lines such as {\sc Niv]}$\lambda$1486 (the magenta diamonds).
The extended galaxies with sizes of $r_\m{e}\sim200-500$ pc do not exhibit prominent rest-UV high ionization emission lines and sometimes show a signature of merger activity (e.g., Gz9p3; \citealt{2024NatAs.tmp...53B}).
In contrast, the compact galaxies exhibit strong high ionization lines such as {\sc Civ}$\lambda$1549, He{\sc ii}$\lambda$1640, and {\sc Niv]}$\lambda$1486 (e.g., $EW^0_\m{NIV]}\simeq10-30\ \m{\AA}$), suggesting compact and intense starburst or AGN activity.
}
\label{fig_re_z}
\end{figure*}

\subsection{Morphologies of Bright Galaxies at $z\gtrsim10$}\label{ss_re_z}

We investigate morphologies of bright galaxies spectroscopically confirmed at $z\gtrsim10$ using JWST/NIRCam images.
Among the galaxies spectroscopically confirmed (Figure \ref{fig_Muv_z}), we select five bright galaxies at $z\geq10.6$ with the UV magnitude brighter than $M_\m{UV}=-20.0$ mag, where the tension between the observed number density and model predictions is seen (Figure \ref{fig_uvlf_model}).
The five selected galaxies are GN-z11 \citep{2023A&A...677A..88B,2023ApJ...952...74T}, CEERS2\_588 \citep{2023ApJ...946L..13F,2024ApJ...960...56H}, Maisie's galaxy \citep{2023ApJ...946L..13F,2023Natur.622..707A}, GHZ2 \citep{2023ApJ...951...72O,2024arXiv240310238C}, GS-z14-0 \citep{2023arXiv231210033R,2024arXiv240518485C}.
Figures \ref{fig_jwstsnap_highz1} and \ref{fig_jwstsnap_highz2} show cutouts of these galaxies. 
We find that these galaxies are classified into two types of galaxies; extended ones with their effective radii in the rest-frame UV band of $r_\m{e}\sim200-500$ pc (Figures \ref{fig_jwstsnap_highz1}; CEERS2\_588, Maisie's galaxy, and GS-z14-0), and compact ones with $r_\m{e}\lesssim 100$ pc (Figure \ref{fig_jwstsnap_highz2}; GN-z11 and GHZ2).
In addition to these five galaxies, Gz9p3 at $z=9.313$ \citep[$r_\m{e}\sim270$ pc][]{2024NatAs.tmp...53B} and GN-z9p4 at $z=9.380$ \citep[$r_\m{e}\sim120$ pc][]{2024arXiv240608408S} can be added into the extended and compact subsamples, respectively. 

In Figure \ref{fig_re_z}, we plot the rest-frame UV effective radii of the seven galaxies as a function of the redshift.
We also plot the radius of a compact galaxy at $z=6.1057$, RXCJ222024MNRAS.535..881J48-ID \citet{2024MNRAS.529.3301T}.
These effective radii are taken from previous studies \citep{2023ApJ...952...74T,2023ApJ...946L..13F,2023ApJ...951...72O,2024arXiv240518485C,2024NatAs.tmp...53B,2024arXiv240608408S,2024MNRAS.529.3301T}.
We find that the sizes of the extended galaxies follow the redshift evolution measured at $z\sim0-8$, while the compact galaxies have significantly smaller sizes than the redshift evolution.

\begin{figure*}
\centering
\begin{minipage}{0.99\hsize}
\centering
\begin{minipage}{0.48\hsize}
\begin{center}
\includegraphics[width=0.85\hsize, bb=5 18 360 321]{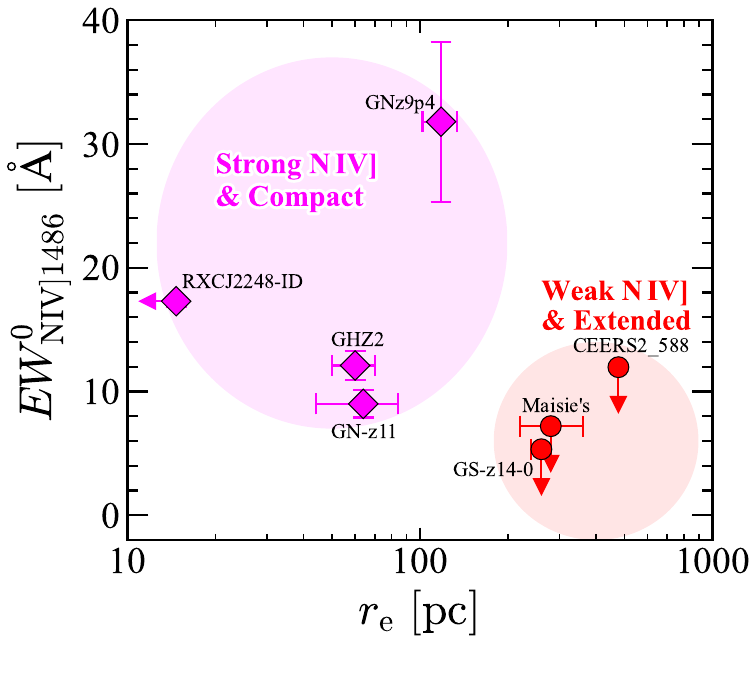}
\end{center}
\end{minipage}
\begin{minipage}{0.48\hsize}
\begin{center}
\includegraphics[width=0.85\hsize, bb=5 18 360 321]{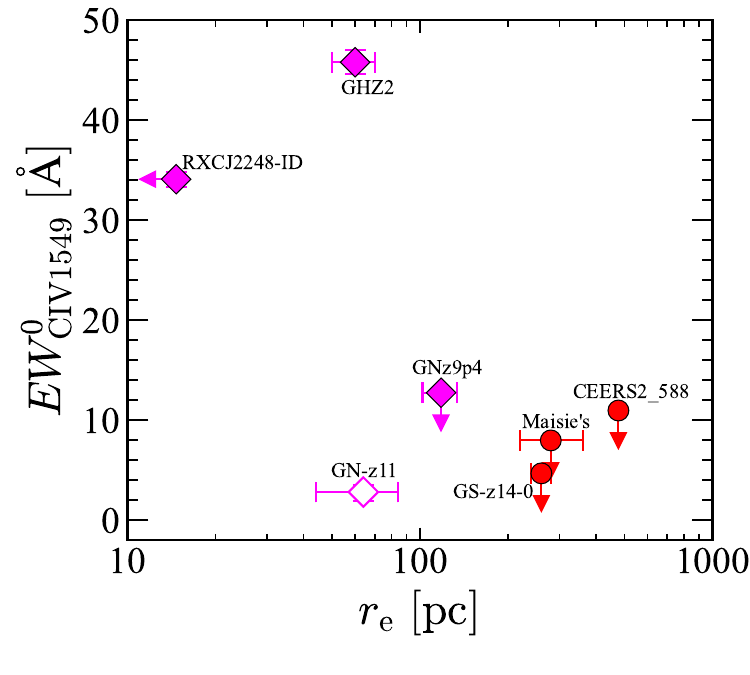}
\end{center}
\end{minipage}
\end{minipage}
\caption{
Rest-frame equivalent widths of {\sc Niv]}$\lambda$1486 (left) and {\sc Civ}$\lambda$1549 (right) as a function of an effective radius.
The magenta diamonds and red circles are compact ($r_\m{e}\lesssim100$ pc) and extended ($r_\m{e}\sim200-300$ pc) galaxies, respectively.
The upper limit is $2\sigma$.
The compact galaxies show strong {\sc Niv}$\lambda$1486 emission lines with $EW^0_\m{NIV]}\gtrsim10\ \m{\AA}$ and sometimes strong {\sc Civ}$\lambda$1549 lines, while these lines are weak in the extended galaxies.
The significance levels of these anti-correlations are 92\% and 87\% for {\sc Niv]}$\lambda$1486 and {\sc Civ}$\lambda$1549, respectively. 
GN-z11 shows a weak {\sc Civ}$\lambda$1549 emission line because of a strong absorption seen in the spectrum \citep{2024Natur.627...59M}.
}
\label{fig_EW}
\end{figure*}

Interestingly, these two types of galaxies also have different emission line properties.
Figure \ref{fig_EW} shows rest-frame equivalent widths of {\sc Niv]}$\lambda$1486 and {\sc Civ}$\lambda$1549 as a function of the rest-UV effective radius.
Rest-frame UV high ionization emission lines such as {\sc Niv]}$\lambda$1486 (47.5 eV), {\sc Civ}$\lambda$1549 (47.9 eV), and He{\sc ii}$\lambda$1640 (54.4 eV) are not significantly detected in NIRSpec spectra of the extended galaxies.
In contrast, the compact galaxies exhibit prominent high ionization lines such as {\sc Civ}$\lambda$1549, He{\sc ii}$\lambda$1640, and {\sc Niv]}$\lambda$1486 (e.g., $EW^0_\m{NIV]}\simeq9-30\ \m{\AA}$), indicating that the interstellar medium of compact galaxies are more nitrogen-enriched and highly ionized compared to that of the extended galaxies.
A strong {\sc Niv]}$\lambda$1486 line with $EW^0_\m{NIV]}\sim20\ \m{\AA}$ is also reported in a compact ($r_\m{e}\sim100$ pc) and bright ($M_\m{UV}\sim-22$ mag) galaxy with an AGN at $z=5.55$, GS\_3073 \citep{2010A&A...513A..20V,2020ApJ...897...94G,2023A&A...677A.145U,2024MNRAS.535..881J}.
CEERS\_1019 at $z=8.68$ also shows strong {\sc Niv]}$\lambda$1486 emission and a compact morphology \citep{2023ApJ...959..100I,2024PASJ...76..219O}, following this anti-correlation trend.
These two types are clearly different both in morphologies and emission line properties, suggesting that at least two different processes are shaping the physical properties of these bright galaxies at $z\gtrsim10$, which will be further discussed in Section \ref{ss_dis_bright}.

\begin{figure*}
\centering
\begin{minipage}{0.99\hsize}
\centering
\begin{minipage}{0.2\hsize}
\begin{center}
\includegraphics[width=0.99\hsize, bb=-1 -20 145 145]{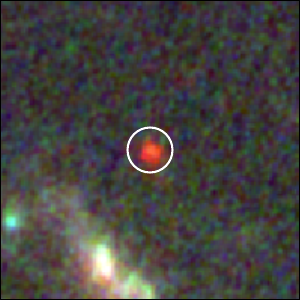}
\end{center}
\end{minipage}
\begin{minipage}{0.35\hsize}
\begin{center}
\includegraphics[width=0.99\hsize, bb=19 0 427 307]{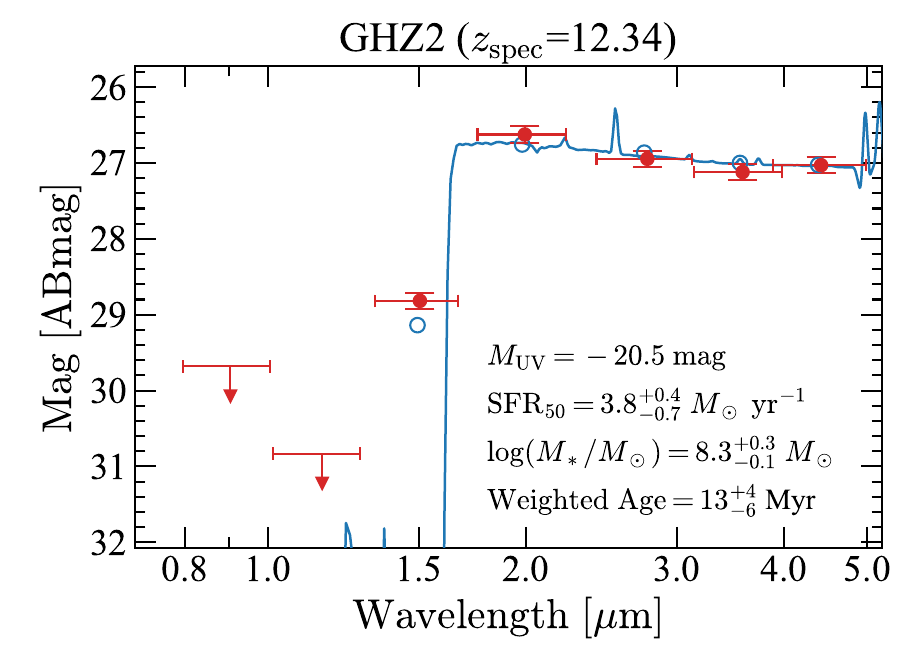}
\end{center}
\end{minipage}
\begin{minipage}{0.118\hsize}
\begin{center}
\includegraphics[width=0.99\hsize, bb=5 5 142 289]{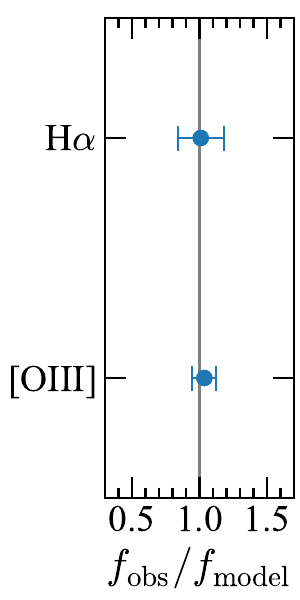}
\end{center}
\end{minipage}
\begin{minipage}{0.24\hsize}
\begin{center}
\includegraphics[width=0.99\hsize, bb=2 0 283 314]{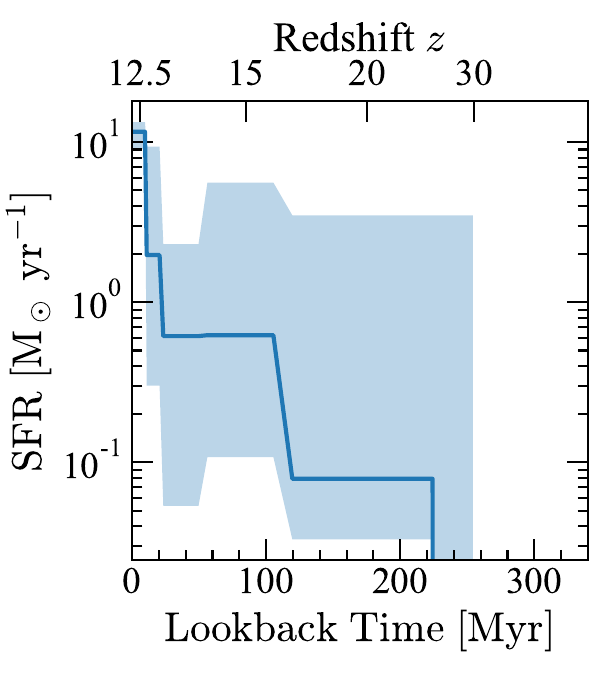}
\end{center}
\end{minipage}
\end{minipage}
\caption{
SED fitting results for GHZ2 at $z_\m{spec}=12.34$.
The leftmost panel shows the $2\arcsec\times2\arcsec$ JWST/NIRCam false-color image made from F115W, F150W, and F277W. The position of GHZ2 is indicated with the white circle whose diameter is 0.\carcsec3.
The second left panel is an SED of GHZ2.
The red circles and arrows show the measured magnitudes and $2\sigma$ upper limits, respectively, and the blue line and open circles denote the best-fit model SED with {\tt Prospector}.
The third left panel shows the comparison between observed and modeled line fluxes of H$\alpha$ and {\sc[Oiii]}$\lambda\lambda$4959,5007.
The rightmost panel shows the star formation history constrained with the SED fitting.
GHZ2 exhibits a bursty star formation history with SFRs increasing by a factor of $\gtrsim10$ within the last $\sim100$ Myr.
}
\label{fig_sed_0}
\end{figure*}

\begin{figure*}
\centering
\begin{minipage}{0.95\hsize}
\centering
\begin{minipage}{0.23\hsize}
\begin{center}
\includegraphics[width=0.99\hsize, bb=-1 -20 145 145]{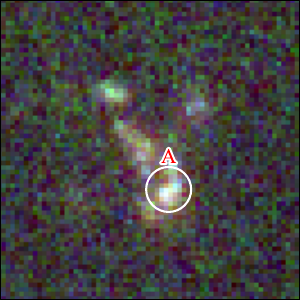}
\end{center}
\end{minipage}
\begin{minipage}{0.4\hsize}
\begin{center}
\includegraphics[width=0.99\hsize, bb=19 0 427 307]{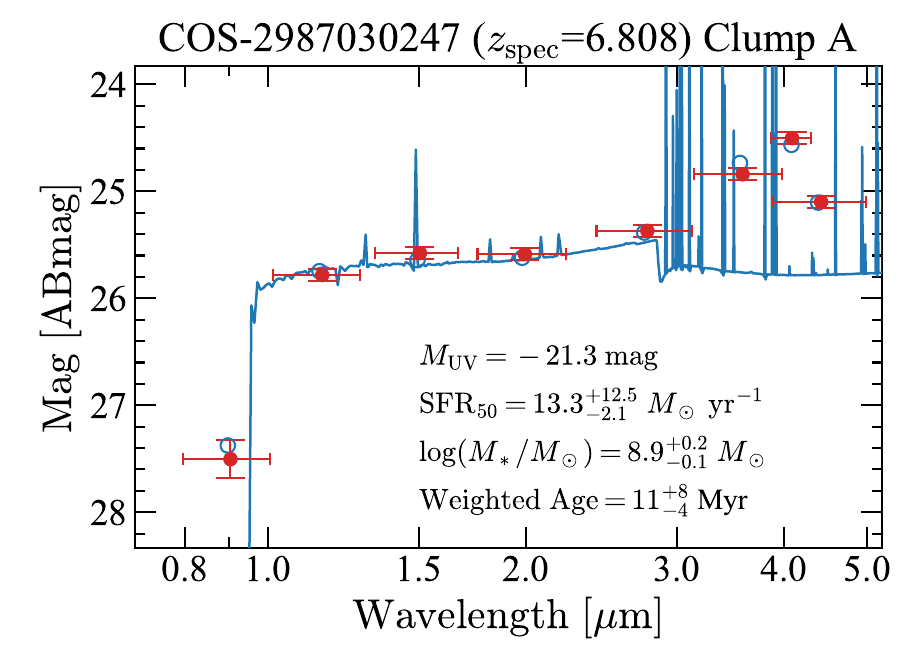}
\end{center}
\end{minipage}
\begin{minipage}{0.27\hsize}
\begin{center}
\includegraphics[width=0.99\hsize, bb=2 0 283 314]{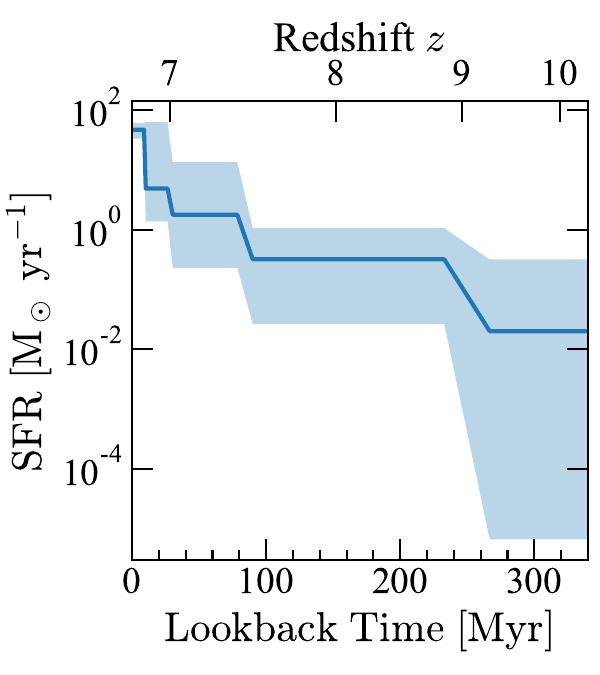}
\end{center}
\end{minipage}
\end{minipage}
\begin{minipage}{0.95\hsize}
\centering
\begin{minipage}{0.23\hsize}
\begin{center}
\includegraphics[width=0.99\hsize, bb=-1 -20 145 145]{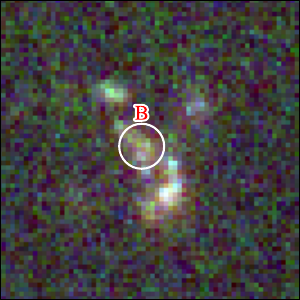}
\end{center}
\end{minipage}
\begin{minipage}{0.4\hsize}
\begin{center}
\includegraphics[width=0.99\hsize, bb=19 0 427 307]{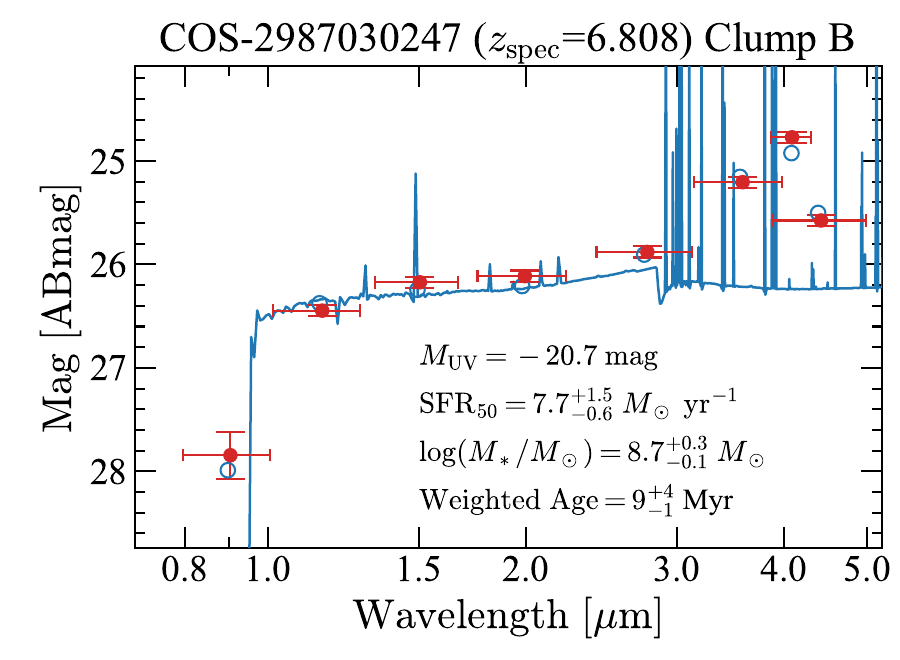}
\end{center}
\end{minipage}
\begin{minipage}{0.27\hsize}
\begin{center}
\includegraphics[width=0.99\hsize, bb=2 0 283 314]{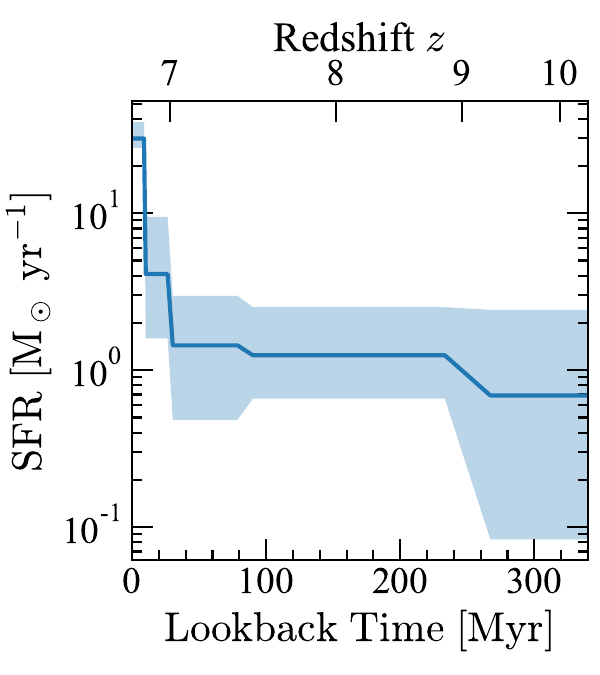}
\end{center}
\end{minipage}
\end{minipage}
\begin{minipage}{0.95\hsize}
\centering
\begin{minipage}{0.23\hsize}
\begin{center}
\includegraphics[width=0.99\hsize, bb=-1 -20 145 145]{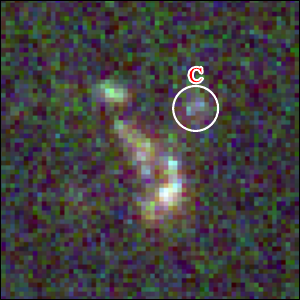}
\end{center}
\end{minipage}
\begin{minipage}{0.4\hsize}
\begin{center}
\includegraphics[width=0.99\hsize, bb=19 0 427 307]{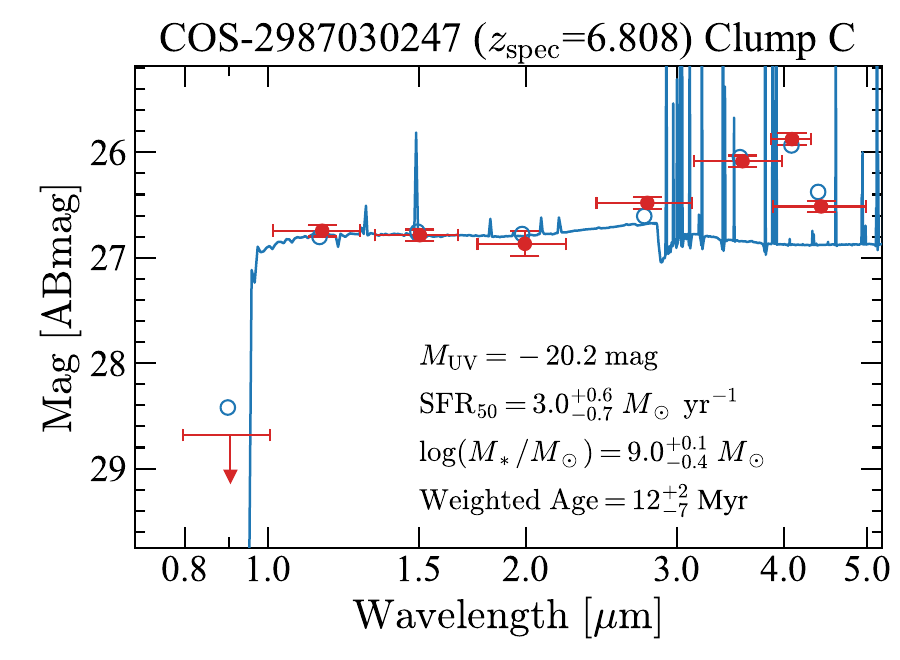}
\end{center}
\end{minipage}
\begin{minipage}{0.27\hsize}
\begin{center}
\includegraphics[width=0.99\hsize, bb=2 0 283 314]{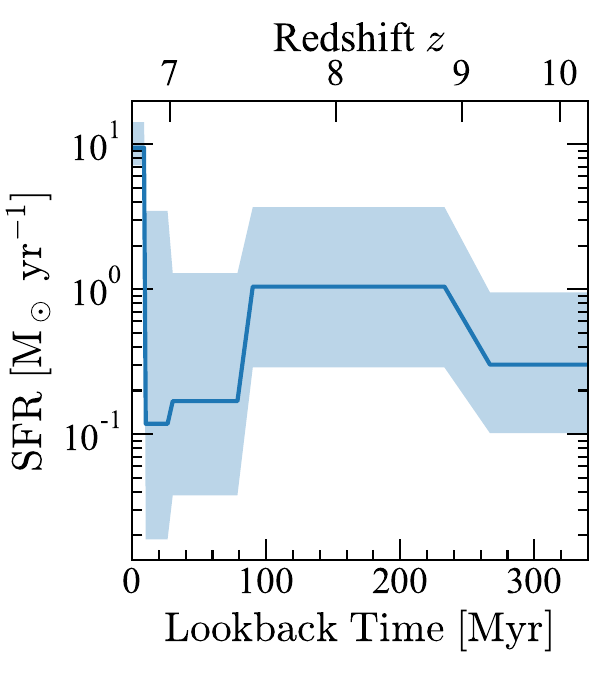}
\end{center}
\end{minipage}
\end{minipage}
\caption{
SED fitting results for sub-components (clumps A-C; from top to bottom) of COS-2987030247 at $z_\m{spec}=6.808$.
The left panels show the JWST/NIRCam false-color image (same as Figure \ref{fig_snapshot}) and the position of each clump.
The middle and right panels are SEDs and the star formation histories, respectively, in the same manner as Figure \ref{fig_sed_0}.
All of the clumps exhibit bursty star formation histories with SFRs increasing by a factor of $\sim10-100$ within the last 100 Myr.
We do not include a northeast clump in our analysis because this clump is detected in the F090W image and is thus considered to be a foreground object.
}
\label{fig_sed_1}
\end{figure*}

\begin{figure*}
\centering
\begin{minipage}{0.95\hsize}
\centering
\begin{minipage}{0.23\hsize}
\begin{center}
\includegraphics[width=0.99\hsize, bb=-1 -20 145 145]{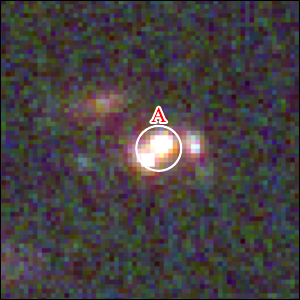}
\end{center}
\end{minipage}
\begin{minipage}{0.4\hsize}
\begin{center}
\includegraphics[width=0.99\hsize, bb=19 0 427 307]{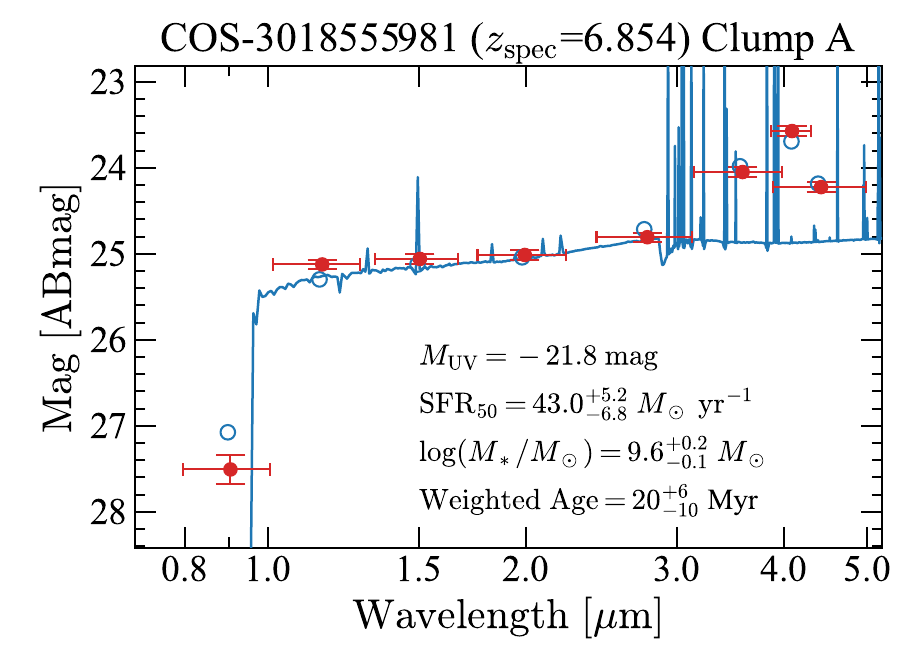}
\end{center}
\end{minipage}
\begin{minipage}{0.27\hsize}
\begin{center}
\includegraphics[width=0.99\hsize, bb=2 0 283 314]{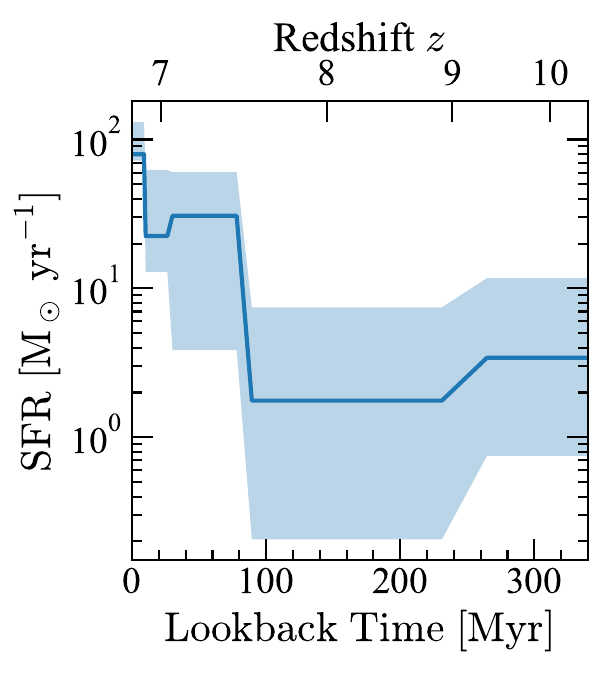}
\end{center}
\end{minipage}
\end{minipage}
\begin{minipage}{0.95\hsize}
\centering
\begin{minipage}{0.23\hsize}
\begin{center}
\includegraphics[width=0.99\hsize, bb=-1 -20 145 145]{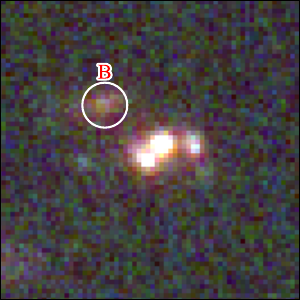}
\end{center}
\end{minipage}
\begin{minipage}{0.4\hsize}
\begin{center}
\includegraphics[width=0.99\hsize, bb=19 0 427 307]{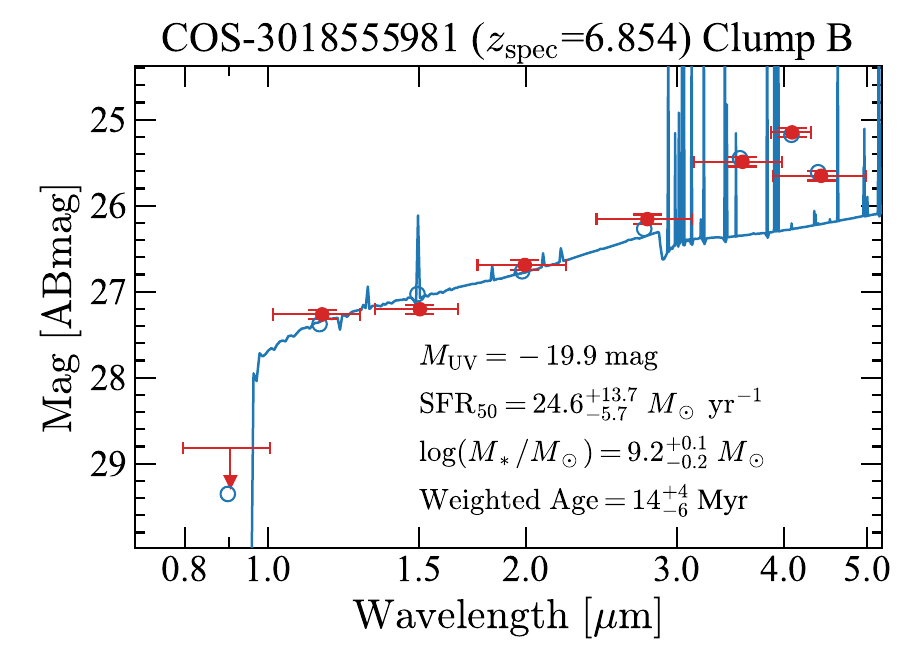}
\end{center}
\end{minipage}
\begin{minipage}{0.27\hsize}
\begin{center}
\includegraphics[width=0.99\hsize, bb=2 0 283 314]{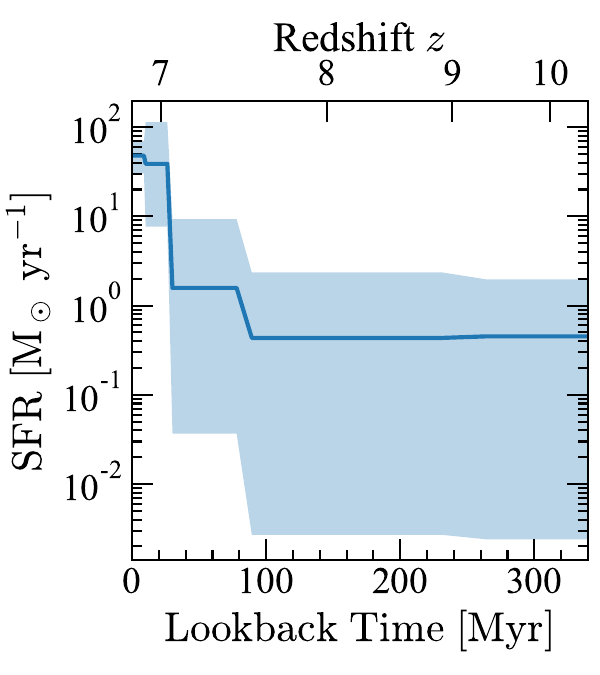}
\end{center}
\end{minipage}
\end{minipage}
\caption{
Same as Figure \ref{fig_sed_1} but for COS-3018555981 at $z_\m{spec}=6.854$.
Since we use the F444W image with a relatively large PSF of $\sim0.\carcsec16$ in the FWHM as the detection image, a few neighboring clumps are identified and analyzed as a single clump (e.g., Clump A).
}
\label{fig_sed_2}
\end{figure*}

\begin{figure*}
\centering
\begin{minipage}{0.95\hsize}
\centering
\begin{minipage}{0.23\hsize}
\begin{center}
\includegraphics[width=0.99\hsize, bb=-1 -20 145 145]{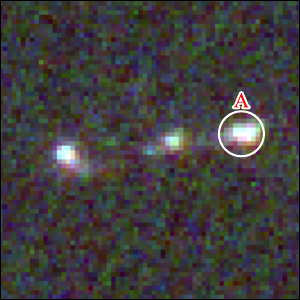}
\end{center}
\end{minipage}
\begin{minipage}{0.4\hsize}
\begin{center}
\includegraphics[width=0.99\hsize, bb=19 0 427 307]{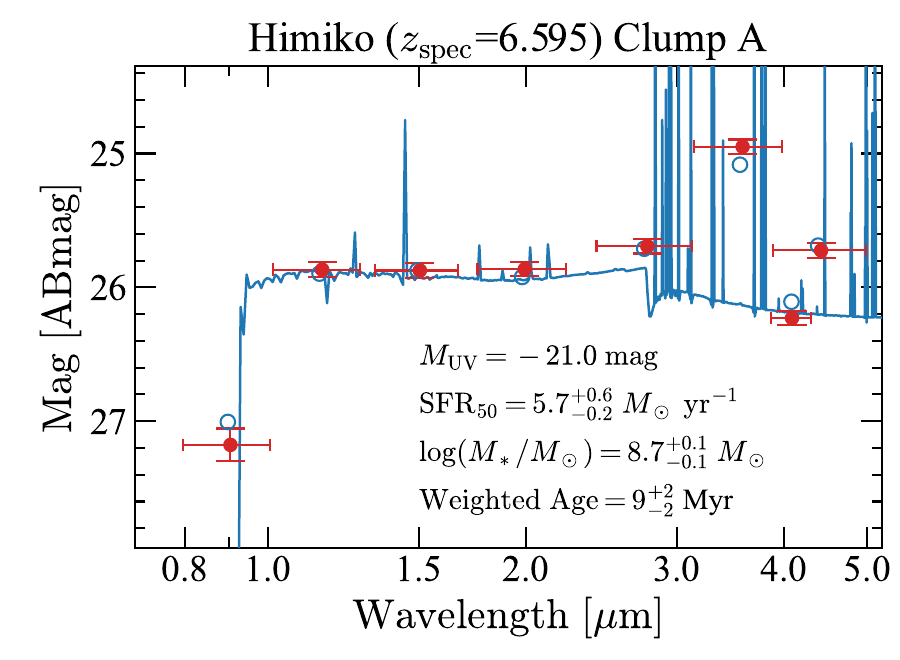}
\end{center}
\end{minipage}
\begin{minipage}{0.27\hsize}
\begin{center}
\includegraphics[width=0.99\hsize, bb=2 0 283 314]{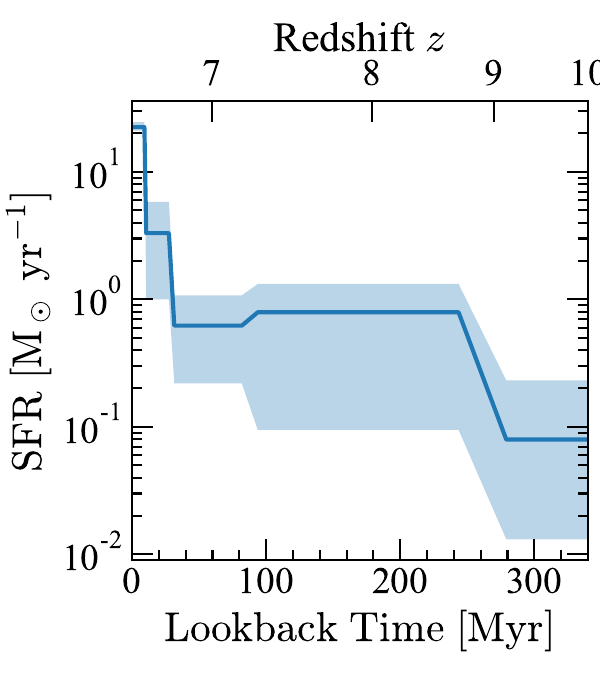}
\end{center}
\end{minipage}
\end{minipage}
\begin{minipage}{0.95\hsize}
\centering
\begin{minipage}{0.23\hsize}
\begin{center}
\includegraphics[width=0.99\hsize, bb=-1 -20 145 145]{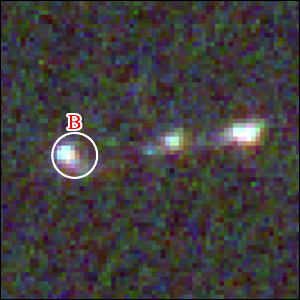}
\end{center}
\end{minipage}
\begin{minipage}{0.4\hsize}
\begin{center}
\includegraphics[width=0.99\hsize, bb=19 0 427 307]{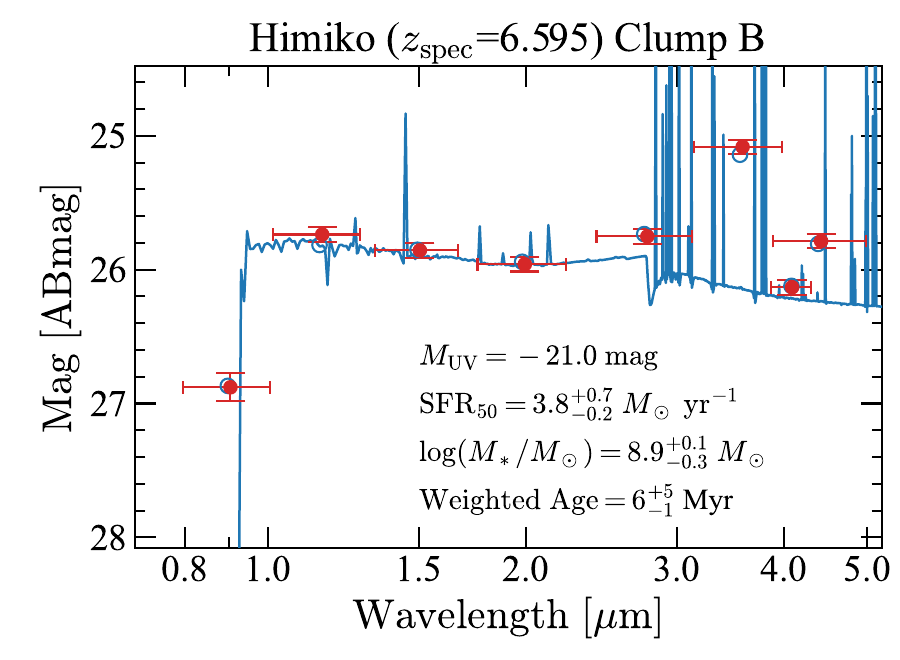}
\end{center}
\end{minipage}
\begin{minipage}{0.27\hsize}
\begin{center}
\includegraphics[width=0.99\hsize, bb=2 0 283 314]{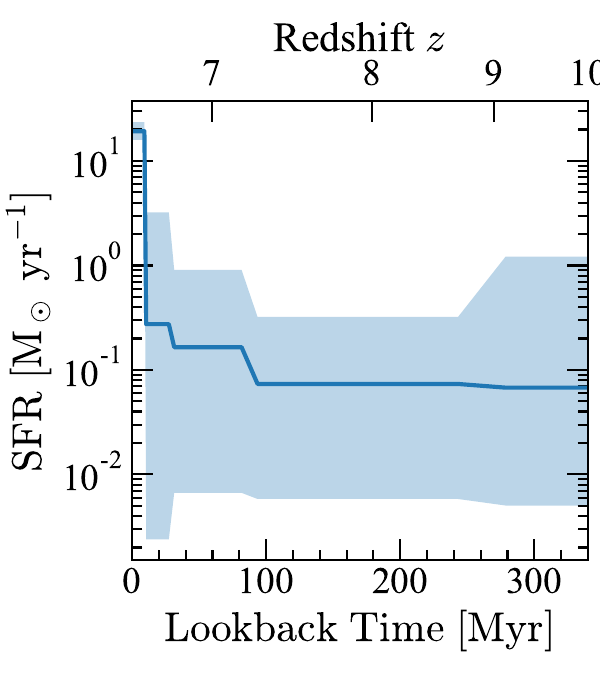}
\end{center}
\end{minipage}
\end{minipage}
\begin{minipage}{0.95\hsize}
\centering
\begin{minipage}{0.23\hsize}
\begin{center}
\includegraphics[width=0.99\hsize, bb=-1 -20 145 145]{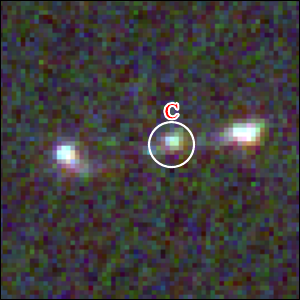}
\end{center}
\end{minipage}
\begin{minipage}{0.4\hsize}
\begin{center}
\includegraphics[width=0.99\hsize, bb=19 0 427 307]{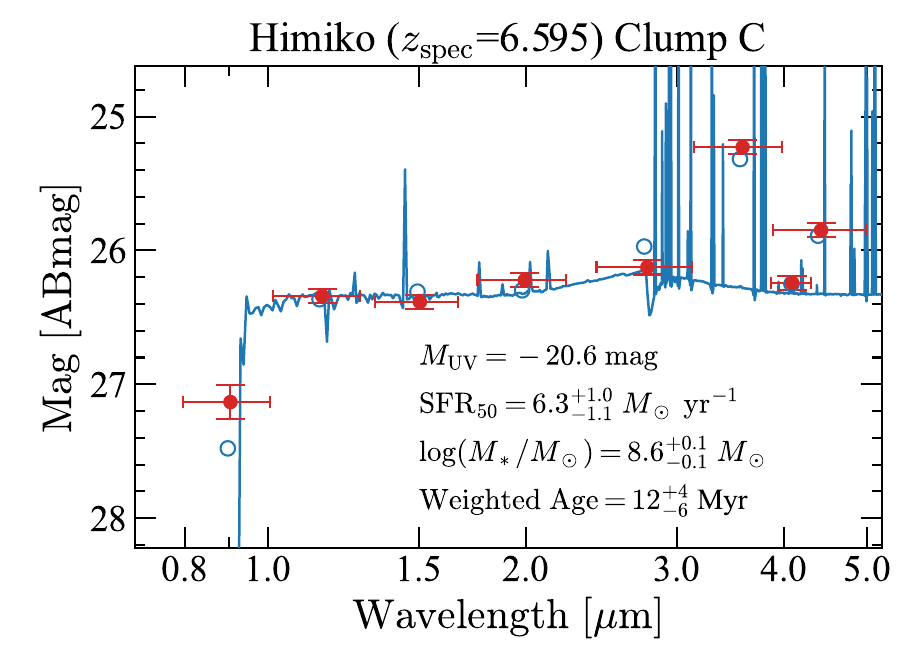}
\end{center}
\end{minipage}
\begin{minipage}{0.27\hsize}
\begin{center}
\includegraphics[width=0.99\hsize, bb=2 0 283 314]{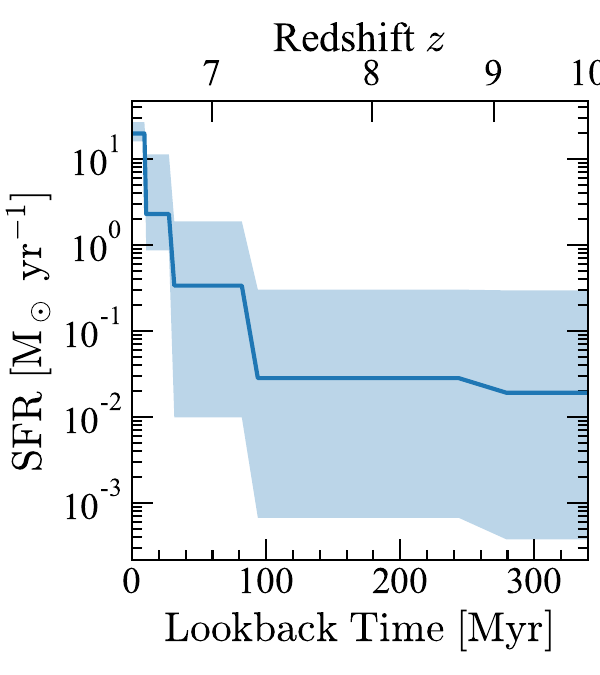}
\end{center}
\end{minipage}
\end{minipage}
\caption{
Same as Figure \ref{fig_sed_1} but for Himiko at $z_\m{spec}=6.595$.
}
\label{fig_sed_3}
\end{figure*}

\begin{deluxetable*}{cccccccccc}
\tablecaption{Summary of the SED Fitting Results}
\label{tab_SEDfit}
\tablehead{\colhead{Name} & \colhead{$z_\m{spec}$} & \colhead{Total/Clump} & \colhead{R.A.} & \colhead{Decl.} & \colhead{$M_\m{UV}$} & \colhead{SFR} & \colhead{$\m{log}M_*$} & \colhead{$\m{Age}$} & \colhead{$E(B-V)$}\\
\colhead{}& \colhead{}& \colhead{}& \colhead{}& \colhead{}& \colhead{(ABmag)}& \colhead{($M_\odot\ \m{yr^{-1}}$)}& \colhead{($M_\odot$)}& \colhead{(Myr)}& \colhead{ABmag}} 
\startdata
GHZ2 & 12.34  & Total & 00:13:59.74 & $-$30:19:29.1 & $-20.5$  & $3.8_{-0.7}^{+0.4}$  & $8.3_{-0.1}^{+0.3}$  & $12.8_{-6.3}^{+3.6}$  & $0.01_{-0.01}^{+0.01}$ \\
COS-2987030247  &  6.808  & Total &  10:00:29.86  &  $+$02:13:02.4  &  $-22.0$  & $23.9_{-2.3}^{+12.6}$  & $9.4_{-0.1}^{+0.1}$  & $10.6_{-2.5}^{+4.3}$  & $0.14_{-0.01}^{+0.01}$ \\
 &  & Clump A & 10:00:29.86 & $+$02:13:02.1 & $-21.3$  & $13.3_{-2.1}^{+12.5}$  & $8.9_{-0.1}^{+0.2}$  & $10.9_{-4.1}^{+8.1}$  & $0.14_{-0.02}^{+0.02}$ \\
 &  & Clump B & 10:00:29.87 & $+$02:13:02.4 & $-20.7$  & $7.7_{-0.6}^{+1.5}$  & $8.7_{-0.1}^{+0.3}$  & $8.8_{-0.6}^{+3.5}$  & $0.15_{-0.01}^{+0.02}$ \\
 &  & Clump C & 10:00:29.84 & $+$02:13:02.7 & $-20.2$  & $3.0_{-0.7}^{+0.6}$  & $9.0_{-0.4}^{+0.1}$  & $12.2_{-6.7}^{+1.8}$  & $0.10_{-0.04}^{+0.01}$ \\
COS-3018555981  &  6.854  & Total &  10:00:30.18  &  $+$02:15:59.7  &  $-22.0$  & $67.6_{-8.8}^{+14.6}$  & $9.8_{-0.1}^{+0.1}$  & $18.9_{-8.7}^{+5.4}$  & $0.20_{-0.02}^{+0.01}$ \\
 &  & Clump A & 10:00:30.17 & $+$02:15:59.7 & $-21.8$  & $43.0_{-6.8}^{+5.2}$  & $9.6_{-0.1}^{+0.2}$  & $19.7_{-10.1}^{+6.3}$  & $0.18_{-0.02}^{+0.01}$ \\
 &  & Clump B & 10:00:30.20 & $+$02:16:00.0 & $-19.9$  & $24.6_{-5.7}^{+13.7}$  & $9.2_{-0.2}^{+0.1}$  & $14.2_{-5.8}^{+3.9}$  & $0.32_{-0.03}^{+0.03}$ \\
Himiko  &  6.595  & Total &  02:17:57.58  &  $-$05:08:44.9  &  $-22.1$  & $15.7_{-1.1}^{+1.3}$  & $9.2_{-0.1}^{+0.1}$  & $8.7_{-1.6}^{+2.1}$  & $0.07_{-0.01}^{+0.01}$ \\
 &  & Clump A & 02:17:57.54 & $-$05:08:44.8 & $-21.0$  & $5.7_{-0.2}^{+0.6}$  & $8.7_{-0.1}^{+0.1}$  & $9.1_{-1.6}^{+1.7}$  & $0.08_{-0.01}^{+0.01}$ \\
 &  & Clump B & 02:17:57.61 & $-$05:08:45.0 & $-21.0$  & $3.8_{-0.2}^{+0.7}$  & $8.9_{-0.3}^{+0.1}$  & $6.3_{-1.3}^{+4.7}$  & $0.04_{-0.01}^{+0.02}$ \\
 &  & Clump C & 02:17:57.57 & $-$05:08:44.9 & $-20.6$  & $6.3_{-1.1}^{+1.0}$  & $8.6_{-0.1}^{+0.1}$  & $11.7_{-5.6}^{+4.0}$  & $0.11_{-0.01}^{+0.02}$ \\
\enddata
\tablecomments{Errors are $1\sigma$. The SFR presented here is the SFR averaged over the past 50 Myr, and the stellar age is the mass-weighted age calculated from the star formation history.
Positions of the clumps in each galaxy (Clumps A, B, and C) can be found in Figures \ref{fig_sed_1}-\ref{fig_sed_3}.\\
}
\end{deluxetable*}

\section{SED fitting}\label{ss_sed}

To understand the physical properties of the bright galaxies at $z\sim7-12$, we conduct SED fitting.
The galaxies to be fitted here are limited to those with well-constrained rest-frame optical emission line fluxes.
We select galaxies within the PRIMER footprint (COS-2987030247, COS-3018555981, and Himiko), where the combination of the F410M medium-band filter and the spectroscopic redshift resolves the degeneracy between the Balmer break and rest-frame optical emission lines \citep[e.g.,][]{2024MNRAS.530.2935D}, and GHZ2 at $z=12.34$ whose {\sc[Oiii]}$\lambda\lambda$4959,5007 and H$\alpha$ emission lines are detected with MIRI \citep{2024arXiv240310491Z}.
The photometric measurements and the {\sc[Oiii]}$\lambda\lambda$4959,5007 and H$\alpha$ emission lines fluxes of GHZ2 are taken from \citet{2022ApJ...940L..14N} and \citet{2024arXiv240310491Z}, respectively.
For the other sources in the PRIMER footprint, we measure the fluxes of F444W-detected sub-components in the PSF-matched NIRCam images using SExtractor in the same manner as \citet{2023ApJS..265....5H}.
We calculate the total magnitude in each band from an aperture magnitude measured in a 0.\carcsec3-diameter circular aperture with an aperture correction.
Since the PSF of the F444W image ($\sim0.\carcsec16$) is larger than that of the F115W image used in Section \ref{ss_sub}, some of the sub-components detected in Section \ref{ss_sub} are not identified here.
In the SED fitting, we use \textsc{prospector} \citep{2021ApJS..254...22J}, with changing the dust optical depth in the $V$-band, metallicity, star formation history, and total stellar mass as free parameters while fixing the redshift to the spectroscopically-determined value.
Model spectra are derived from the Flexible Stellar Population Synthesis \citep[FSPS;][]{2009ApJ...699..486C,2010ApJ...712..833C} package with the modules for Experiments in Stellar Astrophysics Isochrones and Stellar Tracks (MIST; \citealt{2016ApJ...823..102C}).
The boost of ionizing flux production of massive stars due to rotation is included in the MIST isochrones \citep{2017ApJ...838..159C}.
Here we assume the stellar IMF determined by \citet{2003PASP..115..763C}, the \citet{2000ApJ...533..682C} dust extinction law, and the intergalactic medium (IGM) attenuation model by \citet{1995ApJ...441...18M}.
We adopt a flexible star formation history with five bins.
The first bin is fixed at 0-10 Myr and the other bins are spaced equally in logarithmic times between 10 Myr and a lookback time that corresponds to $z=30$, where the SFR within each bin is constant.
We assume a continuity prior for the star formation history, and flat priors for other parameters in the range of $0<\tau_\m{V}<2$, $-2.0<\m{log}(Z/Z_\odot)<0.4$, and $6<\m{log}(M_*/M_\sun)<12$.
We search for the best-fit model to the observed photometric data points (and the {\sc[Oiii]}$\lambda\lambda$4959,5007 and H$\alpha$ fluxes for GHZ2) with the MCMC method by using {\sc emcee} \citep{2013PASP..125..306F}.
Table \ref{tab_SEDfit} summarizes the results of the SED fitting.
The values for the total component of COS-2987030247, COS-3018555981, and Himiko are the luminosity-weighted means of the measurements in individual sub-components (``Clumps").

Figure \ref{fig_sed_0} shows the SED fitting result for GHZ2 at $z=12.34$.
With the blue rest-frame UV slope ($\beta_\m{UV}=-2.4$; \citealt{2024arXiv240310238C}) and the strong {\sc[Oiii]} and H$\alpha$ emission lines detected with MIRI, the stellar age of GHZ2 is very young, about 10 Myr.
The estimated star formation history exhibits a sharp increase by a factor of 10 in the recent $\sim20$ Myr, indicating that GHZ2 is in a starburst phase, although the uncertainty is large at $>20$ Myr.
Figures \ref{fig_sed_1}-\ref{fig_sed_3} present the results for COS-2987030247, COS-3018555981, and Himiko, respectively, at $z\sim7$.
Although the sub-components of these galaxies have various rest-frame UV slopes, the strong {\sc[Oiii]} and H$\alpha$ emission lines inferred from the broad- and medium-band fluxes suggest very young stellar ages of $10-20$ Myr and bursty star-formation histories with the SFR increasing by a factor of $\sim10-100$ within the last 100 Myr, similar to GHZ2, albeit large uncertainties at $>100$ Myr.
Indeed, the observed increase of the SFR in the last $\sim10-100$ Myr is more rapid than an averaged dark matter halo growth history at the same redshifts \citep[e.g.,][]{2010MNRAS.406.2267F}.
Thus these starbursts may be responsible for the UV-bright nature of galaxies at $z\sim7-12$, whose abundances show the tension with the theoretical predictions in the luminosity functions.
More discussions on the physical origin of the overabundance are presented in Section \ref{ss_dis_bright}.

We find that all of the clumps analyzed here show increasing star formation histories at the time of observations, which is in contrast to the results of \citet{2024MNRAS.52711372A} for fainter galaxies at $z\sim5-7$ showing both rising and declining star formation histories.
This difference in the fraction of galaxies with rising star formation histories between this study and \citet{2024MNRAS.52711372A} may be due to the difference in the UV luminosities of the samples used in these two studies.
Given the bright UV magnitudes of our galaxies, it is possible that we are selectively observing the UV-bright phase of galaxies with rising star formation histories, while \citet{2024MNRAS.52711372A} are looking at the UV-bright phase of low-mass galaxies and the UV-faint phase of massive galaxies.
Indeed, \citet{2023arXiv230605295E} find a similar trend of a higher fraction of galaxies with rising histories towards brighter magnitudes.

\section{Discussion}\label{ss_dis}

\subsection{Physical Origin of the Overabundance of Bright Galaxies}\label{ss_dis_bright}

Various studies using JWST have reported that the abundance of bright galaxies at $z\gtrsim10$ is higher than theoretical model predictions.
The physical origin of this overabundance is not clear, but several possibilities are discussed, such as a high star formation efficiency, AGN activity, a top-heavy IMF, bursty star formation, radiation-driven outflows, and a flaw in the current cosmology model (see Section \ref{ss_intro}). 
As shown in Section \ref{ss_uvlf_model}, we have found that the number densities of bright galaxies with $M_\m{UV}\lesssim -21$ mag at $z\sim12-14$ are higher than the model predictions (Figures \ref{fig_uvlf_model} and \ref{fig_uvlf_model_z14}), similar to previous studies using JWST.
In addition, the measured number densities of bright galaxies with $M_\m{UV}\lesssim -23$ mag at $z\sim7$ are also higher than some model predictions.

\subsubsection{$z\sim7$: Merger-Induced Starbursts}

In Section \ref{ss_sed}, we find that the SFR of bright $z\sim7$ galaxies has increased by a factor of 10-100 within the last $\sim100$ Myr, indicating that the recent starburst is contributing to the UV-bright nature of these luminous galaxies.
Since these UV-bright galaxies at $z\sim7$ exhibit clumpy morphologies with multiple sub-components (Section \ref{ss_sub}, Figure \ref{fig_jwstsnap}), the starburst is thought to be triggered by a recent merger event.
Indeed, previous studies have reported evidence of mergers in some galaxies at $z\sim7$ \citep{2019PASJ...71...71H,2024arXiv240317133S}.
Numerical simulation results in \citet{2024NatAs...8..384W} also indicate a starburst during a merger phase at $z>7$. 
These results suggest that the merger-induced starburst can explain the high number density of UV-bright galaxies at $z\sim7$.
The tension between observations and theoretical predictions may be due to the SFRs of model galaxies in the merging phase not increasing sufficiently compared to what has been observed.

\subsubsection{$z\gtrsim10$: Merger-Induced Starburst and Compact Star Formation/AGN}

In contrast to the case at $z\sim7$, the situation for galaxies at $z\gtrsim10$ is not simple.
As described in Section \ref{ss_re_z}, bright galaxies at $z\gtrsim10$ can be classified into two types; extended galaxies with weak high ionization lines, and compact galaxies with strong high-ionization lines.
For the extended galaxies, as discussed in \citet{2024arXiv240518485C}, the weak high ionization lines suggest no strong AGN activity, and the shape of the rest-UV continuum does not support a significant contribution from nebular continuum emission with a top-heavy IMF such as discussed in \citet{2023arXiv231102051C}.
We discuss that a high star formation efficiency, possibly enhanced by a merger-induced starburst similar to the bright $z\sim7$ galaxies studied here, is responsible for their UV-bright nature.
Such a merger-induced starburst is expected to be frequent because the merger rate of bright galaxies at $z\gtrsim10$ is theoretically several times higher than those at $z\sim7$ \citep[e.g.,][]{2010MNRAS.406.2267F,2015MNRAS.449...49R}.
The major merger timescale, the inverse of the merger rate, is $\sim200$ Myr at $z\sim12$ for bright galaxies with $M_*\sim10^9\ M_\odot$ \citep{2015MNRAS.449...49R}, which is shorter than the age of the universe at $z\sim12$ (370 Myr), indicating that most of the bright galaxies at $z\sim12$ have experienced at least one major merger.
Indeed, some bright galaxies at $z\gtrsim9$ show merger signatures \citep{2023ApJ...949L..34H,2024NatAs.tmp...53B}.
Theoretically, \citet{2023ApJ...951...72O} also discuss that simulated galaxies in \citet{2022MNRAS.509.4037Y} whose sizes are $r_\m{e}\sim200-400$ pc and follow the size evolution are experiencing major mergers or tidal interactions.
Note that the bright galaxies at $z\sim7$ showing multiple sub-components also follow the size evolution \citep{2017MNRAS.466.3612B}.
Thus the merger-induced starburst, similar to the bright galaxies at $z\sim7$, is a plausible scenario for the brightness of these extended galaxies at $z\gtrsim10$.
Weak emission lines in these extended galaxies are probably due to their low metallicites or high ionizing photon escape fractions \citep[e.g.,][]{2024arXiv240518485C,2024arXiv240520370F}.
In contrast to the $z\sim7$ galaxies showing clumpy morphologies, some of these $z\gtrsim10$ galaxies do not look clumpy.
This is probably because the smaller angular diameter distance and galaxy's size at $z\gtrsim10$ compared to $z\sim7$ make it difficult to observe clumps in $z\gtrsim10$ galaxies.

The compact galaxies at $z\gtrsim10$ such as GN-z11 and GHZ2 do not show significantly extended or clumpy structures unlike the bright $z\sim7$ galaxies and the extended $z\gtrsim10$ galaxies.
Given the high signal-to-noise ratios of the detection (e.g., $>30\sigma$), if these galaxies exhibit multiple components or extended structures, these components/structures should be detected, indicating that the image depth is not the origin of their apparent compactness.
Rather, their compact morphologies are made by either compact star formation or AGN activity, which enhances the UV luminosity and makes the tension with the model predictions.
Compact star formation can easily happen at high redshifts \citep[e.g.,][]{2015MNRAS.450.2327Z,2016MNRAS.458..242T,2021MNRAS.506.5512F}.
\citet{2023ApJ...951...72O} discuss that such a compact star formation phase with intense accretion of the material makes isolated galaxies whose sizes are less than $\sim100$ pc, similar to GN-z11 and GHZ2.
Interestingly, these compact galaxies are very blue in the rest-UV continuum with slopes of $\beta_\m{UV}\simeq-2.4$ \citep{2023A&A...677A..88B,2024arXiv240310238C}, suggesting negligible/zero dust attenuation \citep{2023MNRAS.520...14C,2024MNRAS.531..997C,2024MNRAS.529.4087T,2024arXiv240410751A,2024ApJ...964L..24M}.
Formation of such compact galaxies with weak dust attenuation is indeed predicted in the feedback-free starburst scenario proposed in \citet{2023MNRAS.523.3201D}.
\citet{2023ApJ...951L...1P} discuss that such compact star formation are not predicted in some cosmological simulations, which may be the origin of the tension between the JWST observations and theoretical predictions.
Such compact star formation may increase the density in a galaxy and the rate of runaway stellar collisions producing supermassive stars and/or tidal disruption events, enhancing the nitrogen production \citep[e.g.,][]{2023MNRAS.523.3516C,2024ApJ...962...50W}. 

AGN activity is also a plausible scenario for the compact galaxies at $z\gtrsim10$.
Several studies claim AGN activity in these compact galaxies such as GN-z11 and GHZ2, based on the detections of high ionization emission lines including Ne{\sc iv}$\lambda$2424 (63.5 eV; \citealt{2024Natur.627...59M}) and a very strong {\sc Civ}$\lambda$1549 line with an equivalent width of $EW_\m{CIV}^0\sim40\ \m{\AA}$ \citep{2024arXiv240310238C}.
AGN activity is also reported in similarly compact and bright galaxies, GS\_3073 at $z=5.55$ \citep{2023A&A...677A.145U} \redc{and COS-zs-1 at $z=7.15$ \citep{2024MNRAS.531..355U}, with broad H$\beta$ and/or H$\alpha$ emission lines}.
To make these galaxies UV-bright, their rest-frame UV continua should be dominated by emission from an accretion disk of the AGN, such as type-1 broad-line AGNs whose accretion disk emission can be directly observed.
Alternatively, GN-z11 and GHZ2 can be narrow-line type-2 quasars \citep[e.g.,][]{2003AJ....126.2125Z}, and their UV continua are significantly contributed by scattered lights of emission from the accretion disk.

\begin{figure*}
\centering
\begin{minipage}{0.49\hsize}
\begin{center}
\includegraphics[width=0.9\hsize, bb=5 9 395 357,clip]{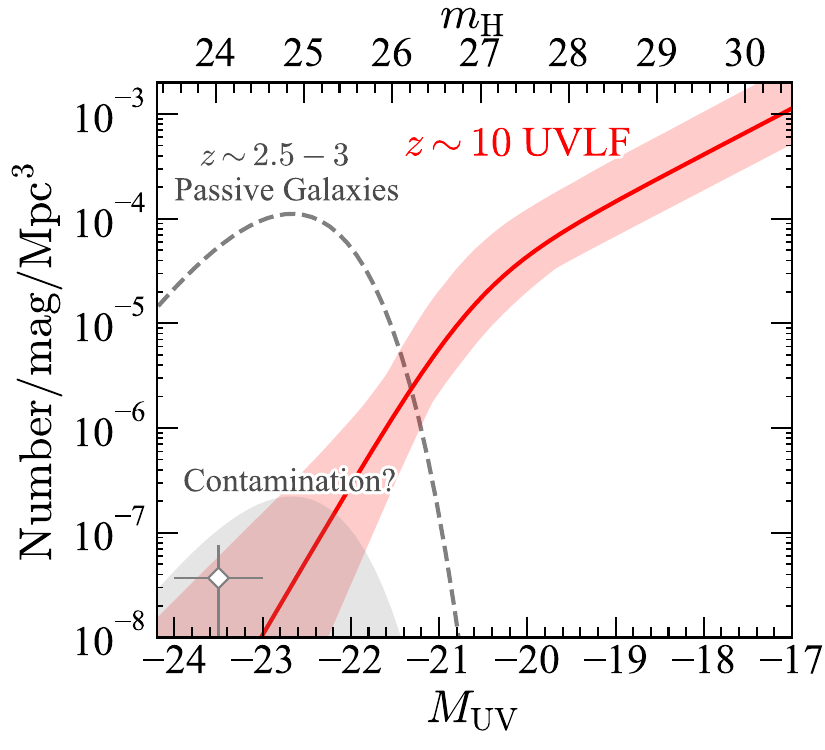}
\end{center}
\end{minipage}
\begin{minipage}{0.49\hsize}
\begin{center}
\includegraphics[width=0.9\hsize, bb=5 9 395 357,clip]{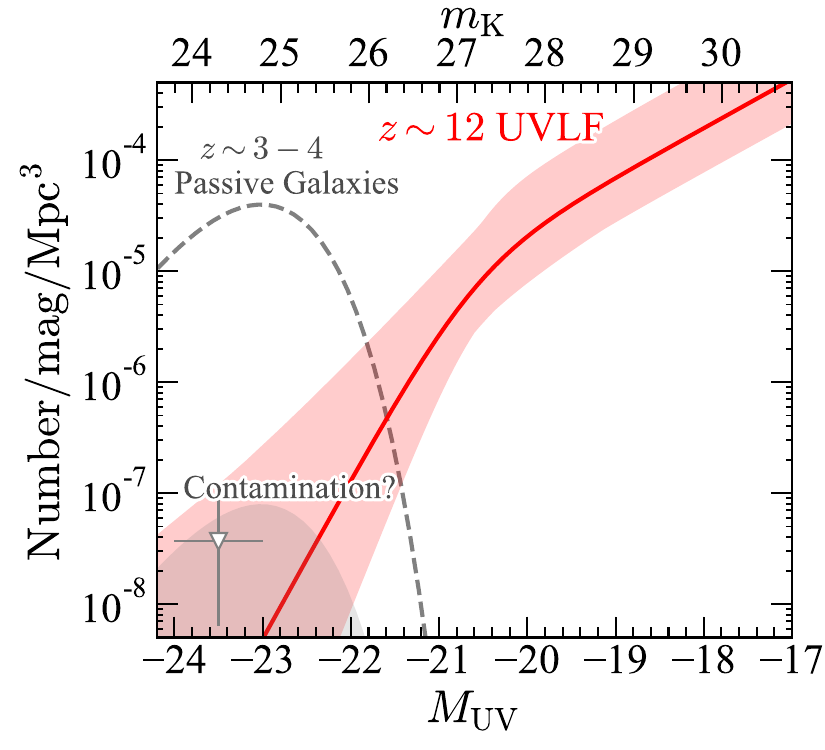}
\end{center}
\end{minipage}
\caption{
Effect of low-redshift interlopers from passive galaxies at $z\sim2.5-4$ (see also \citealt{2023ApJ...955..130F}).
In the left (right) panel, the red solid line represents the best-fit double power-law function at $z\sim10$ ($z\sim12$) obtained in this study, and the gray symbol is the number density of $z\sim10$ ($z\sim13$) galaxy candidate from \citet{2020MNRAS.493.2059B} (\citet{2022ApJ...929....1H}) that is identified as a low-redshift galaxy in this study.
The gray dashed curve is the $H$-band ($K$-band) luminosity function of passive galaxies at $z\sim2.5-3$ ($z\sim3-4$), which are calculated based on the stellar mass functions of passive galaxies in \citet{2017A&A...605A..70D} and the NIRSpec spectrum of XMM3-3085 (HD1) obtained in Section \ref{ss_spec_jwst}.
At these redshifts, passive galaxies contaminate the Lyman break galaxy selections at $z\sim10$ ($z\sim12$) because the Balmer break is redshifted to $\sim1.3\ \m{\mu m}$ ($\sim1.6\ \m{\mu m}$), the wavelength of the Lyman break at $z\sim10$ ($z\sim12$).
The gray shaded region shows the expected number density of low-redshift interlopers assuming that $\leq0.2\%$ of the passive galaxies are selected as $z\sim10$ ($z\sim12$) galaxies.
The passive galaxies at intermediate redshifts can contaminate a sample of bright ($M_\m{UV}\lesssim-22$ mag) galaxy candidates at $z>10$, and strict screening is necessary to select real $z>10$ bright galaxies.
}
\label{fig_uvlf_lowz}
\end{figure*}

\subsection{Low-Redshift Interlopers at the Bright End}\label{ss_dis_lowz}

JWST/NIRSpec spectroscopy has revealed that very luminous ($M_\m{UV}\lesssim -23$ mag) galaxy candidates at $z\sim10-12$ identified in ground-based images before JWST \citep{2020MNRAS.493.2059B,2022ApJ...929....1H} are low-redshift passive galaxies at $z\sim3-4$ whose Balmer break is redshifted to the wavelength of the Lyman break at $z\sim10-12$ (Figures \ref{fig_spec_JWST} and \ref{fig_spec_ALMA_HD1}).
The stellar masses of HD1 and HD2 are estimated to be $\sim10^{10}\ M_\odot$ \citep{2024MNRAS.534.3552S}, lower than typical passive galaxies at similar redshifts \citep[e.g.,][]{2022arXiv220708778C,2024arXiv240502242C}, indicating that HD1 and HD2 are relatively faint and easily affected by photometric scatters.
These results suggest that low-mass passive galaxies at $z\sim3-4$ can be selected as bright Lyman break galaxies at $z\sim10-12$ due to photometric scatters in relatively shallow ground-based and Spitzer images (see Figure \ref{fig_spec_JWST}).
These passive galaxies at intermediate redshifts are important contaminants that should be taken into account in the high redshift galaxy selection, in addition to strong emission line galaxies seen in \citet{2023Natur.622..707A}.
Even more careful galaxy selections are required to select very luminous galaxies at $z\sim10-12$ and remove these low-redshift interlopers in relatively shallow datasets.

To understand how careful selection is needed, we calculate expected number densities of low-redshift interlopers.
We convert the stellar mass functions of passive galaxies at $z\sim2.5-3$ ($z\sim3-4$) in \citet{2017A&A...605A..70D} to the $H$-band ($K$-band) luminosity function using the NIRSpec spectrum of XMM3-3085 (HD1) obtained in Section \ref{ss_spec_jwst}.
Note that the conclusion does not change if we use the stellar mass functions in \citet{2021MNRAS.503.4413M}.
We calculate the expected $H$-band ($K$-band) luminosity function of low-redshift interlopers by assuming that a certain fraction of the passive galaxies are selected as $z\sim10$ ($z\sim12$) galaxy candidates.
We adopt a fraction of $\leq0.2\%$ to match the expected number density to the observed densities of the three low-redshift interlopers we have identified in this study.
In Figure \ref{fig_uvlf_lowz}, we plot the calculated luminosity functions of low-redshift interlopers and the best-fit double-power-law functions of galaxies at $z\sim10-12$ with errors constrained from the spectroscopic results in Section \ref{ss_uvlf_result}.
We find that even if only $0.2\%$ of low-redshift passive galaxies are erroneously selected as high-redshift galaxy candidates, then the number density of low-redshift interlopers becomes higher than that of real high-redshift galaxies at $z\sim10-12$.
In future wide-area bright ($M_\m{UV}\lesssim -22$ mag) galaxy surveys at $z\gtrsim10$ with Euclid, Roman, and GREX-PLUS, it is necessary to devise selection criteria, such as using a more strict dropout color criterion with deeper datasets at the wavelength shorter than the break, to limit the fraction of low-redshift passive galaxies entering into the selection to less than $0.2\%$.
Note that these interlopers are much less of a concern for JWST-selected candidates because they are usually fainter than $M_\m{UV}\sim-22$ mag, where the effect of the contamination is not significant.

\section{Summary}\label{ss_summary}
In this paper, we present the number densities and physical properties of galaxies at $z\sim7-14$, based on the sample of \redc{60} luminous galaxies spectroscopically confirmed at $z_\mathrm{spec}=6.538-14.32$.
Our major findings are summarized below:

\begin{enumerate}
\item
We constrain the UV luminosity functions at $z\sim7-14$.
At $z\sim7$, the bright end of the luminosity function is well described by the double-power-law \redc{or lensed Schecher function rather than the original} Schechter function (Figure \ref{fig_uvlf}).
We find that the number densities of spectroscopically-confirmed bright galaxies at $z\sim7$ and $12-14$ are higher than theoretical model predictions (Figures \ref{fig_uvlf_model} and \ref{fig_uvlf_model_z14}).

\item
Using the high-resolution JWST/NIRCam images, we find that $\sim70\%$ of our bright ($M_\m{UV}\leq-21.5$ mag) galaxy sample at $z\sim7$ exhibit clumpy morphologies with multiple sub-components, suggesting recent merger events (Figures \ref{fig_snapshot} \& \ref{fig_jwstsnap}).

\item
We conduct SED fitting for GHZ2 at $z_\m{spec}=12.34$ and three galaxies at $z_\m{spec}\sim7$, whose {\sc[Oiii]}$\lambda\lambda$4959,5007 and H$\alpha$ emission line fluxes are constrained with MIRI and the NIRCam F410M observations, respectively.
We find that all of the clumps in the four galaxies show bursty star-formation histories with the SFR increasing by a factor of $\sim10-100$ within the last 100 Myr (Figures \ref{fig_sed_0}-\ref{fig_sed_3}).

\item
Based on the clumpy morphologies and the bursty star formation histories revealed in this study, we discuss that a recent merger event has triggered a starburst in the bright galaxies at $z\sim7$.
Such a merger-induced starburst boosts the UV luminosity, resulting in the observed high number density of bright galaxies at $z\sim7$ showing the tension with theoretical models (Section \ref{ss_dis_bright}).

\item
At $z\gtrsim10$, bright galaxies are classified into two types of galaxies; extended ones with weak high-ionization emission lines and compact ones with strong high ionization lines including {\sc Niv}]$\lambda$1486 (Figures \ref{fig_re_z} and \ref{fig_EW}).
These two populations are different in both morphologies and emission line properties, suggesting that at least two different processes are contributing to the overabundance of bright galaxies at $z\gtrsim10$.
We discuss that a merger-induced starburst may be responsible for the UV-bright nature of the extended galaxies, similar to the bright $z\sim7$ galaxies studied here, while the UV luminosity of compact galaxies is enhanced by compact star formation or AGN activity (Section \ref{ss_dis_bright}).

\item
Our JWST/NIRSpec observations have revealed that very bright galaxy candidates at $z\sim10-12$ previously identified from ground-based images are low redshift passive galaxies at $z\sim3-4$ (Figure \ref{fig_spec_JWST}).
These passive low-redshift interlopers are erroneously selected as high redshift galaxies because of 1) large photometric scatters originating from relatively shallow datasets, and 2) their very bright magnitudes.
This result indicates that strict selection criteria that keep the fraction of passive galaxies entering into the selection to less than $0.2\%$ are required in the future wide-area bright galaxy surveys at $z\gtrsim10$ with Euclid, Roman, and GREX-PLUS (Figure \ref{fig_uvlf_lowz}).

\end{enumerate}

\begin{acknowledgments}
\redc{We thank the anonymous referee for careful reading and valuable comments that improved the clarity of the paper.}
We are grateful to Rebecca Bowler and Rohan Varadaraj for providing the latest photometric measurements for XMM3-3085, and to Masayuki Akiyama, Yoshinobu Fudamoto, Takuya, Hashimoto, Taddy Kodama, Mariko Kubo, Masafusa Onoue, and Hannah \"{U}bler for useful comments and discussions.
This paper makes use of the ALMA data obtained in 2019.1.01634.L (REBELS), 2021.1.00207.S, 2021.1.00341.S, and 2022.1.00522.S.
The authors acknowledge the REBELS team led by Richard J. Bouwens for developing their observing program.
This work is based on observations made with the NASA/ESA/CSA James Webb Space Telescope. The data were obtained from the Mikulski Archive for Space Telescopes at the Space Telescope Science Institute, which is operated by the Association of Universities for Research in Astronomy, Inc., under NASA contract NAS 5-03127 for JWST.
The JWST data presented in this paper were obtained from the Mikulski Archive for Space Telescopes at the Space Telescope Science Institute, which is operated by the Association of Universities for Research in Astronomy, Inc., under NASA contract NAS 5-03127 for JWST.
These observations are associated with programs ERS-1324 (GLASS), ERS-1345 (CEERS), GO-1727 (COSMOS-Web), GO-1740, GO-1837 (PRIMER), GO-2561 (UNCOVER), GO-2792, and GTOs-1180, 1181, 1210, and 1286 (JADES). 
The authors acknowledge the GLASS, CEERS, COSMOS-Web, UNCOVER, and JADES 
teams led by Tommaso Treu, Steven L. Finkelstein, Jeyhan Kartaltepe \& Caitlin Casey, Ivo Labbe \& Rachel Bezanson, and Daniel Eisenstein \& Nora Luetzgendorf, respectively, for developing their observing programs.
\redc{The JWST and HST data presented in this article were obtained from the Mikulski Archive for Space Telescopes (MAST) at the Space Telescope Science Institute. The specific observations analyzed can be accessed via \dataset[10.17909/gbkc-2b17]{https://doi.org/10.17909/gbkc-2b17}.}
Some of the data products presented herein were retrieved from the Dawn JWST Archive (DJA). DJA is an initiative of the Cosmic Dawn Center (DAWN), which is funded by the Danish National Research Foundation under grant DNRF140.
This publication is based upon work supported by the World Premier International Research Center Initiative (WPI Initiative), MEXT, Japan, the Japan Society for the Promotion of Science (JSPS) Grant-in-Aid for Scientific Research (20H00180, 21K13953, 21H04467, 22H04939, 23H00131, 23KJ0728, 24H00245), the JSPS Core-to-Core Program (JPJSCCA20210003), and the JSPS International Leading Research (22K21349).
This work was supported by the joint research program of the Institute for Cosmic Ray Research (ICRR), University of Tokyo.
S.F. acknowledges the NASA Hubble Fellowship grant \#HST-HF2-51505.001-A awarded by the Space Telescope Science Institute, which is operated by the Association of Universities for Research in Astronomy, Incorporated, under NASA contract NAS5-26555.
PGP-G acknowledges support from grant PID2022-139567NB-I00 funded by Spanish Ministerio de Ciencia, Innovaci\'on y Universidades MCIU/AEI/10.13039/501100011033, FEDER {\it Una manera de hacer Europa}. 
J.S.D. thanks the Royal Society for support via a Royal Society Research Professorship

\software{Prospector \citep{2021ApJS..254...22J}, PypeIt \citep{pypeit:joss_pub,pypeit:zenodo}, SExtractor \citep{1996A&AS..117..393B}}

\end{acknowledgments}

\bibliography{apj-jour,reference}
\bibliographystyle{apj}


\end{document}